# SKYSURF. X. A Novel Method for Measuring Integrated Galaxy Light

Delondrae D. Carter [1], Timothy Carleton [1], Daniel J. Henningsen [1], Rogier A. Windhorst [1], Seth H. Cohen [1], Scott A. Tompkins [2], Rosalia O'Brien [1], Anton M. Koekemoer [3], Juno Li [4], Zak Goisman [1], Simon P. Driver [4], Aaron S. G. Robotham [4], Rolf A. Jansen [1], Norman A. Grogin [3], Sabine Bellstedt [2], Haina Huang [1], Tejovrash Acharya [1], Jessica M. Berkheimer [1], Haley Abate [1], Connor Gelb [1], Isabela Huckabee [1], and John MacKenty [3]

[1] *School of Earth and Space Exploration, Arizona State University, 1151 S. Forest Ave, Tempe, AZ 85287-1404*
[2] *The University of Western Australia, M468, 35 Stirling Highway, Crawley, WA 6009, Australia*
[3] *Space Telescope Science Institute, 3700 San Martin Drive, Baltimore, MD 21218*
[4] *International Centre for Radio Astronomy Research (ICRAR) and the International Space Centre (ISC), The University of Western Australia, M468, 35 Stirling Highway, Crawley, WA 6009, Australia*

## ABSTRACT

We describe the drizzling pipeline and contents of the drizzled database for Hubble Space Telescope Cycle 27-29 Archival Legacy project "SKYSURF," the largest archival project ever approved for Hubble. SKYSURF aims to investigate the extragalactic background light (EBL) using all 143,914 ACSWFC, WFC3UVIS, and WFC3IR images that have been taken by Hubble since its launch in 2002. SKYSURF has produced 38,027 single-visit mosaics and 7,893 multi-visit mosaics across 28 ACSWFC, WFC3UVIS, and WFC3IR filters using non-standard drizzling methods, which include preserving the lowest sky-level of each visit/group in the drizzled products, applying wider apertures for cosmic ray rejection, correcting effects caused by charge transfer efficiency (CTE) degradation, and removing potential light gradients from input images via sky-map subtraction. We generate source catalogs for all drizzled products with Source Extractor and provide updated star-galaxy separation parameters and integrated galaxy light (IGL) estimates for 25 of the 28 SKYSURF filters (wavelength range 0.2–1.7 $\mu$m) using a novel IGL fitting method made possible by the vast SKYSURF dataset. We discuss the data processing and data analysis challenges encountered, detail our solutions, and offer suggestions that may facilitate future large-scale IGL investigations with Webb, SPHEREx, and Roman.

## 1. INTRODUCTION

Measuring the total amount of light received by our telescopes is one of the most basic measurements we can do in astronomy. The total intensity of all light in the Universe across the entirety of the electromagnetic spectrum is referred to as the extragalactic background light (EBL). The EBL spectrum is divided into different regimes on the basis of wavelength, which span from the low-frequency photons of the cosmic radio background (CRB) to the high-frequency photons of the cosmic gamma-ray background (CGB) (Hill et al. 2018). Direct measurements of the EBL in any wavelength regime should match that of known sources, but despite many decades of research, different techniques arrive at different results in the optical and near-IR, suggesting that our census of galaxies in the Universe is incomplete (McVittie & Wyatt 1959; Partridge & Peebles 1967a,b; Hauser & Dwek 2001; Kashlinsky 2005; Lagache et al. 2005; Finke et al. 2010; Domínguez et al. 2011; Dwek & Krennrich 2013; Khaire & Srianand 2015; Saldana-Lopez et al. 2021). For an overview of the work done in this field, see Cooray (2016).

Hubble Space Telescope Cycle 27–29 Archival Legacy project "SKYSURF" (Windhorst et al. 2022; Carleton et al. 2022) aims to measure the cosmic optical and near-IR integrated galaxy light (IGL), a goal shared by several other missions such as CIBER (Matsuura et al. 2017; Korngut et al. 2022), Pioneer (Matsumoto et al. 2018), New Horizons (Lauer et al. 2021, 2022; Symons et al. 2023), and the recently launched NASA SPHEREx mission (Crill et al. 2020).

The cosmic optical and near-IR background are particularly important to our understanding of the Universe because they are intimately linked to rates of cosmic star formation, with theoretical models producing different EBL predictions based on competing star formation histories (D'Silva et al. 2023; Bellstedt et al. 2020; Driver et al. 2018; Madau & Dickinson 2014). Proposed star formation histories could theoretically be ruled out or confirmed by a collection of robust IGL measurements extracted from galaxy surveys, but this milestone has yet to be achieved due to limited datasets and the challenges encountered in measuring and subtracting astrophysical foregrounds (Cooray 2016).

The background light considered in EBL investigations can be broken up into six different sources: (1) zodiacal light ($ZL$), which is the light from dust within our Solar System, (2) diffuse Galactic light ($DGL$), which is the light within our Galaxy, (3) instrumental emission ($IE$) such as dark current, scattered light, and thermal emission (McIntyre et al. 2025), (4) geocoronal emission ($GE$), which is light emitted by solar UV radiation exciting hydrogen atoms in the outermost layer of the Earth's atmosphere, (5) a possible truly diffuse EBL component ($D_{EBL}$), such as that arising from decaying dark matter (Gong et al. 2016), and (6) integrated galaxy light ($IGL$), which is the area-normalized sum of all galaxy light. The IGL can be further divided into two types: (6a) resolved IGL ($R_{IGL}$), which is the total observed light emitted by all galaxies our telescopes detect, and (6b) extrapolated IGL ($E_{IGL}$), which is an estimate of the total light emitted by galaxies (including galaxies too faint for our telescopes to detect) obtained by using observed galaxy counts to extrapolate to fainter magnitudes. The sum of all six of these terms constitutes a measurement of the sky background ($BKG$) as observed from low Earth orbit:

$$BKG = ZL + DGL + IE + GE + D_{EBL} + IGL. \tag{1}$$

There has been controversy in studies of the cosmic optical and near-IR background, with some studies finding evidence of a truly diffuse background (Wright 2001; Bernstein 2007), while others (Postman et al. 2024; Gréaux et al. 2024) find that the background light is consistent with expectations from detected galaxies.

There are models based on measurements of COBE satellite data that provide estimates of the zodiacal light (Kelsall et al. 1998; Wright 1998) and diffuse Galactic light (Schlegel et al. 1998a). Robust sky background measurements of SKYSURF data were recently completed by O'Brien et al. (2023) (SKYSURF-4), who established robust near-IR EBL upper limits by producing updated zodiacal and diffuse light limits using WFC3IR. Other studies, such as McIntyre et al. (2025), investigate the influence of thermal emission and other instrumental effects on extragalactic background light measurements, and we refer the reader to Windhorst et al. (2022) and Carleton et al. (2022) for a deeper discussion of geocoronal emission in the context of Hubble data. This work focuses primarily on obtaining accurate IGL measurements and mitigating the influence of cosmic variance on such measurements.

An infinitely powerful telescope could, in principle, obtain reasonable direct measurements of the cosmic optical and near-IR EBL by observing every possible galaxy. In reality, however, telescopes are magnitude limited and will only detect a fraction of all galaxies in the Universe. Measurements of the resolved IGL therefore serve as lower limits of the EBL. A significant fraction of the IGL in the Universe is expected to be contained in the magnitude range $18 < m_{\rm AB} < 24$ mag, which is well within the magnitude range over which Hubble excels. In this paper, we quantitatively determine the influence of cosmic variance on IGL measurements by exploring how IGL varies across the thousands of fields in SKYSURF's database as a function of bright objects using a novel IGL fitting method.

This paper is a companion paper to Tompkins et al. (2025) (SKYSURF-9). Both this work and Tompkins et al. (2025) measure the IGL in several Hubble filters using identical drizzled datasets, but apply significantly different methods for object catalog generation, star-galaxy separation, and IGL measurement. The object catalog production of Tompkins et al. (2025) is performed with ProFound (Robotham et al. 2018a), while this work uses Source Extractor (Bertin, E. & Arnouts, S. 1996). While Tompkins et al. (2025) use relatively simple cuts in their star-galaxy separation, this work uses complex fitting techniques to algorithmically identify stellar loci and saturated tracks in object brightness versus size plots. Tompkins et al. (2025) obtain IGL estimates by spline fitting and integrating galaxy flux density as a function of magnitude, as has been done in works such as Driver et al. (2016) and Koushan et al. (2021), while this work applies a novel IGL fitting method to account for the variation in source density among different pointings. Last, in addition to the differing methodologies, this paper outlines the specifics of the SKYSURF image processing and drizzling pipeline in much greater detail.

Since surveys with large spatial volumes probe a more representative portion of the Universe and decrease measurement errors caused by cosmic variance–the variation from uniformity in the distribution of galaxies resulting from linear growth of structure–this makes the SKYSURF database an excellent resource for measuring the cosmic optical and near-IR IGL. SKYSURF contains a large quantity of deep data (to $m_{\rm AB} \lesssim 25-26$ mag on average, much deeper than any other large area survey) that span a wide wavelength range (0.2–1.7 $\mu$m) spread throughout $\gtrsim 1400$ independent fields that cover an area of $\sim 10$ square degrees (Windhorst et al. 2022). The analysis of Tompkins et al. (2025) determined that cosmic variance contributes a 2% error to SKYSURF IGL measurements. This work aims to analyze and account for the effects of cosmic variance on cosmic optical and near-IR IGL measurements using SKYSURF's massive database of 38,027 ACSWFC, WFC3UVIS, and WFC3IR SKYSURF single-visit drizzled (combined) products and 7,893 multi-visit products, which we produced from 143,914 Hubble FLT/FLC (calibrated, CTE-corrected/calibrated, non-CTE-corrected) frames.

Recent developments in the field of EBL studies position this work at a particularly interesting time. The gamma-ray EBL results of Gréaux et al. (2024), obtained from an analysis of over 260 archival TeV gamma-ray spectra from the STeVECat catalog, have resolved decades of tension between gamma-ray, direct, and galaxy counts based measurements in the optical band. They measure an EBL intensity of $6.9\pm1.9$ nW/m$^2$/sr at 600 nm, which is consistent with both IGL measurements and direct measurements taken from beyond Pluto's orbit. This "cosmological optical convergence," as they refer to it, marks a potentially significant milestone in our understanding of the cosmic optical background. Another study conducted by Postman et al. (2024) uses images captured by New Horizons LORRI to arrive at a COB intensity of $11.16 \pm 1.65$ nW/m$^2$/sr and an IGL intensity of $8.17 \pm 1.18$ nW/m$^2$/sr at 608 nm. Given that their non-IGL COB signal of $2.99 \pm 2.03$ nW/m$^2$/sr is consistent with zero, they conclude that galaxy light is likely entirely responsible for the intensity of the COB, which would further reinforce the need for accurate IGL measurements if true. The Postman et al. (2024) LORRI results coupled with the gamma-ray results of Gréaux et al. (2024) constitute two examples of an ongoing effort to harmonize different EBL measurement methods in recent years.

This paper is structured as follows. In Section 2, we detail the SKYSURF drizzling pipeline we developed to combine and process the 143,914 FLC/FLT images in the SKYSURF database. In Section 3, we discuss our Source Extractor object catalog generation for the resulting drizzled products and the star-galaxy separation methods we implemented to classify our detected objects. We also discuss our resulting number counts and compare our counts with GOOD-S (Giavalisco et al. 2004). In Section 4, we describe how we calculate the IGL for individual frames and the new IGL fitting method we developed to obtain an average IGL measurement for each filter. In Section 5, we present and discuss the results of our IGL cosmic variance analysis and provide suggestions for future large-scale IGL research. Last, we summarize key findings in Section 6. Supplemental technical details and additional supporting figures are provided in the Appendix. All magnitudes quoted in this work are measured using the AB system of Oke & Gunn (1983). Throughout, we refer to the Advanced Camera for Surveys Wide Field Channel as ACSWFC, the ultraviolet+visible channel of Wide Field Camera 3 as WFC3UVIS, and the infrared channel of Wide Field Camera 3 as WFC3IR.

## 2. THE SKYSURF DRIZZLED DATABASE

### 2.1. *Image Drizzling*

Image drizzling is a method of combining images that improves image resolution and corrects image defects, including cosmic rays, bad pixels, and geometric distortion (Fruchter & Hook 2002). For SKYSURF, we generated two types of drizzled products: (1) single-visit drizzled products (combining images located on the same tile of the sky), and (2) multi-visit drizzled products (combining partially overlapping images into one mosaic). Drizzled images are both cleaner and deeper than the individual frames that compose them, which is important for minimizing the number of spurious detections during object catalog generation. The development and execution of the SKYSURF drizzling pipeline was started in 2020, and the final drizzled products were finished in 2022. Figure 1 summarizes the SKYSURF drizzling pipeline described below, and all drizzled products are available on the SKYSURF MAST project page[1].

### 2.2. *Grouping Images*

Prior to drizzling, we sorted the FLC/FLT Hubble images in the SKYSURF database into drizzling groups. In the context of Hubble data, FLT images are calibrated images that have had bias, dark, and flat-field corrections applied, and are the finalized calibrated products of WFC3IR. FLC images have the same corrections, but are also corrected for charge transfer efficiency (CTE)–an additional step necessary for the CCD detectors of ACSWFC and WFC3UVIS. For the single-visit drizzling, we sorted these calibrated images into visits based on the first 6 letters of their "ipppssoot" [2] ID. Images whose first six ipppssoot characters are identical belong to the same visit, providing a convenient way to sort all SKYSURF images into single-visit drizzling groups. The multi-visit grouping process, in contrast, involved several steps. For the multi-visit data, single-visit mosaics were first grouped based on the WCS information extracted from their FITS headers. In particular, the convex hull[3] of all pixels in RADEC coordinates were taken as the vertices to define polygons that represented the footprints of each visit. If the polygon of one visit overlapped with the polygon of another visit, the two polygons were merged by determining the convex hull of their vertices. The resulting merged polygon would then be tested for overlap with the polygons of the remaining visits, merged with overlapping polygons into an even bigger polygon, and so on until no more overlapping polygons existed. After grouping visits in this way, we applied additional filtering criteria to exclude star clusters, fields targeting nearby bright galaxies, and fields contaminated by bright Galactic stars. Visits that satisfied one or more of the following criteria were removed from the multi-visit dataset:

1. Fields within ±10 degrees of the Galactic plane.
2. Fields within a 20 degree radius from the Galactic center.
3. Fields near known bright sources. Fields within 3× a bright source's diameter were removed, unless the field was separated from the bright source by more than 10 degrees. We created a merged list of known bright sources using two catalogs: (1) the Revised NGC/IC Catalog[4] and (2) RC3 – Third Reference Catalog of Bright Galaxies[5] (Corwin et al. 1994).

[1] https://archive.stsci.edu/hlsp/skysurf/

[2] https://archive.stsci.edu/hlsp/ipppssoot.html

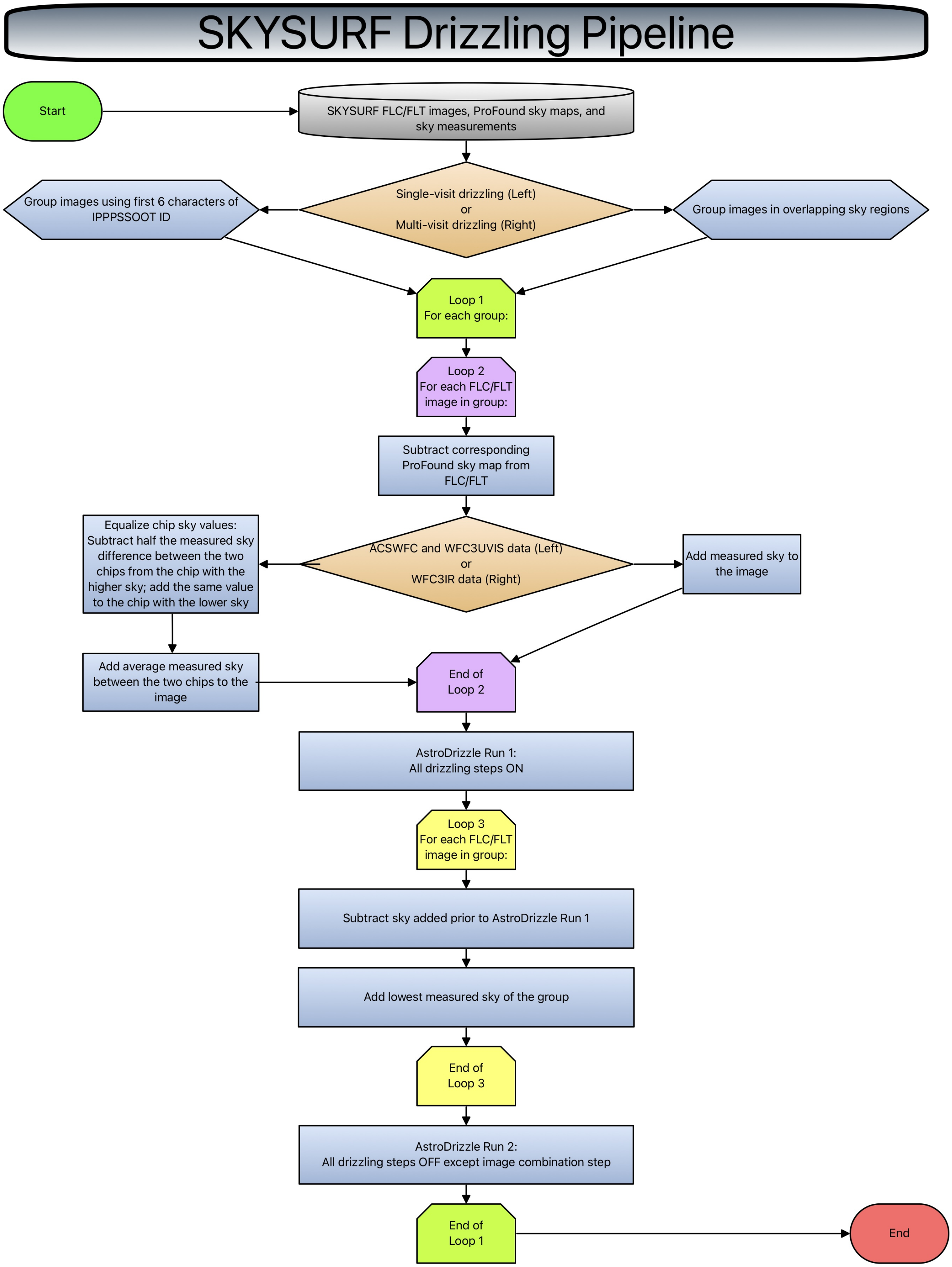

**Figure 1.** Flowchart of all steps performed by the SKYSURF drizzling pipeline.

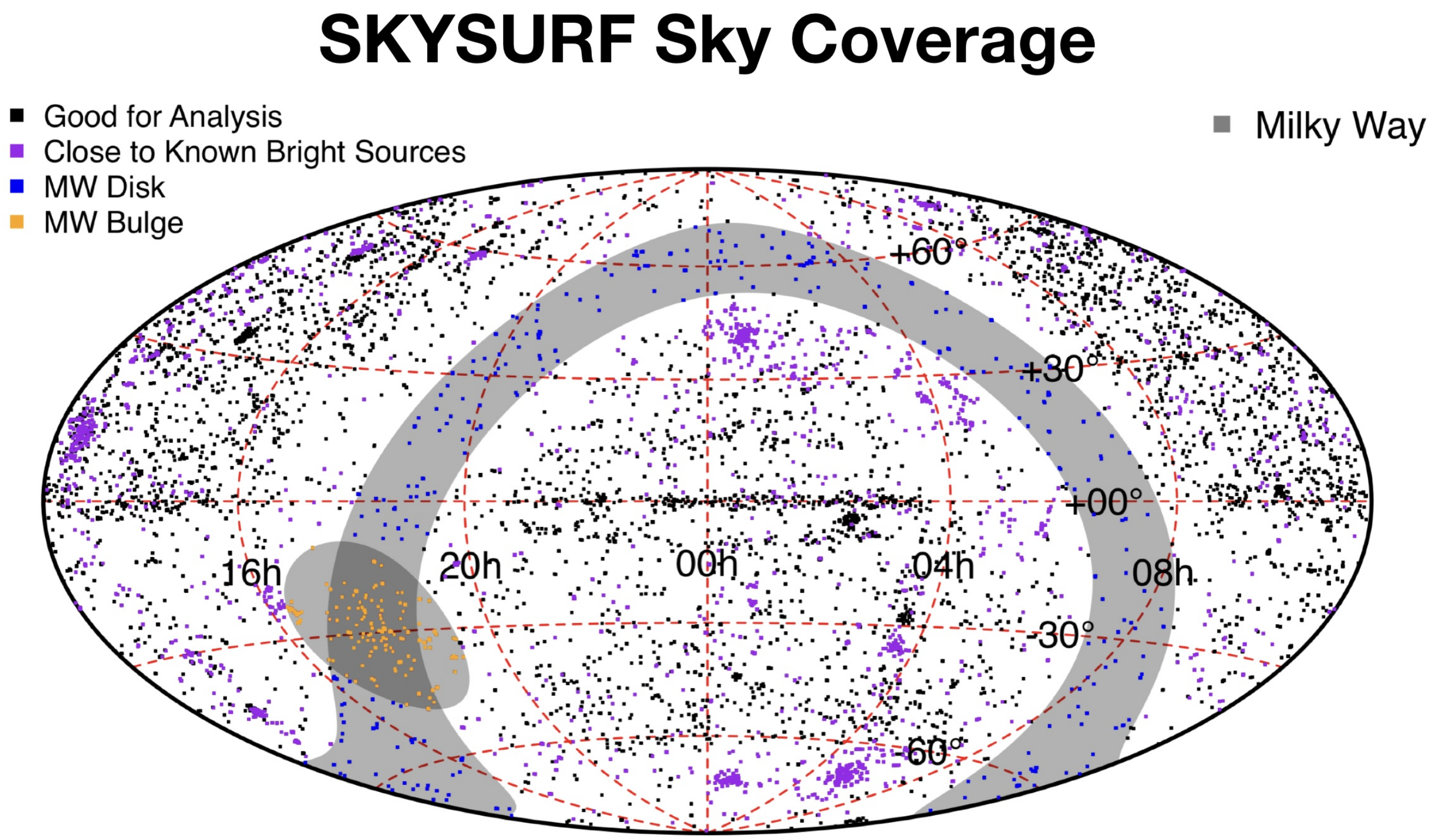

**Figure 2.** Sky coverage of SKYSURF's 38,027 single-visit mosaics. This is an Aitoff Equatorial projection. Each black point represents the location of one single-visit mosaic. The regions shaded grey represent the areas filtered out of the multi-visit analysis according to the filtering criteria discussed in Section 2.2. Visits excluded from the multi-visit analysis are indicated by colored points: purple points indicate that a visit was filtered out due to being too close to a known bright source, blue points indicate that a visit was filtered out due to its location relative to the Milky Way's disk, and orange points indicate that a visit was filtered out due to its location relative to the Milky Way's bulge.

Figure 2 illustrates all fields on the sky, along with those visits excluded for these reasons.

[3] The convex hull of a set of points is the smallest convex polygon that encloses those points.
[4] http://www.klima-luft.de/steinicke/ngcic/rev2000/Explan.htm
[5] https://heasarc.gsfc.nasa.gov/w3browse/all/rc3.html

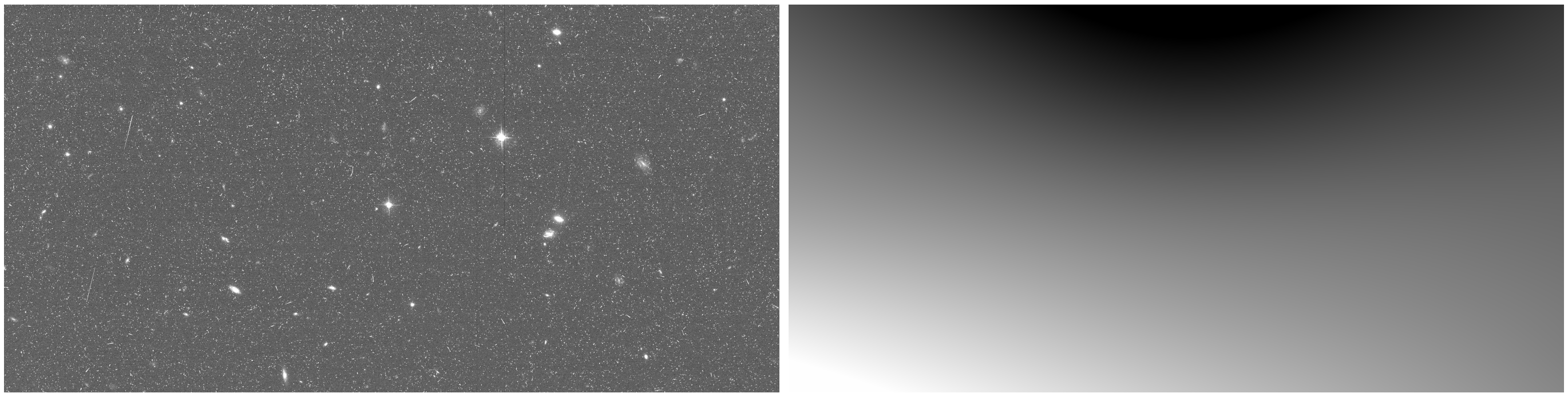

**Figure 3.** Chip 2 of FLC image jb3201hvq_flc.fits (**Left**), which was fed into the drizzling pipeline for single-visit product 67 of ACSWFC F606W, along with its corresponding sky-map generated using ProFound (**Right**). The sky-map shown here has pixel values that range from 55.13 $e^{-}$/s (darkest) to 56.81 $e^{-}$/s (brightest). In step 1 of the drizzling pipeline, each FLC/FLT image had its corresponding ProFound sky-map subtracted to remove potential light gradients.

### 2.3. *The SKYSURF Drizzling Pipeline*

The conventional drizzling pipeline typically subtracts the sky background from each input image prior to image combination, outputting a combined image with a sky background of zero. However, SKYSURF's science goals required a non-standard reprocessing of all HST ACSWFC, WFC3UVIS, and WFC3IR data that entailed producing drizzled products with the lowest estimated sky-level of each visit/group preserved in the combined images. Preserving the lowest estimated sky-level of each visit in the combined products is consistent with SKYSURF's philosophy that since most error sources are positive, the lowest sky-levels should be the most plausible measurements (Carleton et al. 2022). This particular aspect of the drizzling pipeline has no impact on the IGL measurements we discuss in this work since Source Extractor performs local background subtraction when measuring fluxes, but it is important for SKYSURF's diffuse light investigations.

All SKYSURF drizzling was performed using AstroDrizzle (Gonzaga et al. 2012), which is a Python implementation of MultiDrizzle (Koekemoer et al. 2003, 2011) that allows for flexible customization and tuning of all standard drizzling steps. The unconventional nature of SKYSURF's image processing required two separate AstroDrizzle runs for each visit/group: one run with all AstroDrizzle steps turned on for cosmic ray (CR) identification and rejection, and a subsequent run with only the final combination step turned on to produce a combined image containing the lowest sky-level of the visit/group. The details of each of these two runs are discussed in Sections 2.3.1 and 2.3.2, and SKYSURF's specified AstroDrizzle input parameters for runs 1 and 2 are explicitly enumerated in Appendix A. For thorough descriptions of all AstroDrizzle parameters, one may reference the DrizzlePac Handbook (Gonzaga et al. 2012) or the DrizzlePac readthedocs[6].

#### 2.3.1. *AstroDrizzle Run 1: Cosmic Ray Identification and Rejection*

Prior to AstroDrizzle run 1, we generated sky-maps for each FLC/FLT image in the SKYSURF database using ProFound[7] (Robotham et al. 2018a)–a publicly available image analysis package written in R. An example of an input FLC and its corresponding sky-map is shown in Figure 3. These ProFound sky-maps were produced to identify and remove potential light gradients by subtracting each sky-map from its corresponding FLC/FLT prior to drizzling. After sky-map subtraction, the flat measured sky background values detailed in O'Brien et al. (2023) were added to their respective FLCs/FLTs to ensure the input images had accurate sky statistics for each AstroDrizzle step. Since ACSWFC and WFC3UVIS have two sky measurements per image (one sky measurement for each chip), the sky-levels of the chips for these instruments were first equalized by subtracting half the measured sky difference between the two chips from the chip with the higher sky value, and adding half the sky difference to the chip with the lower sky value. After equalizing the chips, we then added the average measured sky between the chips to each input image prior to executing AstroDrizzle run 1[8].

[6] https://drizzlepac.readthedocs.io/en/latest/

[7] https://github.com/asgr/ProFound

[8] **Note:** AstroDrizzle run 1 was skipped for visits/groups containing only 1 input image

To optimize CR rejection and correct effects caused by Charge Transfer Efficiency (CTE) degradation, we applied non-standard wider apertures during AstroDrizzle's CR rejection step. The relevant AstroDrizzle parameters modified for this purpose are `driz_cr_grow` (the radius, in pixels, around cosmic rays within which more stringent CR rejection criteria are applied) and `driz_cr_ctegrow` (the assumed length, in pixels, of the CTE tail to mask). The default AstroDrizzle values for these parameters are `driz_cr_grow = 1` and `driz_cr_ctegrow = 0`. For SKYSURF, we used custom values of `driz_cr_grow = 2` for ACSWFC, WFC3UVIS, and WFC3IR. Increasing the `driz_cr_grow` parameter from its default value of 1 to a wider value of 2 means that pixels where CRs are identified have a surrounding margin of 2 pixels examined for additional flux from the CR, thereby producing cleaner images than the default margin of 1 pixel. Similarly, we used a custom value of `driz_cr_ctegrow = 5` for ACSWFC and WFC3UVIS, where this value enabled the rejection of CTE tails on CRs which can be significant up to this size scale (we used the default value of `driz_cr_ctegrow = 0` for WFC3IR since this detector is not susceptible to CTE effects). We chose all of these non-default values for these parameters after examining the effects of various different values on the data quality, through which we determined that these values provide the best quality data for the cleanest measurements of the sky backgrounds.

#### 2.3.2. *AstroDrizzle Run 2: Image Combination*

After AstroDrizzle run 1, we subtracted the flat sky-levels added prior to run 1 from each input image, then added the lowest measured sky-level of the visit/group to all images in the visit/group. We then performed a second AstroDrizzle run with all drizzling steps turned off except for the image combination step (`driz_combine = True`) to produce a a final combined science image, an inverse variance weight map (`final_wht_type = 'IVM'`), and a cosmic ray mask. We performed all drizzling using a classic square drizzling kernel (`driz_sep_kernel = 'square'` and `final_kernel = 'square'`) and drizzled all data to a pixel scale of 0.06″/pixel (`driz_sep_scale = 0.06` and `final_scale = 0.06`), which we considered a reasonable compromise between the slightly smaller native pixel scales of ACSWFC (0.050″/pixel) and WFC3UVIS (0.0395″/pixel) and the larger pixel scale of WFC3IR (0.135″/pixel). Last, since different AstroDrizzle image combination methods work better for different numbers of input exposures, we specified the `combine_type = 'minmed'` image combination method for visits/groups containing 6 or less input images, and the `combine_type = 'median'` combination method for visits/groups containing 7 or more input images. Example single-visit and multi-visit images are shown in Figure 4. As with the choice of CR rejection parameters, we chose these AstroDrizzle parameters after examining the effects of various different values, where we determined that these values enabled the most robust measurements to be obtained from the final combined images.

#### 2.3.3. *Image Filtering Criteria*

Our drizzling pipeline filtered out images considered "bad" prior to AstroDrizzle run 1 according to specific criteria. Our first cuts were made based on measurements obtained by O'Brien et al. (2023) using their "Per-Clip" method, which divides each FLC/FLT image into multiple square sub-regions ($64 \times 64$ pixel sub-regions for ACSWFC and WFC3UVIS; $39 \times 39$ pixel sub-regions for WFC3IR). O'Brien et al. (2023) flagged image sub-regions in which objects were detected (hereafter referred to as "object sub-regions") and sub-regions whose DQ arrays contained $> 30\%$ of flagged pixels (hereafter referred to as "bad pixel sub-regions") as unreliable for sky measurements. To avoid drizzling crowded fields or images containing an abnormally high amount of bad pixels, we excluded all images whose object sub-regions composed $\geq \frac{3}{4}$ of its total sub-regions or whose bad pixel sub-regions composed $\geq \frac{1}{3}$ of its total sub-regions.

The remaining image cuts were made based on the quality of the ProFound sky-maps and specific FITS header keywords. The quality of each ProFound sky-map can be estimated using the sky chi-squared ($\chi^2$) value that ProFound outputs. We required each input image to have a ProFound sky chi-squared value in the range $0.9 < \chi^2 < 1.1$ to be eligible for drizzling; inputs with $\chi^2$ values outside of this range were omitted from the drizzling process. Last, the FITS header of each input image contains certain keywords that inform how an exposure was taken and whether anything went wrong during a particular observation. To ensure the drizzling pipeline would only process images useful for SKYSURF's science goals, we required all drizzle input images to have the following header keyword values:

1. `EXPFLAG = 'NORMAL'`
2. `FGSLOCK = 'FINE'`
3. `MTFLAG = ''`

These header keyword requirements are essential for ensuring that the FLT/FLC exposures were completed successfully with no known issues (`EXPFLAG = 'NORMAL'`), that the telescope was locked in fine guidance mode for highest precision (`FGSLOCK = 'FINE'`), and that no momentum unload occurred during the exposure that could degrade the exposure's quality (`MTFLAG = ''`). The total number of candidate frames considered for drizzling, the total number of "good" frames that remained after image filtering, and the total number of mosaics produced by the drizzling pipeline for each filter are provided in Table 1.

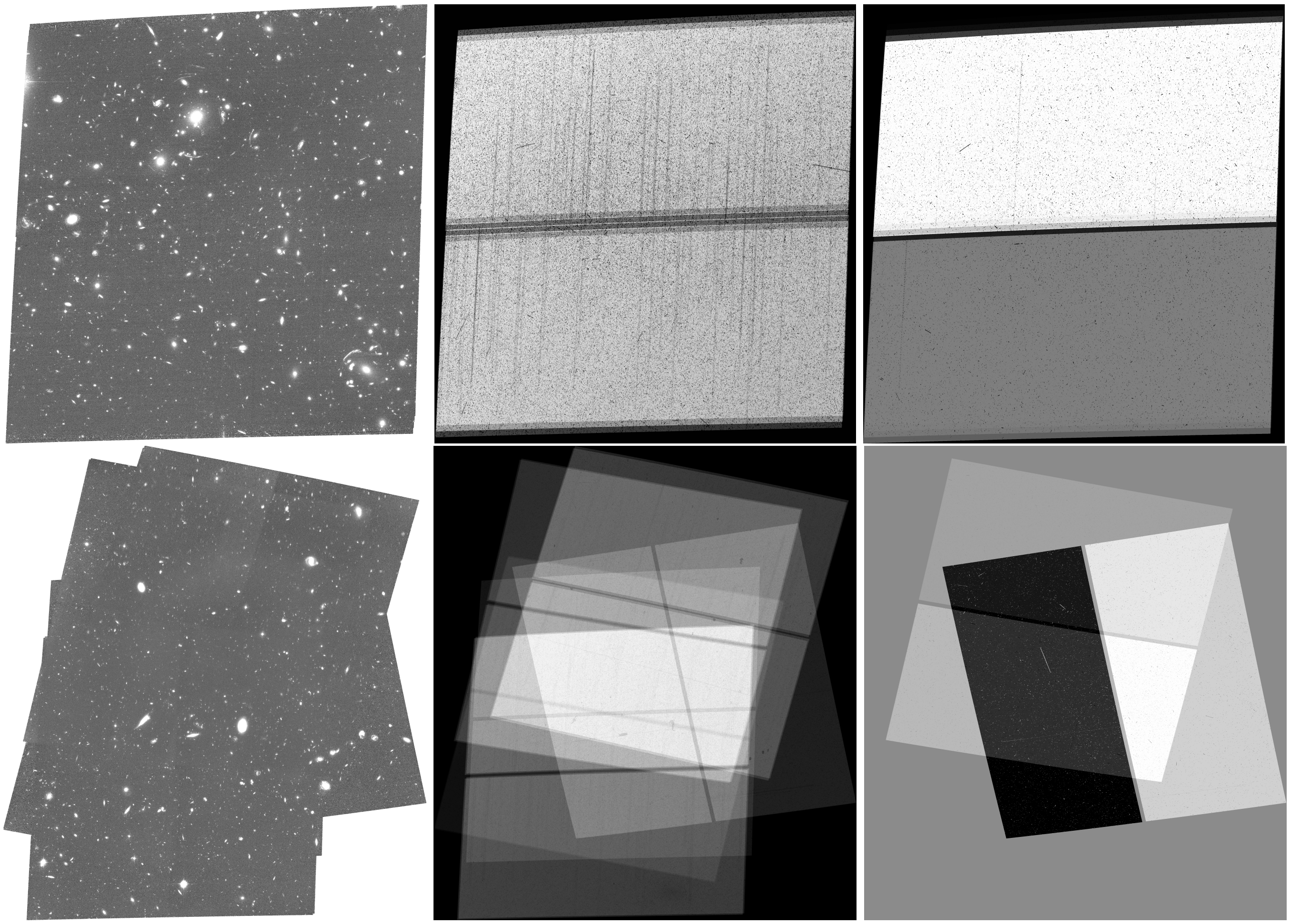

**Figure 4.** The single-visit drizzled image produced for visit 1124 of ACSWFC F606W (**Top Left**), along with its corresponding weight map (**Top Center**) and cosmic ray mask (**Top Right**). (**Bottom Row**) Drizzled products for multi-visit mosaic 28 of ACSWFC F606W.

#### 2.3.4. *SKYSURF FITS Header Keywords*

The SKYSURF drizzling pipeline populated the FITS headers of each drizzled product with several keywords relevant to the image processing described in Sections 2.3.1 and 2.3.2. These keywords, in addition to the detailed image processing notes added to each drizzled image's `HISTORY`, are intended to provide useful information pertaining to the specific sky-maps and sky values used in the processing of each drizzled product so that anyone working with the data can understand the pipeline's steps clearly enough to replicate it. These header keywords are listed below:

1. `SMAP[n]: The image name of the ProFound sky-map subtracted from the nth input image.`
2. `SKY[n]: (electrons/second) For ACSWFC and WFC3UVIS, the average sky value between the two chips of the nth input image. For WFC3IR, the measured sky value of the nth input image.`
3. `MIN_SKY: (electrons/second) Lowest measured sky of all input images in the visit or group.`
4. `DSKY[n]: (electrons) Absolute value of the difference between the sky values of the two chips of the nth ACSWFC or WFC3UVIS input image.`
5. `CSKY[n]: Extension number of the chip with the higher sky value for the nth ACSWFC or WFC3UVIS input image.`

**Table 1.** The number of candidate frames considered for drizzling, the number of "good" frames that remained after applying the image filtering criteria discussed in Section 2.3.3, and the number of mosaics produced for each of SKYSURF's 28 filters (SV = Single-Visit, MV = Multi-Visit).

| **Instrument Filter** | **Candidates** | **SV Good** | **SV Mosaics** | **MV Good** | **MV Mosaics** |
|---|---|---|---|---|---|
| ACSWFC F435W | 5461 | 4117 | 1159 | 1013 | 280 |
| ACSWFC F475W | 5417 | 4679 | 1622 | 862 | 242 |
| ACSWFC F555W | 2317 | 1579 | 507 | 368 | 80 |
| ACSWFC F606W | 15990 | 12841 | 3932 | 3324 | 607 |
| ACSWFC F625W | 1479 | 1294 | 427 | 755 | 200 |
| ACSWFC F775W | 8675 | 7565 | 3215 | 2527 | 594 |
| ACSWFC F814W | 27536 | 22726 | 6885 | 2723 | 357 |
| ACSWFC F850LP | 8586 | 7228 | 2257 | 1515 | 291 |
| WFC3UVIS F225W | 1126 | 847 | 221 | 370 | 91 |
| WFC3UVIS F275W | 3975 | 3078 | 1152 | 404 | 94 |
| WFC3UVIS F300X | 141 | 73 | 34 | 49 | 19 |
| WFC3UVIS F336W | 3999 | 3283 | 1149 | 619 | 115 |
| WFC3UVIS F390W | 912 | 728 | 212 | 494 | 116 |
| WFC3UVIS F438W | 1009 | 729 | 220 | 212 | 44 |
| WFC3UVIS F475W | 905 | 777 | 298 | 292 | 76 |
| WFC3UVIS F475X | 309 | 201 | 82 | 112 | 49 |
| WFC3UVIS F555W | 1334 | 811 | 294 | 99 | 28 |
| WFC3UVIS F606W | 5484 | 2837 | 876 | 1383 | 319 |
| WFC3UVIS F625W | 425 | 350 | 90 | 102 | 25 |
| WFC3UVIS F775W | 279 | 152 | 66 | 83 | 33 |
| WFC3UVIS F814W | 6467 | 3699 | 1302 | 1487 | 387 |
| WFC3UVIS F850LP | 192 | 132 | 50 | 48 | 15 |
| WFC3IR F098M | 1103 | 793 | 472 | 503 | 119 |
| WFC3IR F105W | 4792 | 3641 | 1574 | 1768 | 444 |
| WFC3IR F110W | 6473 | 3720 | 1767 | 2657 | 632 |
| WFC3IR F125W | 5554 | 4074 | 2055 | 1925 | 525 |
| WFC3IR F140W | 4691 | 3587 | 1630 | 2164 | 687 |
| WFC3IR F160W | 19283 | 12109 | 4479 | 6629 | 1424 |
| **ACSWFC** | 75461 | 62029 | 20004 | 13087 | 2651 |
| **WFC3UVIS** | 26557 | 17697 | 6046 | 5754 | 1411 |
| **WFC3IR** | 41896 | 27924 | 11977 | 15646 | 3831 |
| **Total** | 143914 | 107650 | 38027 | 34487 | 7893 |

### 2.4. *Contents of the SKYSURF Drizzled Database*

In total, we produced 38,027 single-visit drizzled products and 7,893 multi-visit drizzled products across SKYSURF's 28 ACSWFC, WFC3UVIS, and WFC3IR filters using the drizzling pipeline described in Section 2.3. Table 1 provides a detailed breakdown of the number of FLC/FLT inputs and drizzled outputs for each SKYSURF filter. The single-visit drizzled mosaics were produced using a total of 107,650 FLC/FLT images (out of 143,914 candidates) that met all of the filtering criteria discussed in Section 2.3.3, while the multi-visit drizzled mosaics were produced using 34,487 FLC/FLT frames that met the more stringent filtering criteria. On average, 3.1, 2.93, and 2.3 input frames were used to produce each ACSWFC, WFC3UVIS, and WFC3IR single-visit mosaic, respectively; averages of 4.94, 4.08, and 4.08 input frames were used to produce each ACSWFC, WFC3UVIS, and WFC3IR multi-visit mosaic. The total area coverage of all drizzled products used for our IGL analysis is provided for each filter in Table 2. Examples of typical single-visit and multi-visit products (science image, weight map, and cosmic ray mask) are shown in Figure 4.

## 3. THE SKYSURF OBJECT IDENTIFICATION PIPELINE

### 3.1. *Source Catalog Generation*

The first step in measuring the influence of cosmic variance on IGL measurements is identifying and cataloging all of the objects (stars and galaxies) contained in each of the drizzled products. To do this, we used Source Extractor (Bertin, E. & Arnouts, S. 1996), which obtains magnitude measurements using a Kron factor (Kron 1980) of 2.5. Source Extractor denotes this specific method of measuring magnitudes as `MAG_AUTO`. There are a number of different ways magnitudes can be measured, and for a detailed overview we refer the reader to Graham & Driver (2005), but we chose to use Source Extractor's `MAG_AUTO` measurements because of their widespread use in other galaxy surveys, which facilitates comparisons with external results.

**Table 2.** The initial number of mosaics available and the final number of mosaics that met the filtering criteria discussed in Section 3.2.1 for each SKYSURF filter (SV = Single-Visit, MV = Multi-Visit). The area columns provided here indicate the total combined area (in $deg^2$) of all mosaics that satisfied all filtering criteria.

| Instrument Filter | SV Initial | SV Filtered | SV Area | MV Initial | MV Filtered | MV Area |
|---|---|---|---|---|---|---|
| ACSWFC F435W | 1159 | 769 | 2.332 | 280 | 212 | 0.599 |
| ACSWFC F475W | 1622 | 637 | 1.928 | 242 | 205 | 0.597 |
| ACSWFC F555W | 507 | 211 | 0.648 | 80 | 74 | 0.225 |
| ACSWFC F606W | 3932 | 2443 | 7.404 | 607 | 539 | 1.501 |
| ACSWFC F625W | 427 | 277 | 0.832 | 200 | 187 | 0.516 |
| ACSWFC F775W | 3215 | 1637 | 4.886 | 594 | 549 | 1.486 |
| ACSWFC F814W | 6885 | 4167 | 12.589 | 357 | 305 | 0.827 |
| ACSWFC F850LP | 2257 | 1824 | 5.577 | 291 | 281 | 0.710 |
| WFC3IR F098M | 472 | 104 | 0.139 | 119 | 107 | 0.147 |
| WFC3IR F105W | 1574 | 899 | 1.186 | 444 | 398 | 0.660 |
| WFC3IR F110W | 1767 | 859 | 1.145 | 632 | 605 | 0.844 |
| WFC3IR F125W | 2055 | 1044 | 1.417 | 525 | 440 | 0.783 |
| WFC3IR F140W | 1630 | 631 | 0.838 | 687 | 515 | 0.835 |
| WFC3IR F160W | 4479 | 1698 | 2.740 | 1424 | 983 | 1.530 |
| WFC3UVIS F336W | 1149 | 531 | 1.065 | 115 | 113 | 0.225 |
| WFC3UVIS F390W | 212 | 182 | 0.368 | 116 | 115 | 0.226 |
| WFC3UVIS F438W | 220 | 100 | 0.198 | 44 | 43 | 0.082 |
| WFC3UVIS F475W | 298 | 192 | 0.387 | 76 | 74 | 0.146 |
| WFC3UVIS F475X | 82 | 64 | 0.130 | 49 | 41 | 0.083 |
| WFC3UVIS F555W | 294 | 170 | 0.341 | 28 | 25 | 0.049 |
| WFC3UVIS F606W | 876 | 613 | 1.237 | 319 | 291 | 0.570 |
| WFC3UVIS F625W | 90 | 48 | 0.094 | 25 | 22 | 0.043 |
| WFC3UVIS F775W | 66 | 35 | 0.070 | 33 | 28 | 0.055 |
| WFC3UVIS F814W | 1302 | 864 | 1.738 | 387 | 358 | 0.699 |
| WFC3UVIS F850LP | 50 | 45 | 0.090 | 15 | 14 | 0.025 |
| **Total** | 36620 | 21155 | 50.879 | 7689 | 6947 | 14.098 |

We performed test runs of Source Extractor on several drizzled images to determine the ideal Source Extractor object identification parameters and the pipeline steps that needed to be implemented before running the source catalog generation pipeline on all drizzled products in the SKYSURF database. The most relevant Source Extractor configuration parameters used for SKYSURF include:

- `DETECT_THRESH 1.5`
- `ANALYSIS_THRESH 1.5`
- `DETECT_MINAREA 4.0`
- `FILTER_NAME gauss_3.0_5x5.conv # For all filters except WFC3UVIS F336W; gauss_6.0_13x13.conv for WFC3UVIS F336W`
- `DEBLEND_MINCONT 0.06 # For ACSWFC and WFC3IR; 0.1 for WFC3UVIS`
- `GAIN 2.0 # For ACSWFC; 2.5 for WFC3IR; 1.6 for WFC3UVIS`

Preliminary testing on single-visit mosaics revealed a common object detection problem in drizzled frames whose input images did not overlap significantly at the outskirts of the mosaics. Because of how these particular drizzled frames were combined, their outskirts contained significantly more noise than their central regions, which Source Extractor spuriously detected as objects. This issue was even more pronounced for the multi-visit mosaics, many of which had widely varying noise levels across a single mosaic due to varying degrees of overlap among input images. To address this problem, we modified the weight images fed into Source Extractor in a manner such that only the deepest regions of these mosaics were considered "good" for source extraction. In practice, this meant identifying the regions with the highest pixel weight values and setting all other pixel weight values to zero.

We used a "double grid technique" to do this that was inspired by methods used in the Per-Clip sky measurement technique of O'Brien et al. (2023). To produce each modified weight map, we first divided the pixels of the original weight map into a grid composed of $128 \times 128$ pixel sub-regions. If a mosaic did not divide evenly into $128 \times 128$ pixel sub-regions, additional rows and/or columns of zero weight pixels were temporarily appended to the bottom and right sides of the mosaic to make the division possible. We calculated the mean weight of each $128 \times 128$ pixel sub-region and set 90% of the highest sub-region mean as the threshold weight value for the image. We then divided the pixels of the weight map into a grid of $8 \times 8$ pixel sub-regions. If an $8 \times 8$ pixel sub-region contained no pixel weight values above the threshold weight value, the sub-region was considered "bad" and all pixel weight values within the sub-region were set to zero; if an $8 \times 8$ pixel sub-region contained at least one pixel value above the threshold weight value, the sub-region was considered "good" and all pixels within the sub-region maintained their original values. The grid of $128 \times 128$ pixel sub-regions enabled a robust determination of the weight value of the deepest and least noisy region within each mosaic, while the grid of $8 \times 8$ pixel sub-regions allowed pixels outside the deepest portion of each mosaic to be filtered out without creating too many "holes" of zero weight within the deepest regions themselves. Examples of these modified weight maps are shown in Figure 5.

We applied this weight map modification method to ACSWFC and WFC3UVIS, but refrained from applying it to WFC3IR. The WFC3IR weight maps contain multiple "blobs"[9] that interfered significantly with the weight map modification method, resulting in some objects having holes of zero weight that artificially reduced flux measurements. For WFC3IR, the original weight maps were used for both detection and measurement when fed into Source Extractor.

Although the modified weight maps removed the spurious detections caused by noise variations across an image, there was an additional group of spurious detections at regions within weight maps where pixel weights abruptly transition from zero to non-zero values (e.g. image boundaries, the empty space between ACSWFC and WFC3UVIS chips, etc.). In plots of object magnitude versus size, these spurious detections form a feature that looks like a diffuse spout that extends well beyond a filter's expected completeness limit (e.g. the magenta objects in the left plot of Figure 6). This spout of objects extends into the main population of galaxies in number density plots, as can be seen in the right plot of Figure 6. To remove these spurious detections, we used Source Extractor's XMIN_IMAGE, XMAX_IMAGE, YMIN_IMAGE, and YMAX_IMAGE measurements to form a box around each detected object in its corresponding modified weight map. We then measured the fraction of non-zero weight pixels within the box and required boxes to contain a non-zero weight pixel fraction of $> 90\%$ for the object to be considered "good." Sources with a non-zero weight pixel fraction of $\leq 90\%$ were flagged as "bad" and omitted from future analysis. Figure 7 illustrates our object identification procedure on an example image.

### 3.2. *Star-Galaxy Separation*

#### 3.2.1. *Removing Crowded Fields, Incomplete Fields, and Other Problematic Mosaics*

After generating source catalogs for all single-visit and multi-visit mosaics, we analyzed the data on a per filter basis. First, we established a set of filtering criteria to determine which source catalogs were scientifically useful. Although we generated source catalogs for all drizzled mosaics, not all mosaics were useful for our IGL analysis. Although most crowded fields were filtered out during the execution of the drizzling pipeline by the image filtering criteria discussed in Section 2.3.3, a handful of star clusters and other problematic fields still made it through. To identify and remove these problematic fields, we counted the number of objects per square arcminute ($N_{obj/arcmin2}$) between $24 < m_{\mathrm{AB}} < 25$ mag and the number of stars per square arcminute ($N_{stars/arcmin2}$) for every source catalog in a given filter, then created histograms of the results to determine acceptable $N_{obj/arcmin2}$ and $N_{stars/arcmin2}$ cutoff values (see Figure 8). The cutoff values we selected were $N_{obj/arcmin2} = 97.22$ and $N_{stars/arcmin2} = 27.78$. Images with $N_{obj/arcmin2}$ and $N_{stars/arcmin2}$ values above these cutoffs were considered outliers and removed from our analysis, as the majority of these were confirmed to be known galaxy clusters or star clusters via visual inspection. This was sufficient to remove several crowded fields that were polluting the dataset. In addition to these cuts, we also automatically rejected all drizzled mosaics that contained only one input frame because at least two input images are required for adequate cosmic ray rejection. Last, we required each mosaic to have an IGL measurement greater than zero. We specified this criterion because there were a handful of outlier mosaics for which no galaxies were detected using the star-galaxy separation parameters discussed below, which resulted in their IGL measurements being exactly zero (to clarify, there were no negative IGL measurements; see Section 4 for how we calculate IGL for each mosaic). After visual inspection, we determined that these zero IGL mosaics were also crowded star and galaxy clusters that evaded previous filtering criteria and for which our Source Extractor settings performed especially poorly.

[9] https://hst-docs.stsci.edu/wfc3dhb/chapter-7-wfc3-ir-sources-of-error/7-5-blobs

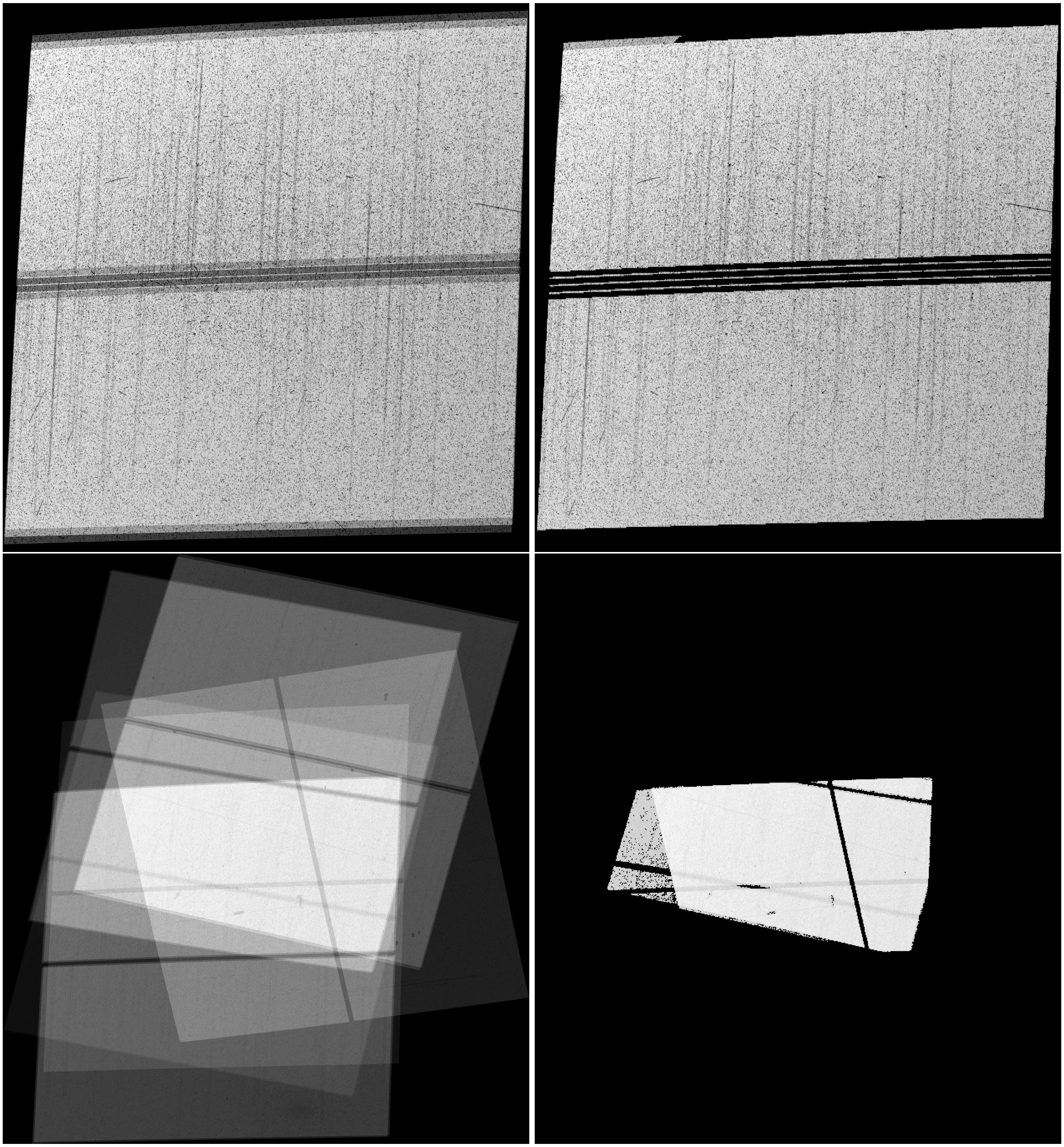

**Figure 5.** The original weight map (**Top Left**) input to Source Extractor as the "measurement" weight map for single-visit mosaic 1124 of ACSWFC F606W, alongside the modified weight map (**Top Right**) used as the "detection" weight map. (**Bottom Left and Bottom Right**) The original weight map and modified weight map for multi-visit mosaic 28 of ACSWFC F606W. The modified weight map captures the deepest regions of the multi-visit mosaics, effectively removing spurious detections in the multi-visit data caused by variations in noise levels across the larger mosaics. This method was not applied to WFC3IR due to the presence of "IR blobs" in the WFC3IR weight maps.

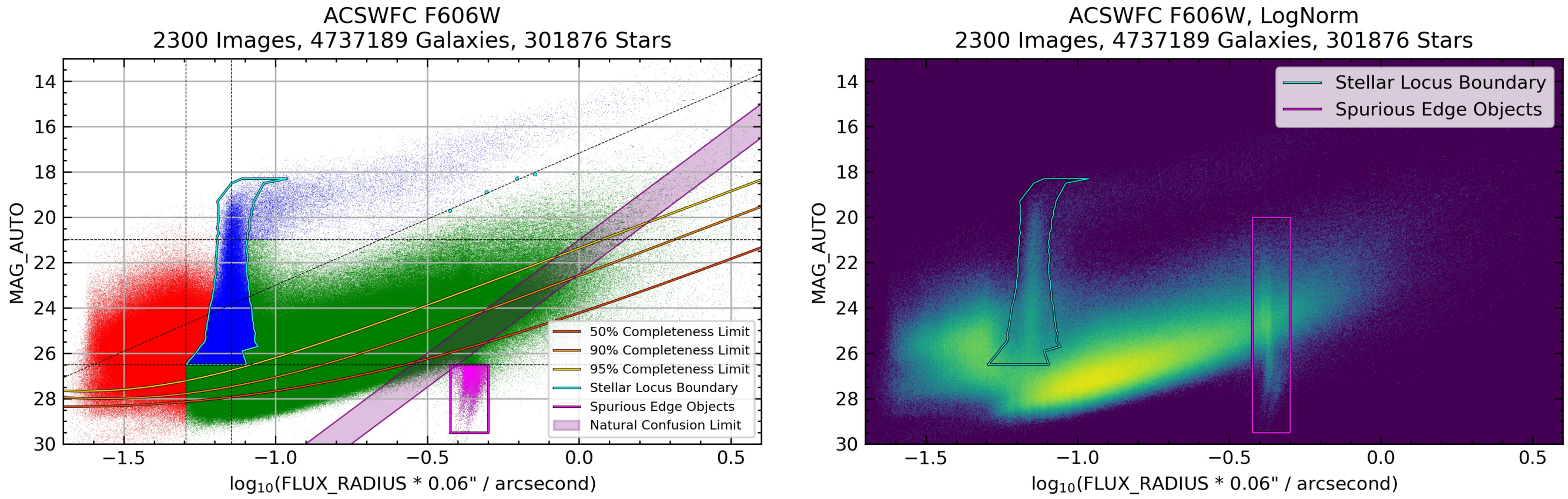

**Figure 6.** (**Left**) Magnitude ($m_{\rm AB}$) versus $\log_{10}$(50% FLUX_RADIUS) for all single-visit sources detected by Source Extractor in ACSWFC F606W with the modified weight map methods discussed in Section 3.1 applied. The dashed red, orange, and yellow lines are the independently modeled ACSWFC F606W 50%, 90%, and 95% completeness limits developed by Goisman et al. (2025) in SKYSURF-8. The purple band is the ACSWFC F606W natural confusion limit from Figure 4 of Kramer et al. (2022). The color scheme used here is: (**Red**) objects classified as noise and filtered out of the analysis due to their unrealistically small size; (**Magenta**) a separate class of noisy objects spuriously detected by Source Extractor at image edges; (**Blue**) stars filtered out of the analysis, which includes the stellar locus and the track of saturated stars; (**Green**) galaxies. (**Right**) A 2D hexagonal bin histogram of the data. The values of each bin are scaled logarithmically. The magenta box indicates the approximate region within which spurious edge detections are located.

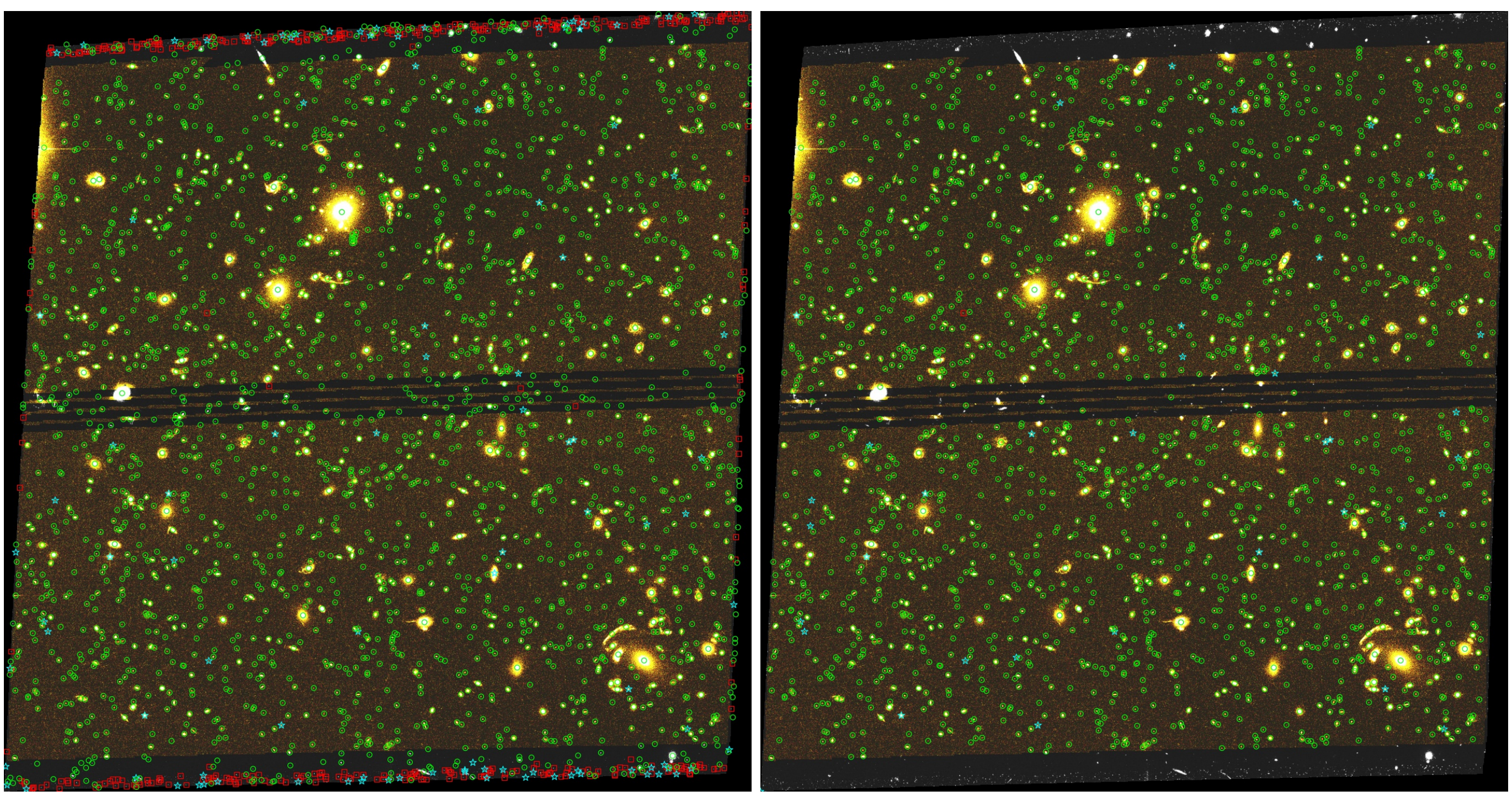

**Figure 7.** Single-visit Source Extractor detections for ACSWFC F606W visit 1124 using the original weight map as the detection image (**Left**) versus the modified weight map as the detection image with the double grid object filtering technique described in Section 3.1 applied (**Right**). Green circles represent galaxies, blue stars represent stars, and red squares represent spurious objects identified through the star-galaxy separation process described in Section 3.2. The regions of nonzero weight in the modified weight map are highlighted orange. The modified weight map coupled with the double grid filtering technique results in significantly less spurious noise detections at the outskirts of drizzled images, most of which were not initially classified as spurious by the star-galaxy separation process for this particular visit.

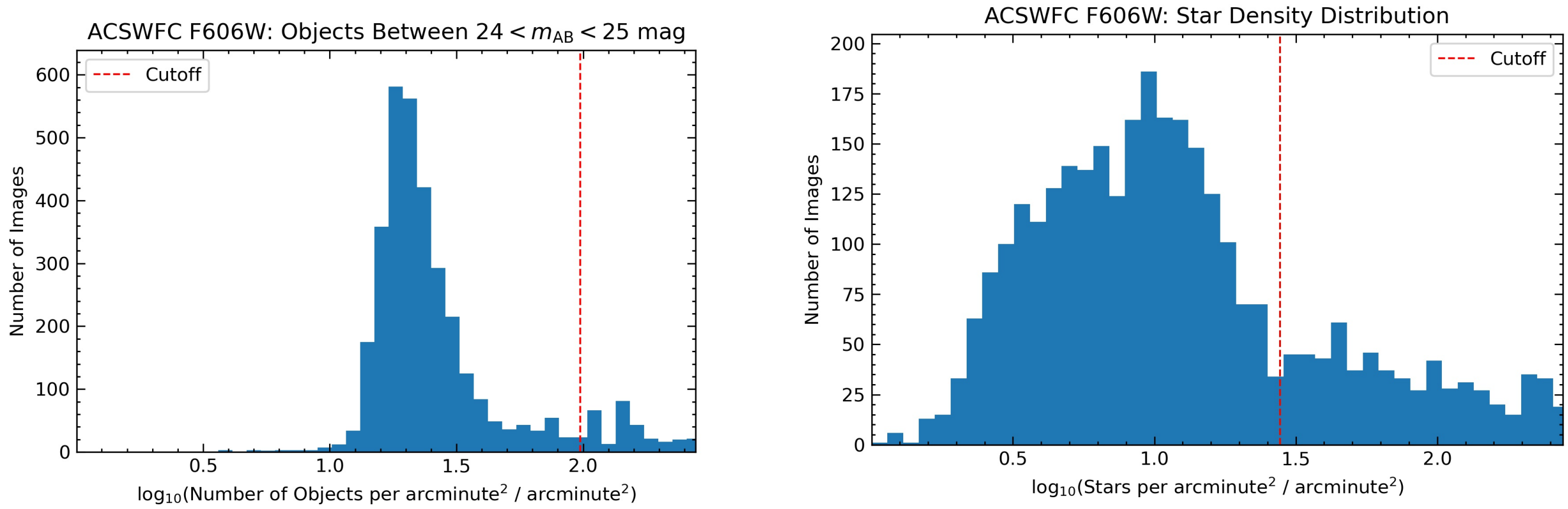

**Figure 8.** (**Left**) Histogram of objects per square arcminute ($N_{obj/arcmin2}$) for objects between $24 < m_{\rm AB} < 25$ mag for all ACSWFC F606W single-visit mosaics. The dashed red line at $N_{obj/arcmin2} = 97.22$ indicates the cutoff used to filter out star clusters and other crowded fields that made it through the drizzling pipeline's filtering criteria. For all SKYSURF filters, all drizzled products with $N_{obj/arcmin2} > 97.22$ were removed from analysis. (**Right**) Histogram of stars per square arcminute ($N_{stars/arcmin2}$) for all ACSWFC F606W single-visit mosaics. The dashed red line at $N_{stars/arcmin2} = 27.78$ indicates an additional cutoff used to filter out star clusters and crowded fields. For all SKYSURF filters, all drizzled products with $N_{stars/arcmin2} > 27.78$ were removed from analysis.

In addition to the filtering criteria above, we also performed a completeness analysis of each mosaic using the recently published Hubble completeness calculator of Goisman et al. (2025)[10], which can estimate the completeness of a Hubble image for an object of a specified magnitude and size from the image's background and exposure time. To measure the background of each drizzled image, we first masked all regions of zero weight from the corresponding science images, then used `astropy.stats.sigma_clipped_stats` with a sigma clip value of 3.0 to obtain median background measurements of the remaining regions. We used these background measurements in conjunction with the FITS header `EXPTIME` values to estimate the completeness of each mosaic for a $0.1''$ 25 mag object. To ensure adequate completeness, we removed any image less than 90% complete at $m_{\rm AB} = 25$ mag from our analysis. This had very little impact on the ACSWFC and WFC3UVIS datasets (and in many cases no impact). The impact on the WFC3IR datasets was more pronounced, but a sufficiently large number of WFC3IR mosaics remained in each filter for analysis after the omissions. The only filter for which these omissions are problematic is WFC3UVIS F850LP, which only contained images 90% complete at $m_{\rm AB} = 24$ mag. Therefore, we refrained from applying this criterion to WFC3UVIS F850LP, and our WFC3UVIS F850LP measurements should be considered with this caveat in mind. The numbers of candidate drizzled frames considered for analysis and the numbers of drizzled frames that remained after applying all criteria described in this section are provided in Table 2.

#### 3.2.2. *Filtering out Spurious Objects*

In addition to the spurious sources discussed in Section 3.2.5, Source Extractor detected several other objects that were obviously nonphysical, such as sources with negative size measurements. Before performing any of the star-galaxy separation methods described here, we required objects to meet all of the following criteria:

1. Magnitude in the range $0 <$ MAG_AUTO $< 99$.
2. Size measurement greater than 0 (50% FLUX_RADIUS $> 0$ and FWHM_IMAGE $> 0$).

#### 3.2.3. *Identifying the Stellar Locus*

After filtering out fields using the above criteria, we merged the catalogs of all images in a given filter into a single combined catalog. We treated the single-visit and multi-visit data separately (i.e. we produced both a single-visit combined catalog and a multi-visit combined catalog for each filter and analyzed the datasets separately). We then applied novel star-galaxy separation techniques to the data in the combined catalogs. We call the first of these techniques the stellar locus Gaussian fitting method, which is an algorithm designed to identify all stars contained within the stellar locus of a magnitude versus size plot. The steps of this algorithm are as follows:

1. Choose an upper and lower bound for object size (in units of $\log_{10}$(50% FLUX_RADIUS/arcsec)) that contains the stellar locus. For ACSWFC and WFC3UVIS, we selected a lower bound of -1.25 dex and an upper bound of -1 dex. For WFC3IR, we used a lower bound of -1.25 dex and an upper bound of -0.8 dex.

2. Choose an upper and lower bound for magnitude that contains the stellar locus. For ACSWFC and WFC3UVIS, we chose a lower bound of 18 mag and an upper bound of 28 mag. For WFC3IR, we chose a lower bound of 17 mag and an upper bound of 28 mag.

3. Bin the objects within the bounds by magnitude. We determined magnitude bin sizes for each filter based on the number of objects contained in each combined catalog. Combined catalogs containing 200000 objects or more were assigned a magnitude bin size of 0.2 mag, whereas objects containing less than 200000 objects were assigned a magnitude bin size of 0.5 mag.

4. For each magnitude bin, create a histogram of the sizes of the objects contained within the bin. We used a bin size of 0.005 dex (units of $\log_{10}$(50% FLUX_RADIUS/arcsec)) for all filters.

5. Fit a Gaussian to each histogram. For our Gaussian fitting, we used the `norm` method contained within `scipy.stats` in conjunction with the `curve_fit` method of `scipy.optimize` (Virtanen et al. 2020). We used the `norm` method to obtain an initial estimate of the standard deviation ($\sigma_{estimate}$) of the Gaussian fit and used the location of the maximum value of the histogram as an initial estimate of the mean ($\mu_{estimate}$)[11]. Then we used `curve_fit` to fit a Gaussian function to all objects with 50% FLUX_RADIUS values contained within $\mu_{estimate} - 2\sigma_{estimate} \leq$ 50% FLUX_RADIUS $\leq \mu_{estimate} + 0.5\sigma_{estimate}$, resulting in a more rigorously calculated mean ($\mu_{final}$) and standard deviation ($\sigma_{final}$).

6. Define the limit at which stars become indistinguishable from galaxies by setting a threshold value for the standard deviations of the Gaussian fits ($\sigma_{threshold}$). We determined via visual inspection of the histograms created in the previous step that stars become indistinguishable from galaxies in all SKYSURF Hubble filters when standard deviations increase beyond $\sigma_{threshold} = 0.06$ dex for magnitude bins fainter than 22 mag.

7. Create a polygon that encloses the stellar locus using the means and standard deviations calculated for each magnitude bin with $\sigma_{final} \leq \sigma_{threshold}$. We obtained the points of our stellar locus polygons for each filter by first calculating the midpoints of each magnitude bin to use as the y-coordinates for each bin. We calculated the x-coordinate of the left-hand side of each magnitude bin using $\mu_{final} - 2.5\sigma_{final}$ and the x-coordinate of the right-hand side of each magnitude bin using $\mu_{final} + 2.5\sigma_{final}$ when $\sigma_{final} < 0.04$ dex; for $\sigma_{final} \geq 0.04$ dex, we used $\mu_{final} + 1\sigma_{final}$ for the right-hand side[12]. By connecting the dots that result from this, a polygon is formed that encloses stars contained within the stellar locus. This polygon is plotted in cyan in Figure 9.

[10] http://skysurf.asu.edu/Completeness/completeness_calculator_gui.html

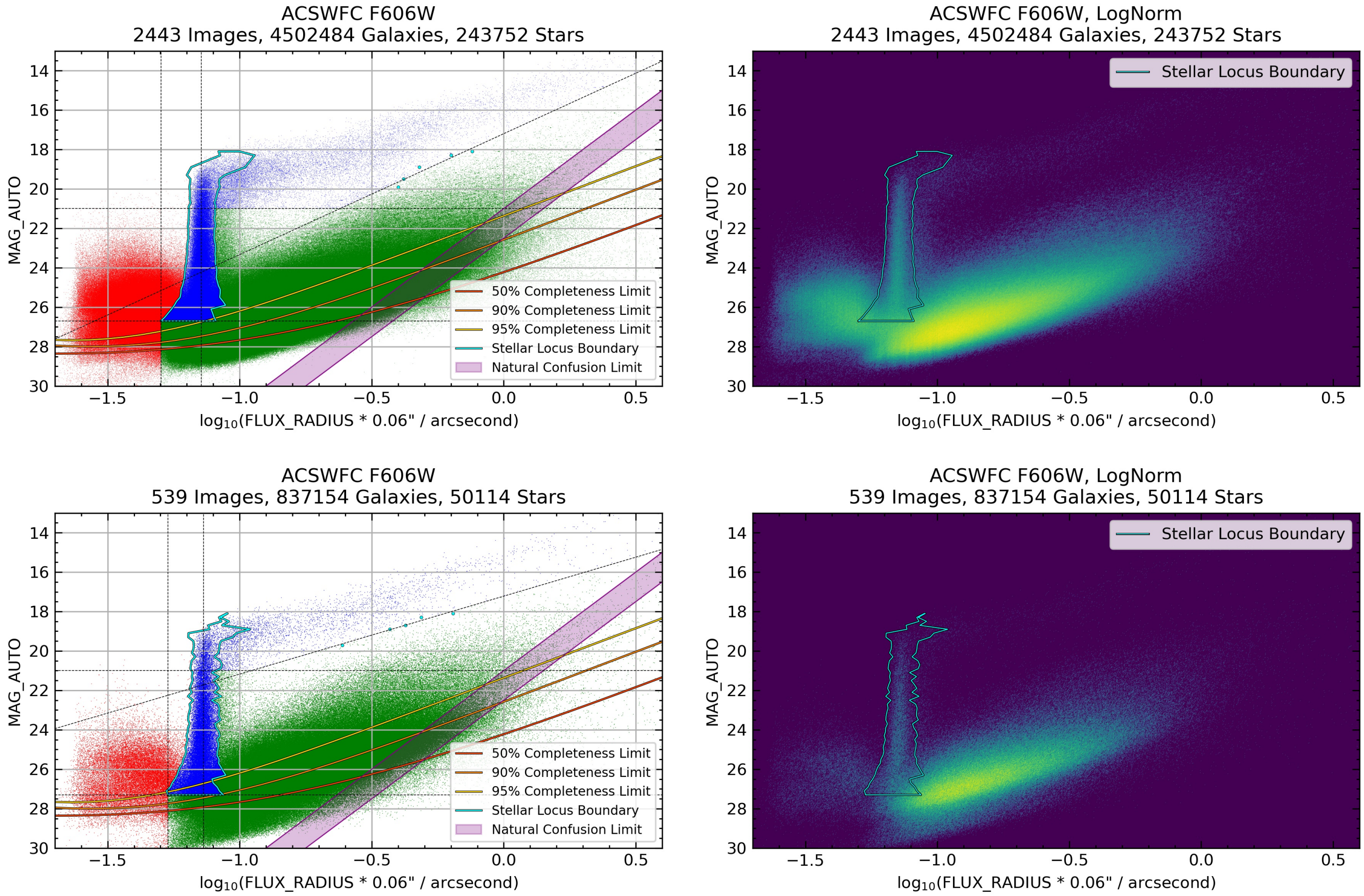

**Figure 9.** (**Top Left**) A cleaned version of the plot shown in Figure 6 with spurious detections at image boundaries removed using the modified weight map methods described in Section 3.1. The stellar locus boundary, plotted in cyan, was determined using the stellar locus Gaussian fitting method described in Section 3.2.3, and the cyan dots indicate the points used to fit a diagonal line to the saturated star track as described in Section 3.2.4. (**Top Right**) A 2D hexagonal bin histogram of the same data. The values of each bin are scaled logarithmically. With the object boxing method described in section 3.1 applied, the spout of spurious objects observed in figure 6 is no longer visible, indicating successful removal of the objects. (**Bottom Left and Bottom Right**) Multi-visit versions of the same plots. The stratified features seen in the single-visit plots toward the upper right portion of the main galaxy population are not present in these multi-visit plots. The stratified features observed in the single-visit plots are caused by certain sources appearing and being measured multiple times in overlapping visits, which is not a problem for the multi-visit products since all overlapping visits are combined into a single mosaic during the multi-visit drizzling.

#### 3.2.4. *Identifying the Saturated Star Track*

The second technique we implemented, which we will refer to as the saturated track fitting method, is an algorithm designed to identify the saturated stars in a magnitude versus size plot. The structure of the steps of our saturated track fitting method are similar to those of the stellar locus Gaussian fitting method discussed in Section 3.2.3, but the concepts involved are applied in a slightly different manner:

1. Choose an upper and lower bound for object size (in units of $\log_{10}$(50% FLUX_RADIUS/arcsec)) that contains the saturated track. For each filter, we set the middle of the stellar locus (defined here as the average value of the means calculated between $22 \leq m_{\rm AB} \leq 24$ mag during the stellar locus Gaussian fitting, which we refer to as $v_{locus}$) as the lower bound and a flat value of 0 dex as the upper bound.

2. Choose an upper and lower bound for magnitude that contains the saturated track. For all SKYSURF filters, we chose a lower bound of 18 mag and an upper bound of 20 mag.

3. Bin the objects within the bounds by magnitude. We determined magnitude bin sizes for each filter based on the number of objects contained in each combined catalog. Combined catalogs containing 200000 objects or more were assigned a magnitude bin size of 0.2 mag, while catalogs containing less than 200000 objects were assigned a magnitude bin size of 0.5 mag.

4. For each magnitude bin, create a histogram of the sizes of the objects contained within the bin. We used different bin sizes (in units of $\log_{10}$(50% FLUX_RADIUS/arcsec)) based on the number of objects contained within a combined catalog. Catalogs containing 3000000 objects or more were assigned a bin size of 0.02 dex; catalogs containing less than 3000000 objects but more than 500000 were assigned a bin size of 0.06 dex; catalogs containing less than 500000 objects were assigned a bin size of 0.1 dex.[13]

[11] For magnitude bins fainter than 24 mag, we used the average value of the means calculated between $22 \leq m_{\rm AB} \leq 24$ mag as the initial mean estimate.

[12] We used two different $\sigma_{final}$ coefficients for our right-hand side x-coordinates to prevent the right side of the polygon from cutting too deeply into the population of galaxies at faint magnitudes.

[13] WFC3UVIS F850LP required a unique bin size of 0.2 dex due to the filter's small number of bright objects.

5. For each histogram, determine the object size that corresponds to the minimum value of the histogram.

6. Using the sizes determined in the previous step as x-coordinates and the centers of each magnitude bin as y-coordinates, fit a diagonal line to the points. We used `scipy.optimize.curve_fit` for our diagonal line fitting. We call the resulting slopes and y-intercepts calculated from this fitting $m_{sat}$ and $b_{sat}$ respectively.

7. Define a "saturation magnitude limit" ($h_{sat}$, in units of magnitude). This is a flat value that, when used in conjunction with the diagonal line calculated previously, separates the stars in the saturated track from the population of galaxies. All objects brighter than both the saturation magnitude limit and the diagonal line are contained within the saturated track. By visually inspecting magnitude versus size plots, we manually specified saturation magnitude limits for each filter. These saturation magnitudes are contained in Table 4 in Appendix H, along with all the other star-galaxy separation parameters discussed here.

#### 3.2.5. *Identifying Spurious Detections*

Besides stars and galaxies, there were a number of sources with size measurements that were smaller than the measured sizes of the stars within the stellar locus. These sources cannot be real because they are smaller than point sources; rather, they are nonphysical spurious detections caused by left over cosmic rays, bad pixels, and noise spikes. We identify spurious detections and remove them from our analysis by first drawing a vertical line at the left point of the faintest magnitude bin of the stellar locus polygon described in Section 3.2.3. We call the location of this vertical line $v_{spurious}$. All objects with sizes smaller than $v_{spurious}$ are automatically considered spurious and excluded from the analysis. In addition, we also draw a horizontal line at the faintest magnitude value of the stellar locus polygon. We call the location of this horizontal line $h_{faint}$. All objects outside of the stellar locus polygon that were brighter than $h_{faint}$ and smaller than $v_{locus}$ were also classified as spurious and omitted from analysis. After identifying and isolating all stellar locus stars, saturated track stars, and spurious detections, all remaining sources were classified as galaxies. The exact numbers of spurious detections, stars, and galaxies identified for each SKYSURF filter are listed in Table 4 and illustrated in Figure 9.

### 3.3. *Number Counts and Quality Checking*

#### 3.3.1. *Measuring Number Counts*

After performing the star-galaxy separation methods described above, we measured the number counts for each SKYSURF filter. To obtain number counts for a filter, we used a magnitude bin size of 0.5 mag and counted the number of combined catalog objects contained within each bin. We limited the magnitude range for the number counts to $18 < m_{\rm AB} < 25$ mag, which is the range within which Hubble is expected to be complete. We did this separately for stars and galaxies to obtain star counts and galaxy counts, and we normalized the counts by dividing by the total area of all drizzled frames used to create the combined catalog for a given filter. We calculated the area of each drizzled frame by multiplying the number of pixels with non-zero weight by the drizzled pixel scale ($0.06''$/pixel). The area-normalized star counts and galaxy counts for ACSWFC F606W are shown in Figure 10, and for ACSWFC F435W, ACSWFC F606W, WFC3UVIS F606W, and WFC3IR F160W in Appendices B and C. The corresponding plots for all other SKYSURF filters are available as an electronic supplement.

#### 3.3.2. *Comparing With GOODS South*

To test the quality of our image processing and object cataloging, we compiled a subsample of our single-visit ACSWFC F606W SKYSURF catalogs located in the GOODS-S field and compared the results of our analysis to the GOODS South V-Band catalog (Giavalisco et al. 2004). For the comparison, we matched the objects in the GOODS-S V-Band catalog to our ACSWFC F606W single-visit combined catalog using a separation threshold of $0.4''$ and applied our SKYSURF star-galaxy separation cuts directly to the GOODS-S V-Band catalog. The brightness versus size plots for the SKYSURF ACSWFC F606W and GOODS-S V-Band catalogs used for the comparison are shown in Figure 11, and the corresponding number counts plots and partial IGL measurements are provided in Figure 12. For both magnitude and size, the galaxy and star counts are consistent between SKYSURF ACSWFC F606W and the GOODS South V-Band. A notable difference is the slightly higher measured partial IGL value for the GOODS-S V-Band data (3.68 nW/m$^2$/sr) relative to our SKYSURF subsample (3.52 nW/m$^2$/sr), which may be due to differences in specific Source Extractor parameters, such as DETECT_THRESH (see Section 4 for how we measure IGL).

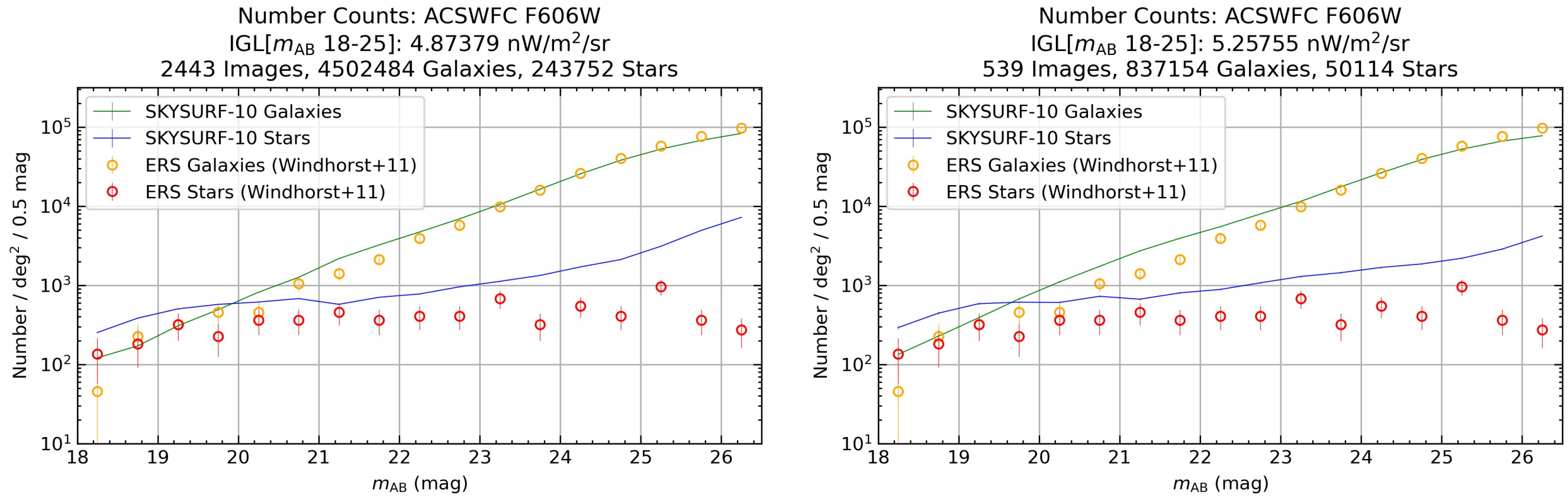

**Figure 10.** SKYSURF number counts for all ACSWFC F606W single-visit (**Left**) and multi-visit (**Right**) stars and galaxies. The ACSWFC F606W ERS counts from Windhorst et al. (2011) are also included for comparison. In total, 2,443 drizzled mosaics (4,502,484 galaxies and 243,752 stars) were used to obtain the ACSWFC F606W single-visit counts and 539 drizzled mosaics (837,154 galaxies and 50,114 stars) were used to obtain the ACSWFC F606W multi-visit counts. The final IGL measurement for this filter was 4.90 nW/m$^2$/sr for the single-visit dataset and 5.29 nW/m$^2$/sr for the multi-visit dataset.

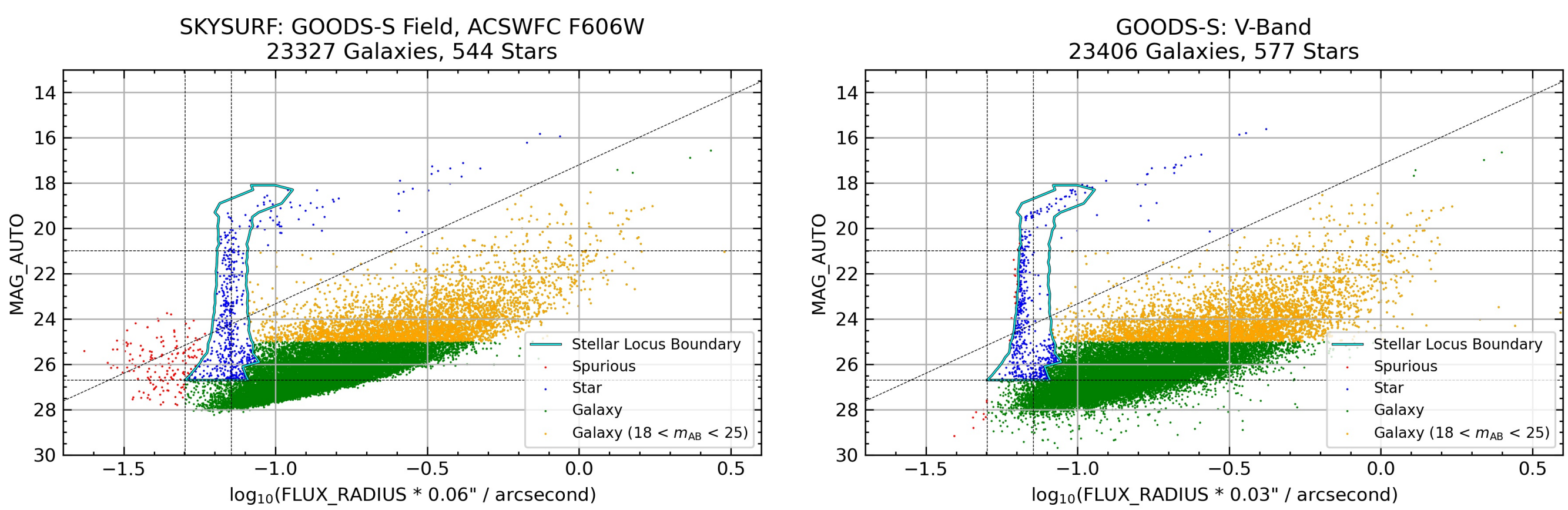

**Figure 11.** Magnitude versus size for a subsample of SKYSURF ACSWFC F606W objects from the GOODS South field (**Left**), alongside the same plot for the GOODS South V-Band data (**Right**). To ensure a proper photometric comparison, we used a separation threshold of 0.4″ to match objects in our ACSWFC F606W catalog to objects in the GOODS-S V-Band catalog. Stars are plotted in **blue** , galaxies in **green** , and spurious detections in **red** . The **orange** points indicate the galaxies in the range $18 < m_{\rm AB} < 25$ mag that we used to obtain the R50 size counts and IGL estimates included in Figure 12.

In addition, to test the validity of our star-galaxy separation, we compared the photometric distributions of SKYSURF's stars and galaxies in the GOODS South field to the photometric distributions of matched objects in the GOODS-S V-Band catalog and the 3D-HST photometric catalog (Brammer et al. 2012; Momcheva et al. 2016). We used a separation threshold of 0.4″ for the object matching between our SKYSURF measurements and these external catalogs. We used the `class_star` parameter to separate stars from galaxies in the GOODS-S V-Band catalog (`class_star` $\geq 0.9$ for stars; `class_star` $< 0.9$ for galaxies). For the 3D-HST catalog, we used the provided `star_flag` parameter to separate stars from galaxies (`star_flag` $= 1$ for stars; `star_flag` $= 0$ for galaxies). Magnitude histograms of the matched objects between SKYSURF and GOODS-S and between SKYSURF and 3D-HST are provided in Figure 13. The photometric distribution of SKYSURF's stars and galaxies exhibits significant overlap with the photometric distributions of both GOODS-S and 3D-HST, which provides two independent validations of our star-galaxy separation and photometry. All comparison plots described in this Section were performed using the SKYSURF single-visit dataset.

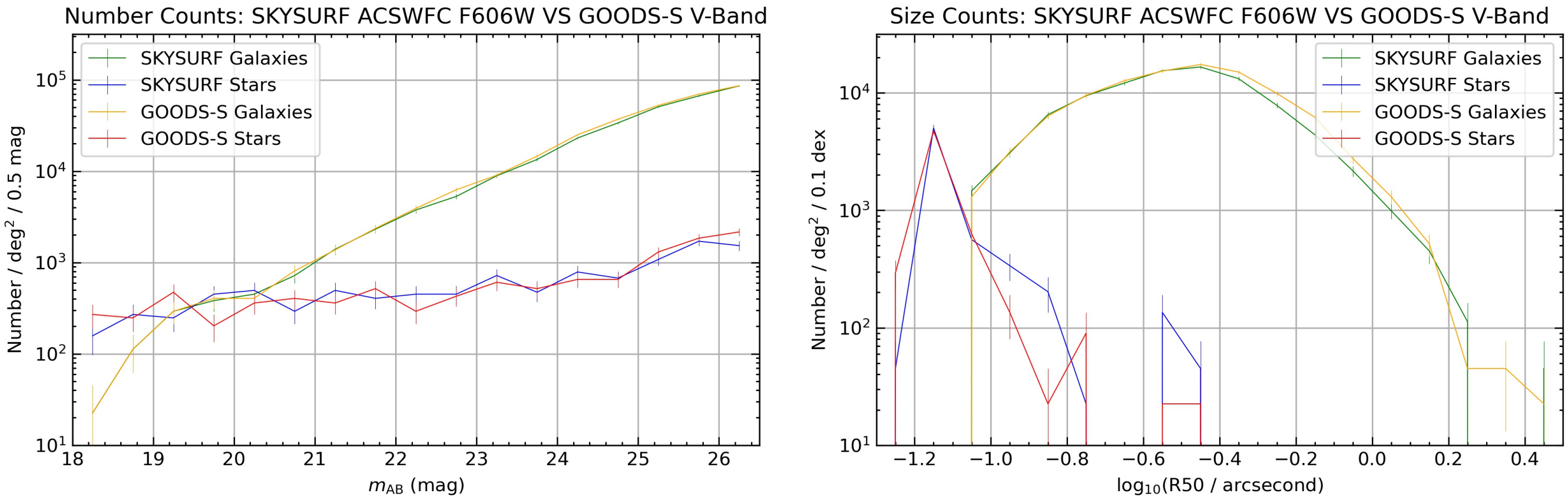

**Figure 12.** SKYSURF number counts for a subsample of SKYSURF ACSWFC F606W objects from the GOODS South field versus the GOODS South V-Band number counts for magnitude (**Left**) and R50 size (**Right**). For the R50 size comparisons, both the SKYSURF and GOODS data were magnitude limited to a range of $18 < m_{\mathrm{AB}} < 25$ mag to ensure completeness.

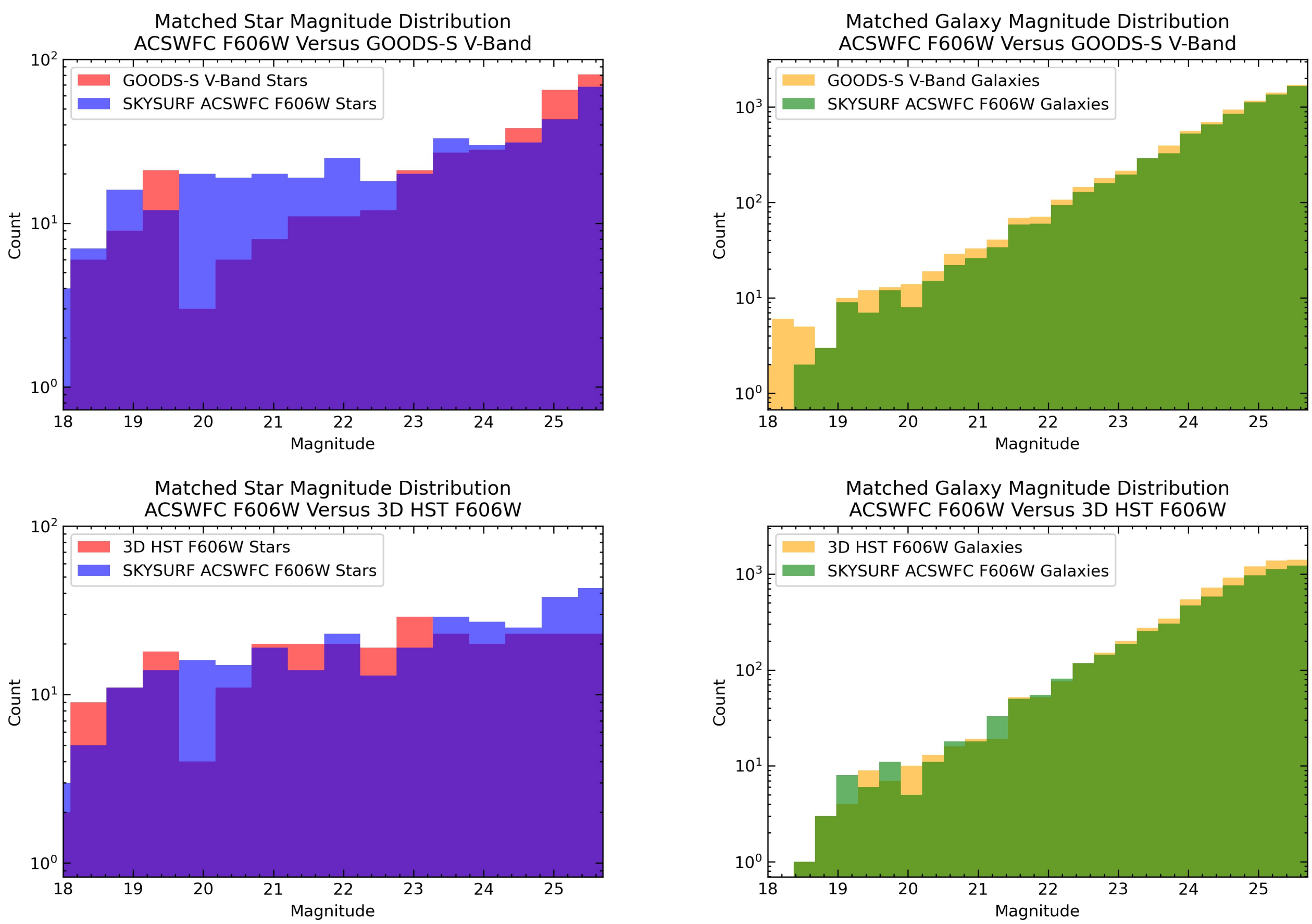

**Figure 13.** Distribution of star (**Left**) and galaxy (**Right**) photometric measurements for matched objects between SKYSURF and GOODS-S (**Top**) and SKYSURF and 3D-HST (**Bottom**) in the GOODS South field. Objects were matched using a separation threshold of 0.4″

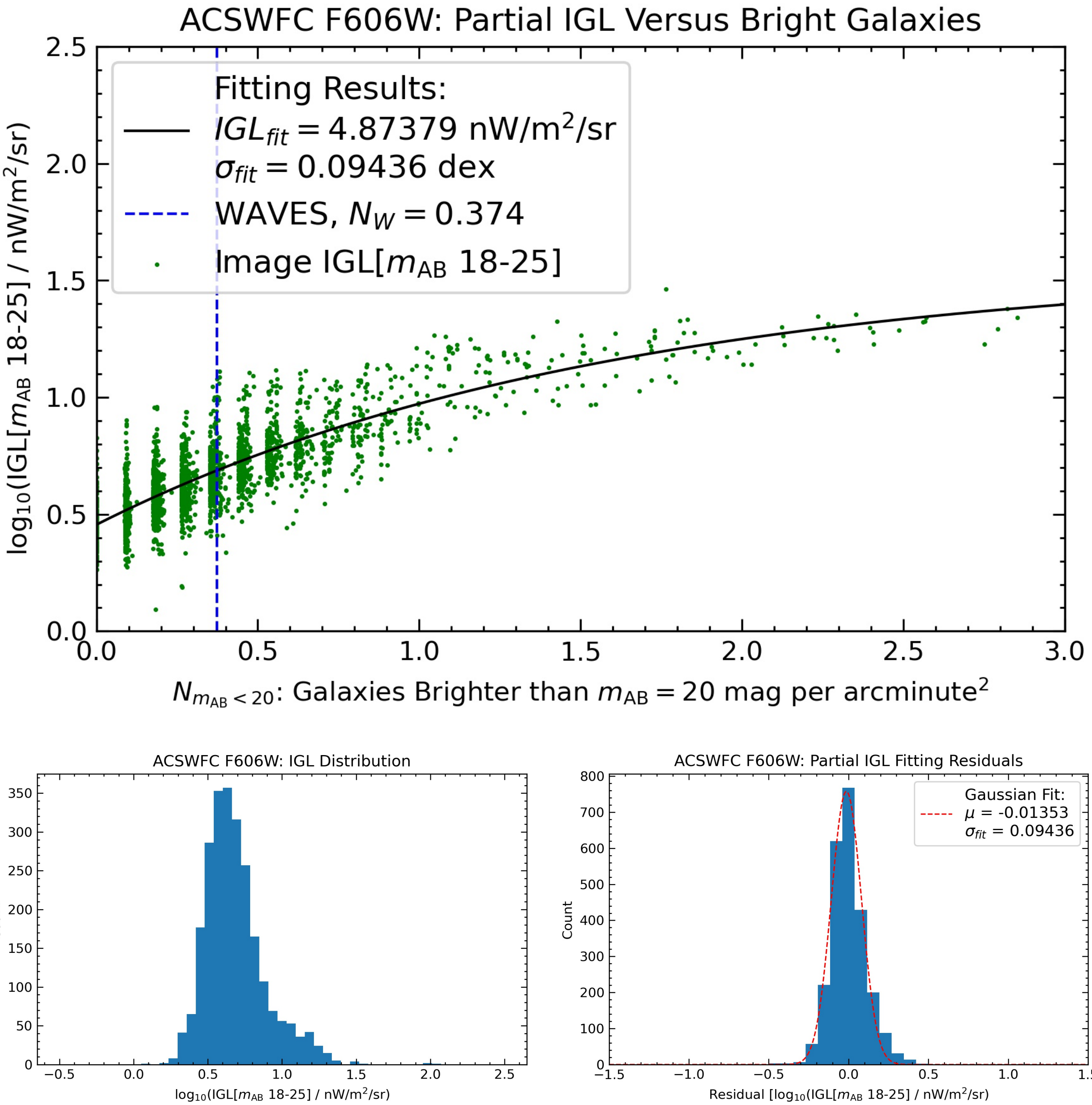

**Figure 14.** (**Bottom Left**) Histogram of the partial IGL values calculated for each single-visit drizzled image for ACSWFC F606W. (**Top**) Partial IGL versus the number of galaxies brighter than $m_{\rm AB} = 20$ mag for the single-visit drizzled products of ACSWFC F606W. Each green point represents one single-visit frame. The solid black line is the fit to the data, and the dashed blue line is the WAVES baseline value ($N_W$) obtained from the WAVES r band counts, which was used to calculate the partial IGL for this filter. The quantized pattern here is caused by the fact that all SKYSURF mosaics have nearly identical area measurements, but contain varying numbers of galaxies brighter than $m_{\rm AB} = 20$ mag (e.g. 0 galaxies, 1 galaxy, 2 galaxies, etc.). (**Bottom Right**) Histogram of the partial IGL versus $N_{m_{\rm AB}<20}$ fitting residuals for ACSWFC F606W. The dashed red curve is the Gaussian fit from which the standard deviation of the residuals was obtained for this filter. Standard deviation was calculated in this way to reduce the influence of outliers.

## 4. MEASURING THE IGL

This Section outlines our IGL measurement methods. In Section 4.1, we introduce a new concept called "partial IGL" and describe how partial IGL is calculated for each Hubble mosaic. In Section 4.2, we describe our process for measuring partial IGL for the entire sky in each filter using a new IGL fitting method in conjunction with external galaxy counts data from the Wide-Area VISTA Extragalactic Survey (WAVES) (Driver et al. 2019). In Section 4.3, we provide additional details regarding how the WAVES galaxy counts were calculated. Section 4.4 details our uncertainties and error propagation, and Section 4.5 evaluates the impact that spurious edge detections have on IGL measurements. Last, Section 4.6 describes how we used the IGL results from Tompkins et al. (2025) in conjunction with our partial IGL results to obtain extrapolated IGL estimates, which we used to compare our IGL results with previous IGL measurements and theoretical models.

### 4.1. *Measuring Partial IGL for an Image*

Because there is so much variation in the number of stars and galaxies (and in effect, variation in the IGL) across individual pointings, we performed partial IGL measurements on a frame-by-frame basis. By partial IGL, we mean an IGL measurement obtained using a magnitude-limited subset of the data. In this work, the magnitude-limited subset we selected for our partial IGL measurements was $18 < m_{\rm AB} < 25$ mag. The reason we measured partial IGL was to ensure consistency in the measurements on a frame-by-frame basis by preventing issues caused by sporadic bright objects and incompleteness. To obtain these partial IGL measurements for each frame in a given filter, we first performed Galactic extinction corrections to all galaxies within a given image. We used the SVO Filter Profile Service (Rodrigo et al. 2012; Rodrigo & Solano 2020) to obtain the effective wavelengths for all SKYSURF HST filters, then used the `IrsaDust` module within `astroquery` to retrieve a Galactic extinction table corresponding to the Galactic coordinates of an image. Our Galactic extinction correction for an image entailed obtaining a linearly interpolated `A_SandF` extinction magnitude (Schlegel et al. 1998b; Schlafly & Finkbeiner 2011) from the extinction table and subtracting that extinction magnitude from all galaxies in the image.[14]

After correcting for Galactic extinction, we calculated the area-normalized flux of the total galaxy light contained in each image as follows:

$$I_\nu = \frac{\sum_{n=1}^{n=N} 10^{\frac{m_n}{-2.5}} \times 3631 \times 10^{-17}}{A_{image}}, \tag{2}$$

where $I_\nu$ is flux in nW/m$^2$/sr/Hz, $m_n$ is the magnitude of the $n$th galaxy in the image, $N$ is the total number of galaxies in the image, and $A_{image}$ is the total area of the image in steradians. We then calculated the effective frequency ($\nu_{eff}$) of each SKYSURF filter using $\nu_{eff} = \frac{c}{\lambda_{eff}}$, where $c$ is the speed of light and $\lambda_{eff}$ is the effective wavelength of the filter. Last, we calculated the partial IGL for each image by measuring the area-normalized flux for galaxies in the magnitude range $18 < m_{\rm AB} < 25$ mag and multiplying by the filter's effective frequency:

$$\mathrm{IGL} = \nu I_\nu. \tag{3}$$

A histogram of our partial IGL measurements for each drizzled frame in ACSWFC F606W is shown in Figure 14.

### 4.2. *Measuring Partial IGL for a Hubble Filter*

In the small field of view of HST, and in the SKYSURF sample that is biased toward objects, significant variation in IGL is essentially guaranteed, and we empirically found higher field-to-field IGL variation than expected (Tompkins et al. 2025). To account for this variation, the IGL can be measured as a function of environment, and an empirical way to do this is to plot partial IGL as a function of bright galaxies. In this work, when we refer to "bright galaxies," we mean any galaxy brighter than $m_{\rm AB} \leq 20$ mag. Plotting $\log_{10}(IGL[m_{\rm AB}\ 18-25])$, where $IGL[m_{\rm AB}\ 18-25]$ is the partial IGL, against the number of bright galaxies per square arcminute ($N_{m_{\rm AB}<20}$) revealed a consistent trend across all SKYSURF Hubble filters: for $N_{m_{\rm AB}<20} < 3.0$ in ACSWFC and WFC3UVIS, and $N_{m_{\rm AB}<20} < 4.0$ in WFC3IR, $\log_{10}(IGL[m_{\rm AB}\ 18-25])$ initially grows rapidly before leveling off as the number of bright galaxies per square arcminute increases, suggesting an exponential saturation pattern. This is seen in the top plot of Figure 14, which shows $\log_{10}(IGL[m_{\rm AB}\ 18-25])$ plotted against $N_{m_{\rm AB}<20}$ for all single-visit frames with $N_{m_{\rm AB}<20} < 3.0$.

[14] We did not include WFC3UVIS F225W, WFC3UVIS F275W, or WFC3UVIS F300X in our IGL analysis because these filters are too far below the typical range of wavelengths available in the IrsaDust module. For WFC3UVIS F336W specifically, since it is near the lower end of wavelengths provided in the extinction tables, we used the extinction magnitude corresponding to the lowest available wavelength in a table instead of an interpolated value when necessary.

We used `scipy.optimize.curve_fit` to fit a saturated exponential model of the form $f(x) = a_{fit}(1 - e^{-b_{fit}x}) + c_{fit}$ between $IGL[m_{\rm AB}\ 18 - 25]$ and $N_{m_{\rm AB}<20}$ for each filter, as illustrated in the top plot of Figure 14 for ACSWFC F606W. Though the exponential saturation trend may hold beyond $N_{m_{\rm AB}<20} < 3.0$ (ACSWFC, WFC3UVIS) and $N_{m_{\rm AB}<20} < 4.0$ (WFC3IR), we selected these as upper bounds to ensure sufficient data for fitting in all filters and to prevent excessively dense data beyond these ranges from negatively impacting the fits in the range representative of the average Universe (for a deeper look into how fitting bounds impact the fitting measurements, see Appendix F). Given this relationship between partial IGL and $N_{m_{\rm AB}<20}$, partial IGL can be estimated using an expected baseline value of $N_{m_{\rm AB}<20}$ for the entire sky. To obtain this baseline for each filter, we summed the galaxy counts bin data for bins brighter than $m_{\rm AB} = 20$ mag from the Wide-Area VISTA Extragalactic Survey (WAVES) (Driver et al. 2019). The WAVES bands corresponding to each SKYSURF filter are listed in Table 3. By plugging the $N_W$ baseline values into the fitting functions for each filter, we obtained IGL measurements ($IGL_{fit}$) that are far more resistant to outlier image IGL values than the standard area-normalized summation of total galaxy light across all images ($IGL_i$). Furthermore, since $N_{m_{\rm AB}<20}$ is a quantity that can be easily measured, this enables extrapolations of the total IGL expected for any pointing that falls within the range over which our IGL fitting method works. We note that not all ACSWFC and WFC3UVIS filters have enough data to span the full fitting range, but there is sufficient data in most filters for the fits to be reliable up to $N_{m_{\rm AB}<20} \approx 1.0$ or $N_{m_{\rm AB}<20} \approx 1.5$. The WFC3IR datasets span the entire fitting range and are reliable up to $N_{m_{\rm AB}<20} \approx 4.0$.

### 4.3. *WAVES Number Counts*

Since our average partial IGL measurements depend on the WAVES $N_W$ measurements, we consider it appropriate here to describe the WAVES number counts in more detail. The WAVES counts were measured by Bellstedt et al. (in prep), who perform the star-galaxy separation for the WAVES counts by using a combination of two photometric redshift codes: scaled flux matching (SFM) (Baldry et al. 2021) and Tartu Observatory Photo-z (TOPz) (Laur et al. 2022; Tempel et al. 2025). Both photo-z codes calculate a classification probability between 0 and 1 for each object, with SFM providing a probability of an object being a galaxy ($P_{SFM}(Galaxy)$) and TOPz providing a probability of an object being a star ($P_{TOPz}(Star)$). Bellstedt et al. (in prep) classified a WAVES object as a star if $P_{SFM}(Galaxy) < 0.1$ and $P_{TOPz}(Star) > 0.9$. They classified an object as ambiguous if $P_{SFM}(Galaxy) > 0.1$ and $P_{TOPz}(Star) > 0.9$, or if $P_{SFM}(Galaxy) < 0.1$ and $P_{TOPz}(Star) < 0.9$. All other objects were classified as galaxies.

Bellstedt et al. (in prep) constructed the WAVES counts using the above criteria. They calculated lower and upper limits for the WAVES galaxy counts using the objects classified as ambiguous, with the lower limit being the resulting galaxy counts if all ambiguous objects are classified as stars and the upper limit being the resulting galaxy counts if all ambiguous objects are classified as galaxies. They then used the mean of the lower and upper limits as their final galaxy counts measurements (denoted as $N_W$ throughout this work) and used the range as the object classification uncertainty (denoted as $N_{WC}$ in Section 4.4).

### 4.4. *Partial IGL Error Analysis*

To calculate the error of our partial IGL measurements for a filter, we first histogrammed the residuals of the fit and used `scipy.optimize.curve_fit` to fit a Gaussian to the result (see bottom right of Figure 14). To create the histogram, we used 40 equally sized bins spread across -1.5 to 1.5 dex. During the Gaussian fitting of the residuals, we applied reasonable bounds for specific parameters to ensure that the fits would accurately represent the data. We bounded the standard deviation parameter by requiring it to be larger than the histogram bin size (in this case, larger than 0.769) and smaller than half the range of the data. We also required the amplitude to be between zero and the maximum bin value. Fitting a Gaussian curve to the bins in the histogram with non-zero values allowed us to accurately measure the standard deviation of the residuals, which we denote here as $\sigma_{fit}$[15].

We then used $\sigma_{fit}$ to calculate the log-scaled standard error ($SE_{iglfit}$) of the IGL fit using:

$$SE_{iglfit} = \frac{\sigma_{fit}}{\sqrt{N_{images}}}, \tag{4}$$

where $N_{images}$ is the number of images with $N_{m_{\rm AB}<20} < 3.0$ (or $N_{m_{\rm AB}<20} < 4.0$ for WFC3IR) used to calculate the fit.

[15] We calculated standard deviation in this way to make the error measurements more resistant to outliers.

In addition to the standard error of the IGL fits, it is also necessary to account for the uncertainty in the $N_W$ WAVES measurements. To do this, we first added, in quadrature, the Poisson errors of all WAVES magnitude count bins brighter than 20 mag, which we will refer to as $N_{WP}$ (the total Poisson error of $N_W$). Then, by propagation of uncertainty:

$$N_{WE} = N_{WM} f'(N_W), \tag{5}$$

where $N_{WE}$ is the log-scaled WAVES error contribution to the partial IGL fitting measurement, $f'(N_W) = a_{fit} b_{fit} e^{-b_{fit} N_W}$ is the derivative of our IGL fitting function at $N_W$, and $N_{WM}$ is the total uncertainty of the WAVES measurement, composed of the WAVES Poisson error ($N_{WP}$) and WAVES object classification error ($N_{WC}$) added in quadrature:

$$N_{WM} = \sqrt{N_{WP}^2 + N_{WC}^2}. \tag{6}$$

Further details regarding the WAVES number counts measurements and uncertainties will be available in Bellstedt et al. (in prep).

We have also devoted significant attention to evaluating the impact of target selection on our measurements, which is particularly important given the inhomogeneous nature of Hubble's archival data. Having both a single-visit and a multi-visit dataset for each filter allows us to compare partial IGL measurements between the two datasets, and any notable differences in the partial IGL measurements between the two datasets serve as indicators of potential systematic errors. One such difference is the 0.39 nw/m$^2$/sr difference between the ACSWFC F606W single-visit and multi-visit partial IGL measurements, and in Appendix F, we consider a variety of different potential error sources that could be responsible for this, including target selection, fitting bounds, star-galaxy separation, and weight map modification. We refer the reader to Appendix F for the specific details of the tests we performed, but to summarize, we determined that fitting bounds, star-galaxy separation, and weight map modification have a negligible impact on the difference between the two datasets; furthermore, the difference vanishes when we perform our measurements using only matched fields between the datasets. We therefore attribute the difference to a systematic error caused by the differing distribution of fields available in the single-visit and multi-visit datasets. To account for this, we have added an additional term to our error budget, $\sigma_{SVMV} \approx 0.0325$ dex, which is the log-scaled standard deviation of the differences between the single-visit and multi-visit partial IGL measurements. This is the last of our error terms, and putting everything together yields a total log-scaled error ($E_{iglfit}$) of:

$$E_{iglfit} = \sqrt{SE_{iglfit}^2 + N_{WE}^2 + \sigma_{SVMV}^2}. \tag{7}$$

Last, to calculate the linearly-scaled total error ($E_{fit}$), we used the following equation obtained by applying propagation of uncertainty:

$$E_{fit} = E_{iglfit} \times \log(10) \times 10^{S_{IGL}}, \tag{8}$$

where $S_{IGL}$ is the log-scaled IGL value obtained from the fit.

The IGL and standard error values calculated for each filter are provided in Table 5 in Appendix I along with several of the other parameters discussed here, including our model fitting parameters ($a_{fit}$, $b_{fit}$, $c_{fit}$). The IGL fitting plots for ACSWFC F435W, ACSWFC F606W, WFC3UVIS F606W, and WFC3IR F160W and their corresponding residuals histograms are included in Appendices D and E, and the corresponding plots for the remaining SKYSURF filters are available as an electronic supplement for those interested in visually inspecting the quality of the fits.

For completeness, we have also quantitatively investigated the impact of our sky sub-region and ProFound sky chi-squared cuts (discussed in Section 2.3.3) on our partial IGL measurements to determine whether these choices change our results significantly. Appendix G contains plots of partial IGL as a function of different sky sub-region and ProFound sky chi-squared cuts for the single-visit datasets of ACSWFC F435W, ACSWFC F606W, WFC3UVIS F606W, and WFC3IR F160W. To create these plots, we varied the cut thresholds (x-axis), excluded any drizzled mosaics that contained FLC/FLT images invalidated by the cuts, and performed the partial IGL fitting discussed in Section 4.2 on the remaining mosaics. The left plots of Appendix G show the impact of varying the object sub-region cut, where the x-axis corresponds to the object sub-region cuts and the y-axis is the partial IGL measured for the cuts. Similarly, the right plots show the impact of varying the ProFound sky chi-squared bounds, where the x-axis represents symmetric sky chi-squared bounds about 1 (e.g. $x = 0.1$ corresponds to the bounds $0.9 < \chi^2 < 1.1$) and the y-axis is the partial IGL measured for the bounds. As the plots show, there is a wide range of acceptable values for these cuts. The partial IGL measurements change by only $\approx 0.1$ nW/m$^2$/sr between $x = 0.1$ and $x = 0.7$ for the object sub-region cuts and between $x = 0.06$ and $x = 0.1$ for the ProFound sky chi-squared cuts for the vast majority of SKYSURF filters. We therefore do not add additional error terms to account for these choices, as their impact on our IGL fitting method is negligible relative to other error sources. The bad pixel sub-region cut could be analyzed similarly, but given that only a handful of images are removed by our bad pixel sub-region cut ($\lesssim 5$ FLT/FLC images per filter), this cut is largely inconsequential.

**Table 3.** SKYSURF filters and the corresponding WAVES filters used to obtain the baseline $N_{m_{\rm AB}<20}$ values ($N_W$) discussed in Section 4.2.

| **SKYSURF Filter** | **WAVES Filter** |
|---|---|
| ACSWFC F435W | g |
| ACSWFC F475W | g |
| ACSWFC F555W | g |
| ACSWFC F606W | r |
| ACSWFC F625W | r |
| ACSWFC F775W | i |
| ACSWFC F814W | i |
| ACSWFC F850LP | Z |
| WFC3UVIS F390W | u |
| WFC3UVIS F438W | g |
| WFC3UVIS F475W | g |
| WFC3UVIS F475X | g |
| WFC3UVIS F555W | r |
| WFC3UVIS F606W | r |
| WFC3UVIS F625W | r |
| WFC3UVIS F775W | i |
| WFC3UVIS F814W | i |
| WFC3UVIS F850LP | Z |
| WFC3IR F098M | Z |
| WFC3IR F105W | Y |
| WFC3IR F110W | Y |
| WFC3IR F125W | J |
| WFC3IR F140W | J |
| WFC3IR F160W | H |

### 4.5. *The Impact of Spurious Edge Sources on Partial IGL Measurements*

As can be seen in Figure 6, the spurious sources discussed in Section 3.1 that Source Extractor detects at regions where weight image pixel values abruptly transition from zero to non-zero values extend into the main galaxy population. To evaluate the impact that these spurious sources would have on our IGL measurements if not excluded, we performed a test using the ACSWFC F606W single-visit dataset.

The filtering criteria discussed in Section 3.2.1 resulted in slightly different numbers of analyzed mosaics for ACSWFC F606W for the cleaned catalogs and raw catalogs (2,443 mosaics for the cleaned catalogs, 2300 mosaics for the raw catalogs). To ensure a proper comparison, we compared the raw and cleaned measurements for the 2,276 mosaics in common between the two versions. The measurements performed using the 2,276 raw catalogs resulted in 4,648,532 galaxies, 295,086 stars, and 882,366 spurious sources. The measurements performed using the 2,276 cleaned catalogs resulted in 4,221,759 galaxies, 210,503 stars, and 417,550 spurious sources. The partial IGL measured using our IGL fitting method was 5.28 nW/m$^2$/sr for the raw catalogs and 4.89 nW/m$^2$/sr for the cleaned catalogs. This suggests an upward bias of up to $\approx 0.4$ nW/m$^2$/sr to the partial IGL measurements if the spurious edge sources are not excluded from the analysis, and highlights the importance of the Section 3.1 methods implemented to remove these sources.

### 4.6. *Estimating Extrapolated IGL*

In addition to partial IGL measurements, Tompkins et al. (2025) also performed extrapolations to estimate total IGL. In this work, since we analyzed the same datasets as Tompkins et al. (2025), we used their extrapolated IGL results to estimate our own extrapolated IGL values. We calculated our extrapolated IGL ($IGL_e$) by multiplying our partial IGL fitting measurements by the ratio of the extrapolated to partial IGL measurements ($IGL_{Ratio}$) of Tompkins et al. (2025):

$$IGL_e = IGL_{fit} \times IGL_{Ratio} \tag{9}$$

Here, $IGL_{Ratio} = \frac{IGL_T}{IGL_{TP}}$, where $IGL_T$ and $IGL_{TP}$ are the extrapolated and partial IGL measurements of Tompkins et al. (2025) respectively. We then calculated the error for each of our extrapolated IGL measurements ($ERR_e$) as follows:

$$ERR_e = \sqrt{E_{fit}^2 \times IGL_{Ratio}^2 + ERR_{extrap}^2 \times IGL_{fit}^2} \tag{10}$$

where $ERR_{extrap}$ is the uncertainty of the $IGL_{Ratio}$ measurement, obtained by propagation of error:

$$ERR_{extrap} = \sqrt{\frac{ERR_T^2}{IGL_{TP}^2} + ERR_{TP}^2 \frac{IGL_T^2}{IGL_{TP}^4}} \tag{11}$$

Here, $ERR_{TP}$ is the error of the partial IGL measurement of Tompkins et al. (2025). Two of the 25 SKYSURF filters analyzed in this work were omitted from the analysis of Tompkins et al. (2025) and therefore have no extrapolated IGL estimate corresponding to the provided partial IGL measurements in this work: WFC3UVIS F475X and WFC3IR F098M. Both of these filters are used less frequently relative to other Hubble filters, and the available data is difficult to work with, so Tompkins et al. (2025) opted not to include these filters in their analysis. The extrapolated IGL measurements for the remaining 23 filters are provided in Table 6 in Appendix J. They are also plotted alongside the extrapolated IGL measurements of Tompkins et al. (2025) and several IGL models resulting from different star formation histories (D'Silva et al. 2023; Bellstedt et al. 2020; Driver et al. 2018; Madau & Dickinson 2014) in Figure 18.

We emphasize here that the purpose of this extrapolation method is not to arrive at the most rigorously determined total IGL measurements, but rather to enable an approximate comparison of our IGL fitting method to the extrapolated results of previous IGL investigations. The main focus of this work is the IGL fitting method itself, not our extrapolated results, but given that partial IGL is a new concept, our partial IGL measurements cannot be directly compared with external IGL measurements without some form of normalization.

## 5. RESULTS AND DISCUSSION

### 5.1. *Star-Galaxy Separation Results*

#### 5.1.1. *Updated Hubble Star-Galaxy Separation Parameters*

The various empirically determined horizontal, vertical, and diagonal cuts used for our star-galaxy separation are displayed in Table 4 in Appendix H for each filter, and the stellar locus polygon points for each filter can be downloaded from the SKYSURF MAST project page. Combined source catalogs for each SKYSURF filter are also included on the SKYSURF MAST project page, with each catalog containing a column specifying the object type (stellar locus star, saturated track star, galaxy, or spurious) for each object identified by Source Extractor.

#### 5.1.2. *New Features Observed in Magnitude Versus Size Plots*

The star-galaxy separation cuts for each filter are illustrated in the brightness versus size plots contained in Appendices B and C for the single-visit and multi-visit data. As expected, the standard features of these types of plots are present: nonphysical spurious detections, the stellar locus, the saturated track, and the main body of galaxies. Furthermore, because the SKYSURF dataset is so large, new features can be seen in the brightness versus size plots that have never been observed in previous smaller-scale projects. The first of these features is the small collection of extended objects that are slightly bigger than the stars contained in the stellar locus, which could potentially be active galactic nuclei, double stars, or something else entirely. The second feature is the dense "bubble" of objects observed at the beginning of the saturated track, which is most visible in the brightness versus size plots of the reddest ACSWFC filters (F775W, F814W, F850LP). These two new features may provide additional opportunities to further group and classify the objects observed by Hubble. An in-depth analysis of these objects is beyond the scope of this work and would not significantly impact our IGL measurements, but would be an interesting topic for a separate project.

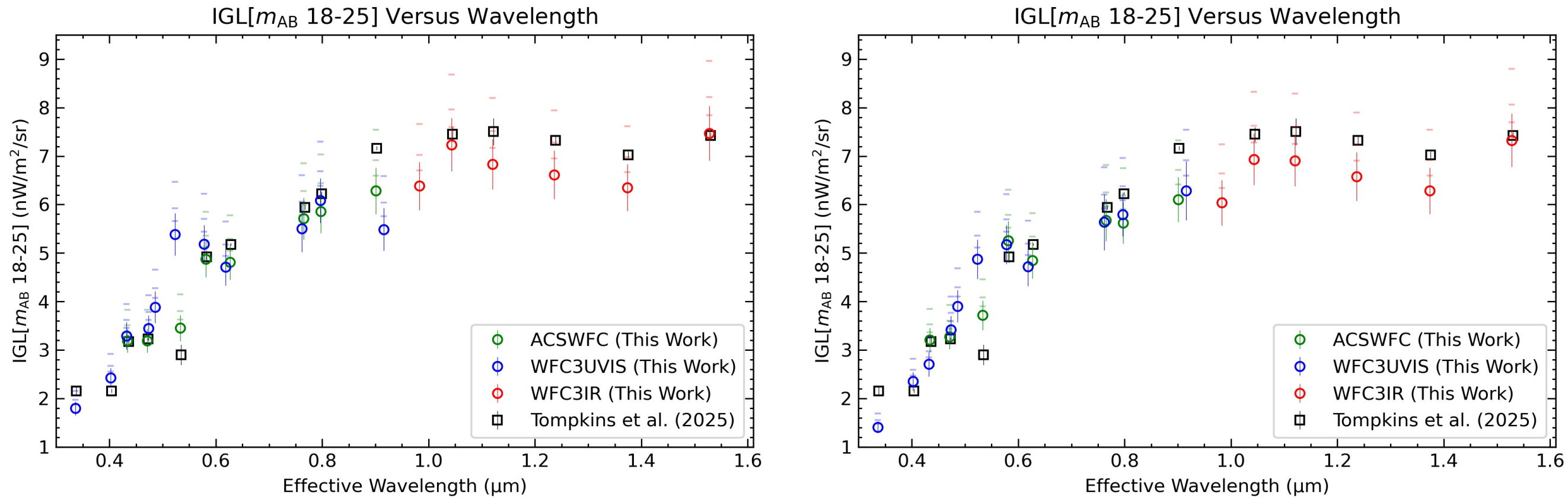

**Figure 15.** Partial IGL[$m_{\rm AB}$ 18-25] versus wavelength for 25 SKYSURF Hubble filters for the single-visit (**Left**) and multi-visit (**Right**) data. The black points are the results of the Tompkins et al. (2025) ProFound analysis of the SKYSURF data. The green, red, and blue points are the ACSWFC, WFC3IR, and WFC3UVIS results of this work obtained using `SExtractor`. The transparent colored dashes above each of this work's measurements show where the measurements would land if a 5% (lowest dash), 10% (middle dash), and 20% (highest dash) correction were performed to account for potential flux lost during source detection. Despite the different methods and techniques used in Tompkins et al. (2025), our partial IGL measurements are in general agreement, which provides strong evidence for the validity of these results.

### 5.2. *Object Number Counts and SKYSURF Completeness*

The star and galaxy number counts plots for ACSWFC F435W, ACSWFC F606W, WFC3UVIS F606W, and WFC3IR F160W are also included in Appendices B and C alongside their corresponding brightness versus size plots (the corresponding plots for other filters are available as an electronic supplement). The galaxy counts in almost all filters follow a roughly linear trend in the range $18 < m_{\rm AB} < 25$ mag (the known exception being WFC3UVIS F850LP discussed previously), which provides further qualitative evidence that these filters are complete within the magnitude range analyzed in this work. We have plotted the ERS number counts of Windhorst et al. (2011) alongside the counts of this work to enable an easy comparison with previously published results. Our galaxy counts match nicely with the ERS galaxy counts, but our star counts tend to be slightly higher than those of Windhorst et al. (2011), which is expected given the range of galactic latitudes included in our sample. For the purposes of our IGL analysis, ensuring that our galaxy counts are correct is of primary importance, but Ingram et al. (in prep) will be investigating the nature of our star counts more thoroughly.

### 5.3. *IGL Fitting Performance*

#### 5.3.1. *Impact of Cosmic Variance on IGL Measurements*

Extended partial IGL versus $N_{m_{\rm AB}<20}$ plots for the single-visit and multi-visit ACSWFC F606W data are shown in Figure 16. Interestingly, in this particular case, the exponential saturated model seems to capture the few data points beyond the fitting range rather well, but this is not necessarily true for all filters. The baseline values of WAVES ($N_W$) for each SKYSURF filter all fall well within the $N_{m_{\rm AB}<20} < 3.0$ range over which our IGL fitting is performed. Notice that there are a number of frames with $N_{m_{\rm AB}<20}$ measurements significantly beyond $N_W$, which suggests that these frames are not representative of the average Universe. This is important, because it implies that obtaining IGL measurements via the typical area-normalized summations of galaxy light across all frames ($IGL_i$) may be biased toward higher values for some filters without additional field-selection criteria. This is evident from the $IGL_i$ and $IGL_{fit}$ measurements included in Table 5, which generally show our $IGL_i$ measurements to be higher than the $IGL_{fit}$ measurements.

Within our $N_{m_{\rm AB}<20} < 3.0$ (or $N_{m_{\rm AB}<20} < 4.0$ for WFC3IR) fitting range, there is significant variation in partial IGL measurements due to cosmic variance. Some fields have many bright galaxies while other fields are sparser relative to the $N_W$ baseline values, which affects where the expected partial IGL values fall along the fits. Furthermore, at any given point along the fit, there is a spread within which partial IGL values are likely to fall. These spreads are well captured in the measured standard deviations of the fitting residuals ($\sigma_{fit}$), which range from 0.075 to 0.15 dex for most SKYSURF filters. The average spread of IGL values across all filters is 0.1044 dex for the single-visit data and 0.1028 dex for the multi-visit data, both of which constitute a $> 40\%$ improvement relative to the average spreads without fitting ($\sigma_{nofit}$) of 0.2023 dex and 0.1861 dex for the single-visit and multi-visit data, respectively. Some individual filters do much better than the average, with many achieving up to two times thinner spreads with IGL fitting relative to their spreads without fitting. For the exact values of $\sigma_{fit}$ and $\sigma_{nofit}$ for each filter, see Table 5 in Appendix I.

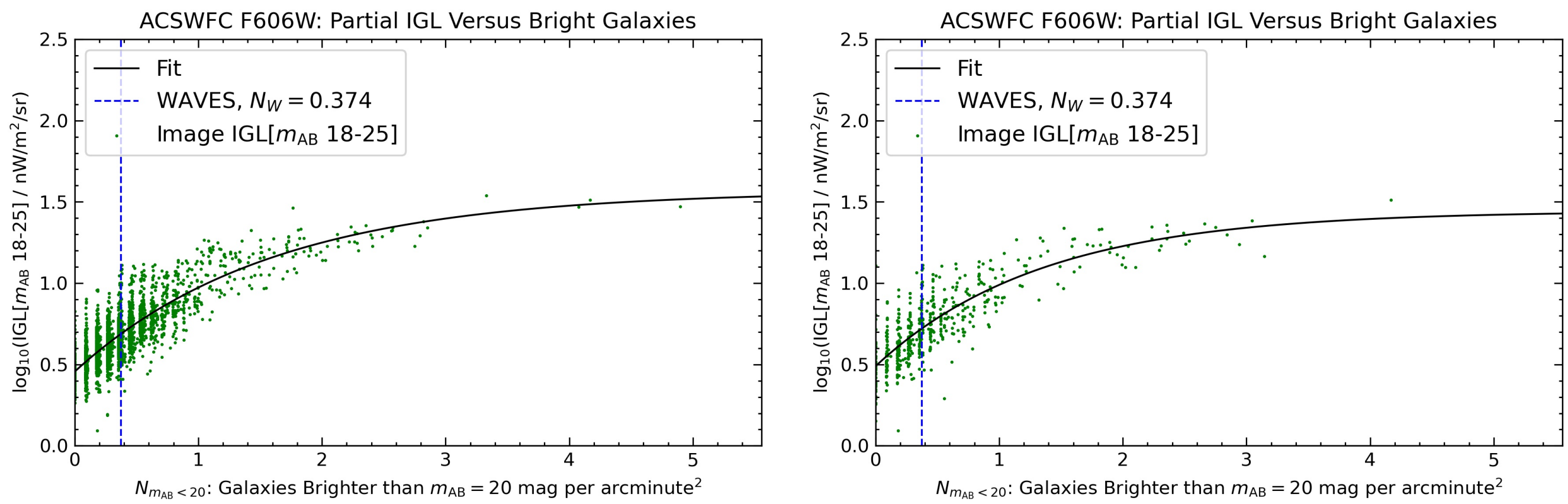

**Figure 16.** Partial IGL versus the number of galaxies brighter than $m_{\rm AB} = 20$ mag for the single-visit (**Left**) and multi-visit (**Right**) drizzled products of ACSWFC F606W with the x-axis extended out to $N_{m_{\rm AB}<20} = 5.56$. Each green point represents one drizzled frame, and the solid black line is the fit to the data.

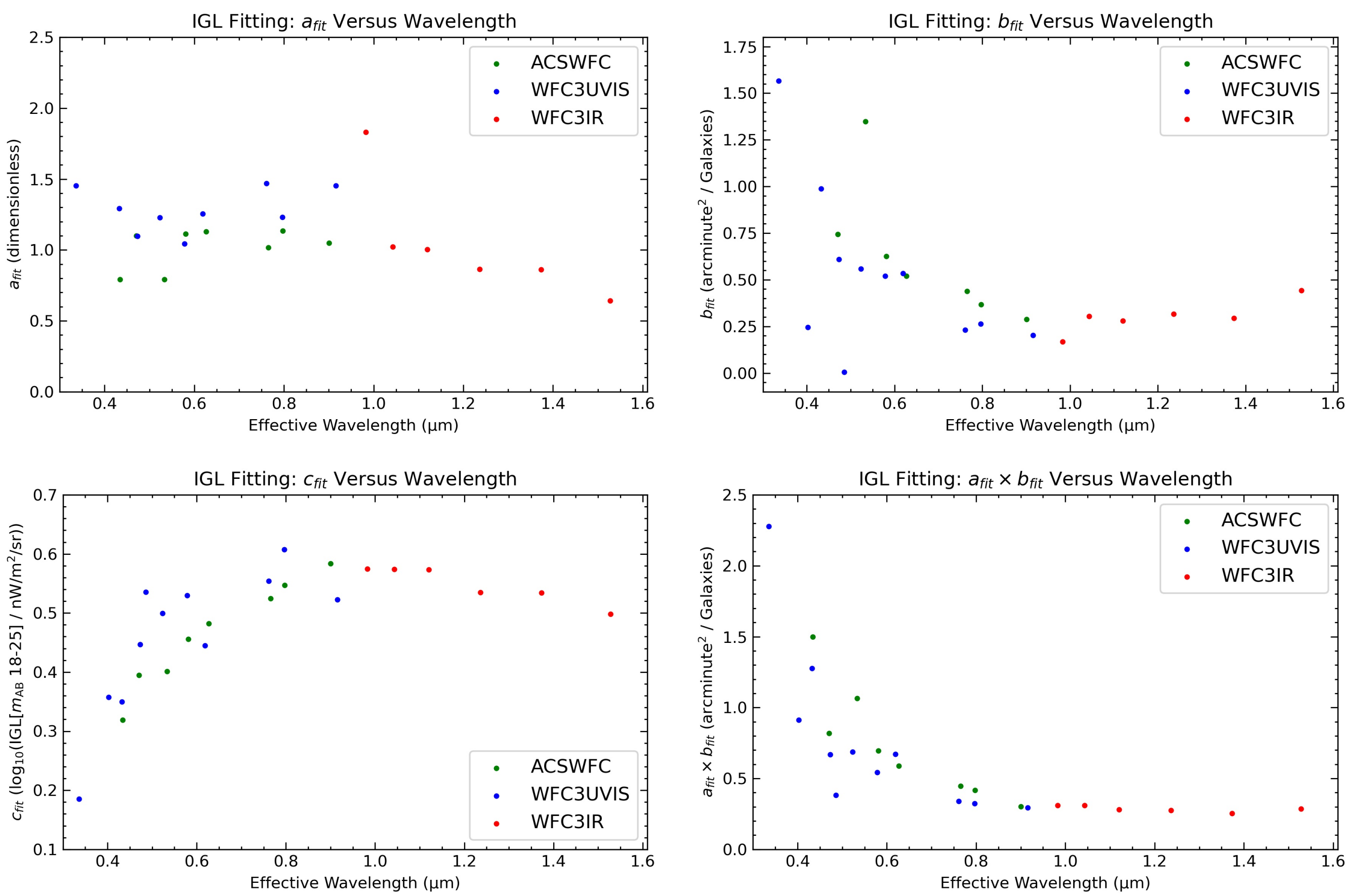

**Figure 17.** Single-visit IGL fitting parameters ($a_{fit}$, $b_{fit}$, and $c_{fit}$) as a function of wavelength. The coefficient of the IGL fitting function's derivative ($a_{fit} \times b_{fit}$) is also plotted here as function of wavelength (bottom right) and shows a well-defined trend.

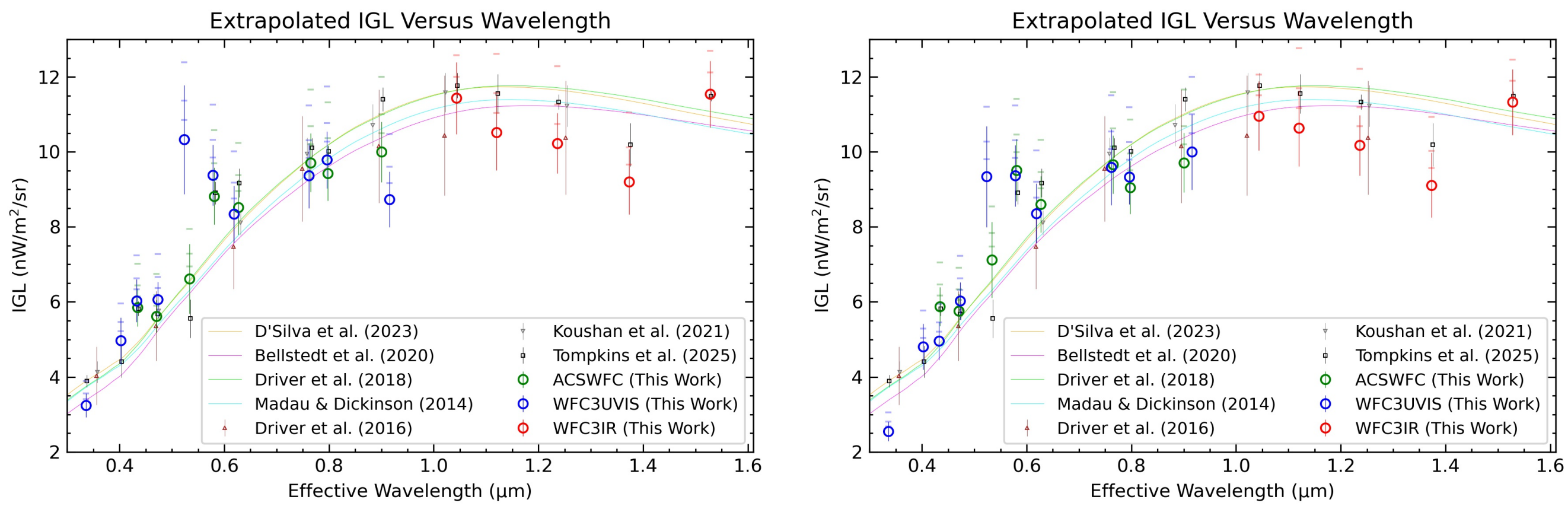

**Figure 18.** Extrapolated IGL versus wavelength for 23 SKYSURF Hubble filters for the single-visit (**Left**) and multi-visit (**Right**) data. The black points are the results of the Tompkins et al. (2025) ProFound analysis of the SKYSURF data. The green, red, and blue points are the ACSWFC, WFC3IR, and WFC3UVIS results of this work obtained by multiplying our partial IGL[$m_{\rm AB}$ 18-25] values by the ratio of the extrapolated over partial IGL[$m_{\rm AB}$ 18-25] measurements of Tompkins et al. (2025). The transparent colored dashes above each of this work's measurements show where the measurements would land if a 5% (lowest dash), 10% (middle dash), and 20% (highest dash) correction were performed to account for potential flux lost during source detection. The orange, magenta, lime, and cyan lines are IGL models from the cosmic star formation history analyses of D'Silva et al. (2023), Bellstedt et al. (2020), Driver et al. (2018), and Madau & Dickinson (2014) respectively. It's important to note that ACSWFC F555W and WFC3UVIS F555W are frequently used for specific stellar population studies and often capture dense star fields, which makes the available data challenging to work with for both this work and Tompkins et al. (2025). Although our IGL measurements for the F555W filters are included here, they should be considered with this in mind.

As further validation of our IGL fitting method, we provide plots of our IGL fitting parameters ($a_{fit}$, $b_{fit}$, $c_{fit}$, and $a_{fit} \times b_{fit}$) as a function of wavelength in Figure 17. Interestingly, the coefficient of the fitting model's derivative ($a_{fit} \times b_{fit}$) can be observed decreasing at a decreasing rate as wavelength increases–a well-defined trend that has never been noted in previous IGL investigations, but can be seen here due to the nature of our IGL measurement method. This suggests a more gradual growth in partial IGL as a function of $N_{m_{\rm AB}<20}$ at redder wavelengths. We hypothesize that this trend may be caused by higher numbers of satellite galaxies appearing in low redshift galaxies at bluer wavelengths, which may become less visible in images as wavelength increases. However, this has not been confirmed, and we leave a determination of the specific cause of this trend for a future study.

#### 5.3.2. *Comparing IGL Fitting Results With Previous Measurements and Models*

Figure 15 shows the partial IGL results of this work plotted against wavelength and Figure 18 shows the extrapolated (total) IGL as a function of wavelength. The results of Tompkins et al. (2025), Koushan et al. (2021), and Driver et al. (2016) are also included in both figures, while IGL models generated using the ProSpect (Robotham et al. 2020) SED fitting+generation code and the cosmic star formation histories of D'Silva et al. (2023), Bellstedt et al. (2020), Driver et al. (2018), and Madau & Dickinson (2014) are included with our extrapolated IGL results in Figure 18. The results of this work, obtained using our novel IGL fitting method, are in general agreement with the independent SKYSURF IGL analysis of Tompkins et al. (2025), the previous IGL measurements of Koushan et al. (2021) and Driver et al. (2016), and theoretical IGL models. The error bars of our IGL measurements often overlap with the Tompkins et al. (2025) error bars, the Driver et al. (2016) error bars, or both simultaneously. A notable difference between the results of this work and the Tompkins et al. (2025) results are the somewhat lower IGL measurements of this work relative to the corresponding measurements of Tompkins et al. (2025), which is more pronounced for the redder WFC3IR filters. This may partially be due to a difference in photometry between Source Extractor and ProFound, as simulations show that Source Extractor misses about 10-20% of total galaxy flux while ProFound misses 10-15% of total galaxy flux on average in the magnitude range $18 < m_{\rm AB} < 23$ mag (Robotham et al. 2018a, see Figure 14).

#### 5.3.3. *Limitations of the IGL Fitting Method*

Although the IGL fitting method used in this work is an excellent tool for analyzing patterns and trends in IGL data, it is difficult to obtain fits in the manner performed in this work without sufficient data. Smaller scale IGL investigations that consider only a handful of images should not expect to obtain robust results via fitting, as the reliability of the fits is dependent on having a sizable collection of images from different parts of the sky, which is uniquely the case for project SKYSURF's massive database. The same can also be said of the unique star-galaxy separation methods discussed in Section 3.2, which has thrived on the SKYSURF dataset. Appendix F illustrates the variation in IGL measurements as a function of randomly selected mosaics with $N_{m_{\rm AB}<20} < 3.0$ used for analysis. Ideally, an IGL investigation attempting to use the IGL fitting measurement method should have $\gtrsim 30$ or so mosaics with $N_{m_{\rm AB}<20} < 3.0$ for a given filter to mitigate the influence of cosmic variance. The main benefit of this work lies not in the implementation of the techniques themselves, but in the application of the results obtained here. Regardless of scale, any IGL investigation seeking to calibrate for cosmic variance can do so by using the fits obtained in this work, which are available in Table 5 and illustrated in Appendices D and E.

In addition, it is important to remain cognizant of the fact that these IGL measurements are indeed lower limits on the total galaxy light due to missing light from faint objects and undetected light in galaxy outskirts. Some studies such as Cheng et al. (2021) suggest that a non-negligible amount of galaxy light is contained in galactic halos (e.g. intra-halo light, or IHL), but this work does not perform an assessment of IHL or other missing light in galactic outskirts, which places it in the context of other investigations such as Tompkins et al. (2025), Koushan et al. (2021), and Driver et al. (2016) that also refrain from such an assessment. Similarly, studies such as Cooray et al. (2012) and Cheng & Chang (2022) perform masking in their analysis that assumes no corrections to the IGL.

Quantifying and correcting for undetected light in the outskirts of galaxies is a challenging problem in and of itself. Table 1 of Graham & Driver (2005) suggests that a correction of $\approx 10\%$ is warranted for elliptical galaxies (sersic index $n = 4$), but only $\approx 4\%$ for disk galaxies ($n = 1$) and $< 1\%$ for some dwarf galaxies ($n \approx 0.5$), and this is assuming perfectly symmetrical galaxies. However, as Figure 14 of Robotham et al. (2018a) demonstrates, corrections to the magnitudes of more realistic galaxies can be uncertain by up to $\approx 1$ magnitude near the detection limit, which highlights the non-trivial nature of this problem. Therefore, like several previous IGL investigations, we treat undetected light in galaxy outskirts as a separate problem that is beyond the scope of this work. To transparently reflect this limitation in our results, we have added faint dashes above our IGL measurements in Figures 15 and 18 to show where our IGL measurements would fall given 5%, 10%, and 20% upward corrections.

### 5.4. *Suggestions for Future IGL Investigations*

One of the main goals of IGL investigations is to measure the IGL at multiple wavelengths accurately enough to determine which theoretical IGL models fit best to the Universe in which we exist, which would enable us to distinguish between competing cosmic star formation histories. This work elucidates many of the obstacles that can be encountered in IGL investigations of this scale and presents a variety of unique solutions to pave the way for future projects that may be able to further constrain the IGL. In particular, there are at least three major space telescopes that are well poised to achieve the accurate IGL measurements sought after in this field:

**(1) The James Webb Space Telescope:** Project SKYSURF-IR, formerly called DARK-SKY (Windhorst et al. 2024), is a direct sequel SKYSURF. It will perform the same rigorous data processing on a large collection of JWST archival data in an effort to further scrutinize the EBL. Since JWST sees the Universe in the infrared, SKYSURF-IR will obtain multiple new IGL measurements in a wavelength range almost entirely invisible to Hubble. This work foreshadows the exciting new JWST results that await us in the near future, and nearly all of the methods herein are directly applicable to the ongoing successor project. Furthermore, far from being rendered obsolete, the Hubble results obtained in this work function as a valuable probe of the optical IGL that will perfectly complement the SKYSURF-IR JWST infrared IGL analysis.

**(2) SPHEREx:** With the recent launch of SPHEREx on March 11, 2025, the astrophysics community will soon

have another great dataset for an IGL investigation. Though it will not peer as deeply into the Universe as Hubble or Webb, SPHEREx will cover far more area. It will map the entire sky with an unprecedented color resolution of 102 different wavelength bands in the range 0.75–5.0 $\mu$m. When SPHEREx completes its two-year mission, the resulting dataset will be virtually free of field selection biases that are intrinsic to archival Hubble and Webb IGL investigations. It will also enable a detailed probe of the IGL in 102 different colors–more than Hubble and Webb combined. Given its wavelength span, color resolution, and area coverage, the SPHEREx dataset will be one of the best datasets for EBL research in the near future. Moreover, the wavelength range of SPHEREx partially overlaps with Hubble's redder filters, which would enable comparisons with the IGL results of this work.

**(3) The Nancy Grace Roman Space Telescope:** Looking further ahead, the Roman Space Telescope, formerly known as WFIRST (Wide Field Infrared Survey Telescope), is NASA's next flagship mission, scheduled to launch by May 2027 (Spergel et al. 2013). Roman will have 8 filters that span a wavelength range of 0.48–2.3 $\mu$m, which overlaps significantly with the Hubble IGL results presented in this work. The mission will capture data that is just as deep as Hubble data while covering significantly more area. Like SPHEREx, Roman is a survey telescope, albeit with a much lower color resolution than SPHEREx. Though Roman has lower color resolution, its superior wavelength coverage and field of view will significantly improve studies of the IGL, and applying the new techniques described in this work to Roman data would be a fascinating way to further probe the IGL several years down the road.

## 6. CONCLUSION

This work in conjunction with its companion paper, Tompkins et al. (2025), is the largest IGL investigation ever performed. About 140,000 images spread across three Hubble cameras and 28 filters were grouped into approximately 38,000 single-visit mosaics and 8,000 multi-visit mosaics, which were subsequently analyzed with Source Extractor to generate thousands of object catalogs. The massive scale of the SKYSURF dataset enabled novel star-galaxy separation methods and revealed interesting features in object magnitude versus object size plots that have never been seen before. The SKYSURF dataset also led to the development of a novel IGL fitting method. The IGL fitting method produces measurements that are largely in agreement with theoretical models and the results of Tompkins et al. (2025). This work paves the way for future large-scale IGL investigations by illustrating solutions to some of the most difficult obstacles that can be encountered in this type of work, and our IGL fitting results provide a novel way for other IGL investigations to calibrate for cosmic variance. Perhaps someday in the future, the methods described here will facilitate the process of obtaining IGL measurements that are even more robust using data from other space telescopes, such as JWST, SPHEREx, and the upcoming Nancy Grace Roman Space Telescope.

## 7. DATA ACCESS

All of the FLC/FLT images used in our analysis were obtained from the Mikulski Archive for Space Telescopes (MAST). The IPPPSSOOT ID lists of the specific observations downloaded for this project are available on MAST via doi:10.17909/ajha-4009. All important SKYSURF data products discussed in this work can also be retrieved from MAST via doi:10.17909/ajha-4009, including our ProFound sky-maps, star-galaxy separation parameters, IGL fitting parameters, and drizzled mosaics (WHT, SCI, CTX, and modified WHT).

## 8. ACKNOWLEDGEMENTS

We thank Annalisa Calamida, Phil Korngut, and Tod Lauer for helpful discussions. In addition, we thank John Mather for his helpful comments regarding his suggestion of a spherical distribution of Sun-approaching comets from SOHO/STEREO and his reference to the Sungrazer project. We thank HST Archive staff at STScI for their expert advice on HST component temperatures.

This project is based on observations made with the NASA/ESA Hubble Space Telescope and obtained from the Hubble Legacy Archive, which is a collaboration between the Space Telescope Science Institute (STScI/NASA), the Space Telescope European Coordinating Facility (ST-ECF/ESA), and the Canadian Astronomy Data Centre (CADC/NRC/CSA).

We thank Ms. Desiree Crawl, Prof. Thomas Sharp, and the NASA Space Grant Consortium in Arizona for consistent support of our many undergraduate SKYSURF researchers at ASU during the pandemic. We acknowledge support for HST programs AR-09955 and AR-15810 provided by NASA through grants from the Space Telescope Science Institute, which is operated by the Association of Universities for Research in Astronomy, Incorporated, under NASA contract NAS5-26555.

We also acknowledge the indigenous peoples of Arizona, including the Akimel O'odham (Pima) and Pee Posh (Maricopa) Indian Communities, whose care and keeping of the land has enabled us to be at ASU's Tempe campus in the Salt River Valley, where this work was conducted.

We are grateful to the anonymous referee, whose suggestions greatly improved this paper.

*Software:*
Astropy: http://www.astropy.org (Astropy Collaboration et al. 2013, 2018, 2022)
IDL Astronomy Library: https://idlastro.gsfc.nasa.gov/ (Landsman 1993)
Photutils: https://photutils.readthedocs.io/en/stable/ (Bradley et al. 2020)
`ProFound`: https://github.com/asgr/ProFound (Robotham et al. 2018a)
`ProFit`: https://github.com/ICRAR/ProFit (Robotham et al. 2018b)
Source Extractor: https://www.astromatic.net/software/sextractor/ or https://sextractor.readthedocs.io/en/latest/ (Bertin, E. & Arnouts, S. 1996).

*Facilities:*
Hubble Space Telescope Mikulski Archive: https://archive.stsci.edu
Hubble Legacy Archive (HLA): https://hla.stsci.edu
Hubble Legacy Catalog (HLC): https://archive.stsci.edu/hst/hsc/
SVO Filter Profile Service: This research has made use of the SVO Filter Profile Service "Carlos Rodrigo", funded by MCIN/AEI/10.13039/501100011033/ through grant PID2020-112949GB-I00.
This work is based on observations taken by the 3D-HST Treasury Program (GO 12177 and 12328) with the NASA/ESA HST, which is operated by the Association of Universities for Research in Astronomy, Inc., under NASA contract NAS5-26555.

# APPENDIX

## A. ASTRODRIZZLE INPUT PARAMETERS

Here we present the complete list of AstroDrizzle parameters described in Section A for both run 1 and run 2.

**AstroDrizzle Run 1 Parameters**

driz_sep_kernel = 'square'
final_kernel = 'square'
crbit = 4096
resetbits = 4096
driz_sep_pixfrac = 1.0
final_pixfrac = 1.0
driz_sep_scale = 0.06
final_scale = 0.06
final_wht_type = 'IVM'
skysub = True
median = True
blot = True
driz_cr = True
driz_cr_corr = True
static = True
driz_combine = True
driz_cr_grow = 2
driz_cr_ctegrow = 5 (ACSWFC and WFC3UVIS)
driz_cr_ctegrow = 0 (WFC3IR)
combine_type = 'minmed' (6 or less input images)
combine_type = 'median' (7 or more input images)

**AstroDrizzle Run 2 Parameters**

driz_sep_kernel = 'square'
final_kernel = 'square'
crbit = 4096
resetbits = 0
driz_sep_pixfrac = 1.0
final_pixfrac = 1.0
driz_sep_scale = 0.06
final_scale = 0.06
final_wht_type = 'IVM'
skysub = False
median = False
blot = False
driz_cr = False
driz_cr_corr = False
static = False
driz_combine = True
driz_cr_grow = 2
driz_cr_ctegrow = 5 (ACSWFC and WFC3UVIS)
driz_cr_ctegrow = 0 (WFC3IR)
combine_type = 'minmed' (6 or less input images)
combine_type = 'median' (7 or more input images)

## B. MAGNITUDE VERSUS SIZE PLOTS AND NUMBER COUNTS (SINGLE-VISIT)

Here we present the single-visit ACSWFC F435W, ACSWFC F606W, WFC3UVIS F606W, and WFC3IR F160W magnitude vs size plots and number counts. Plots for other filters are available as an electronic supplement.

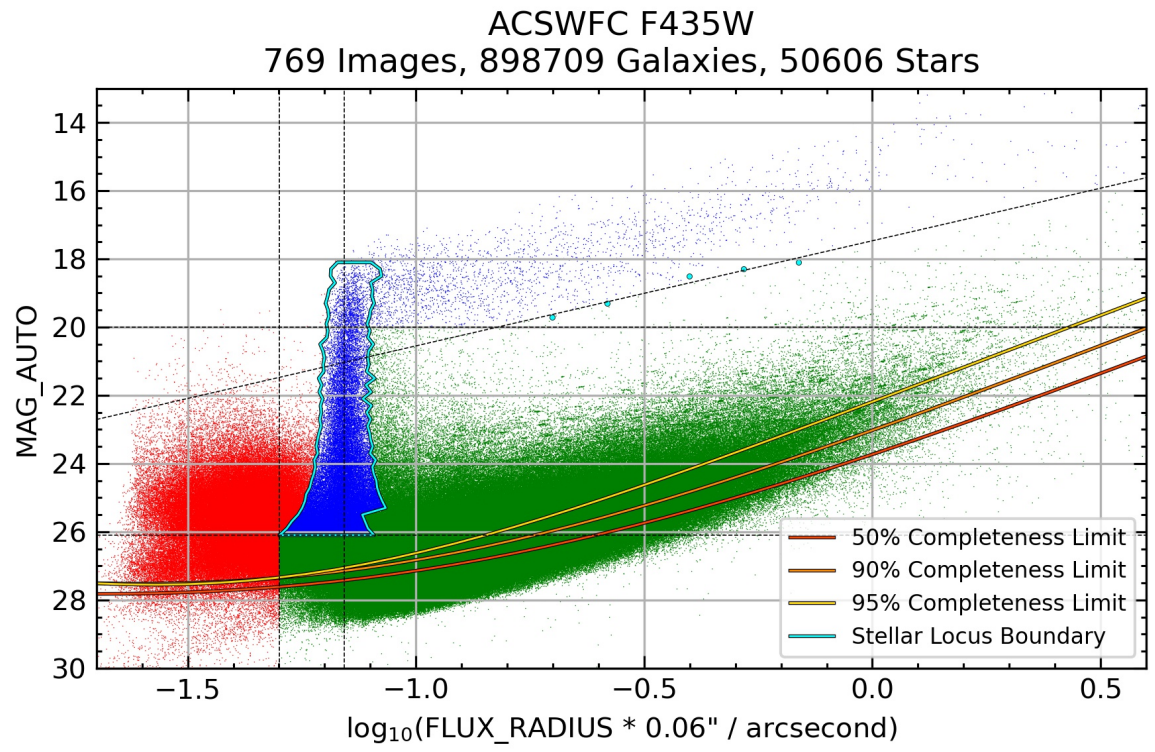

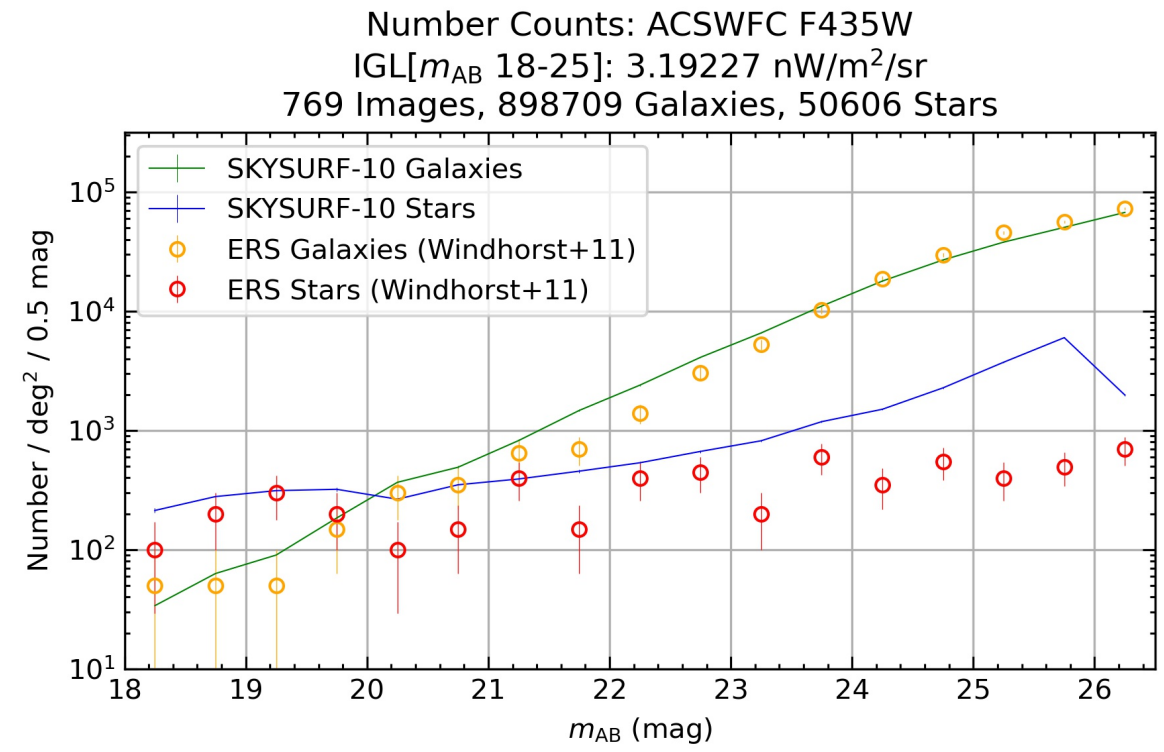

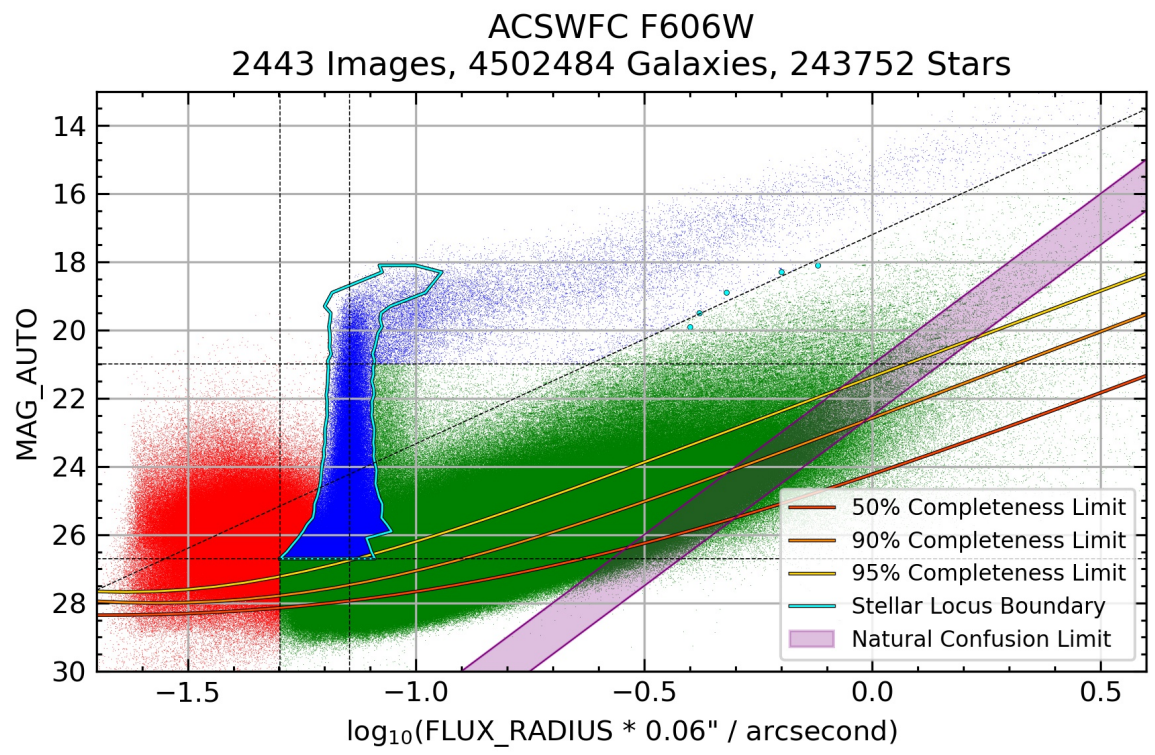

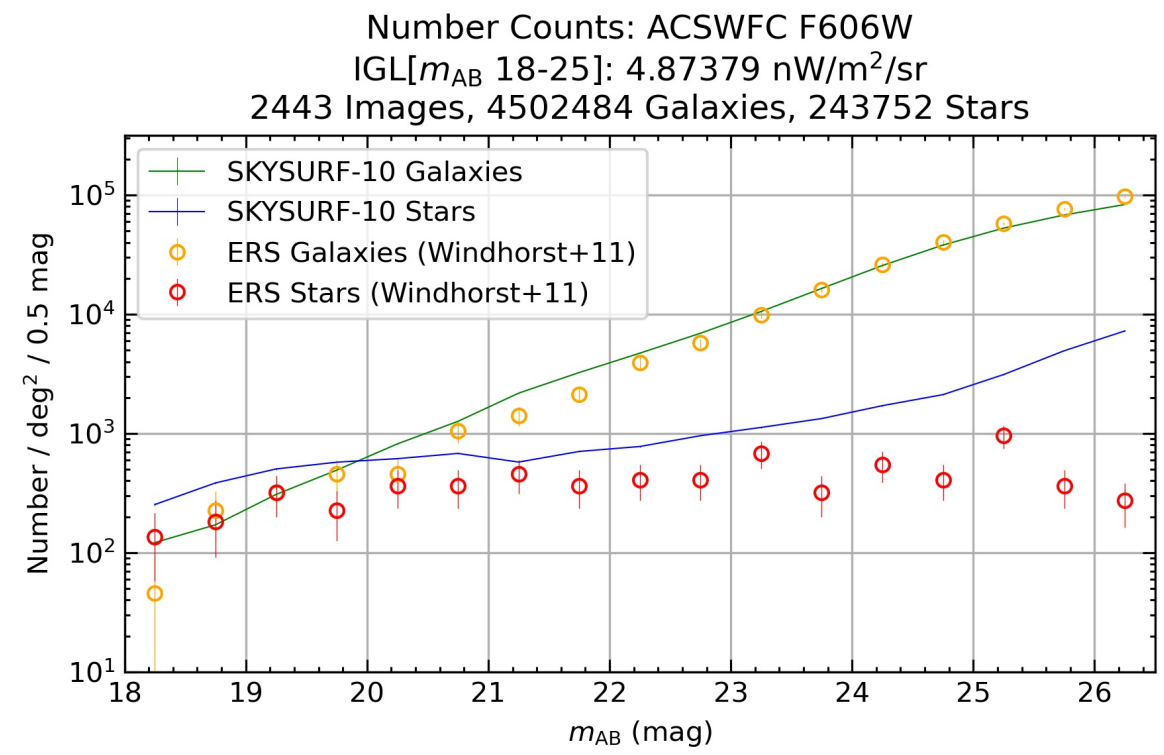

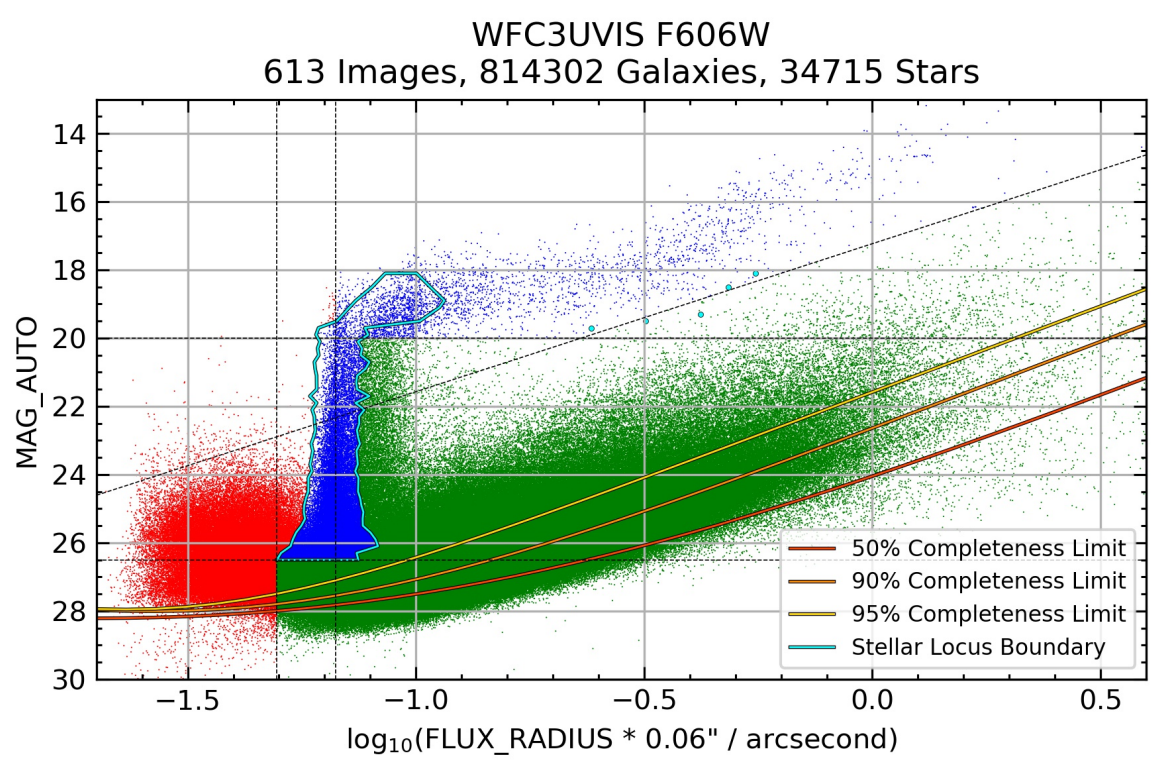

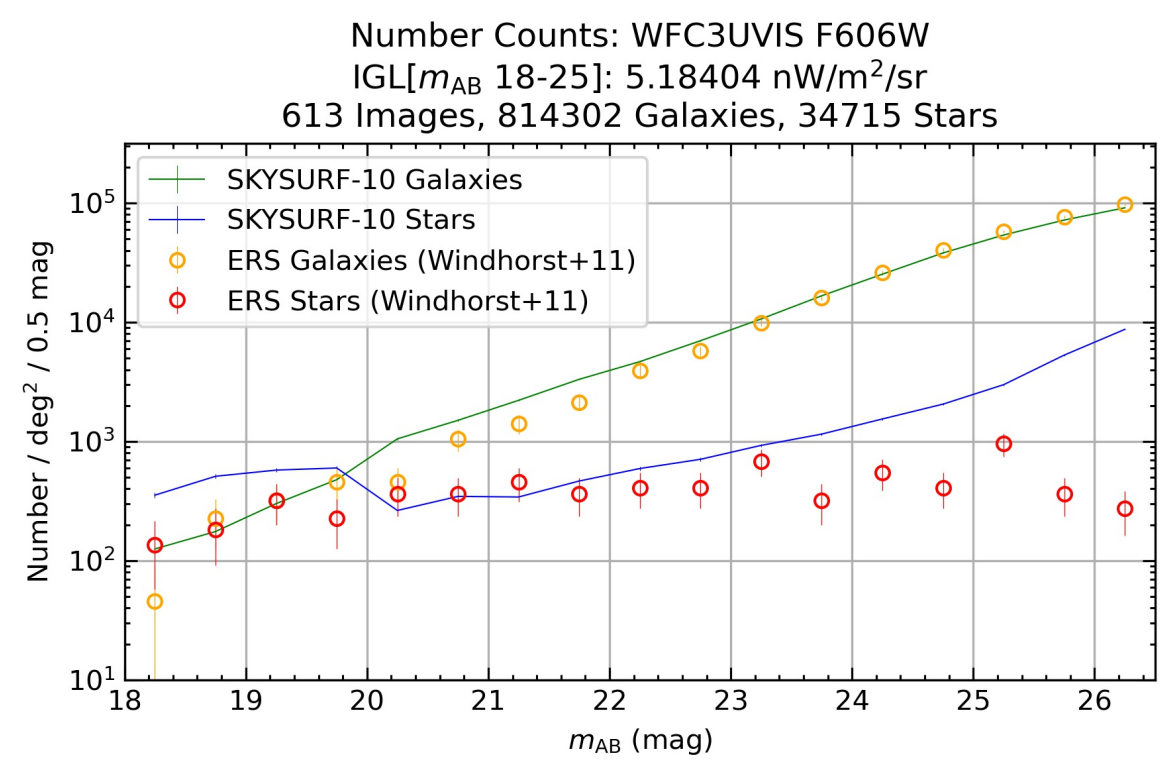

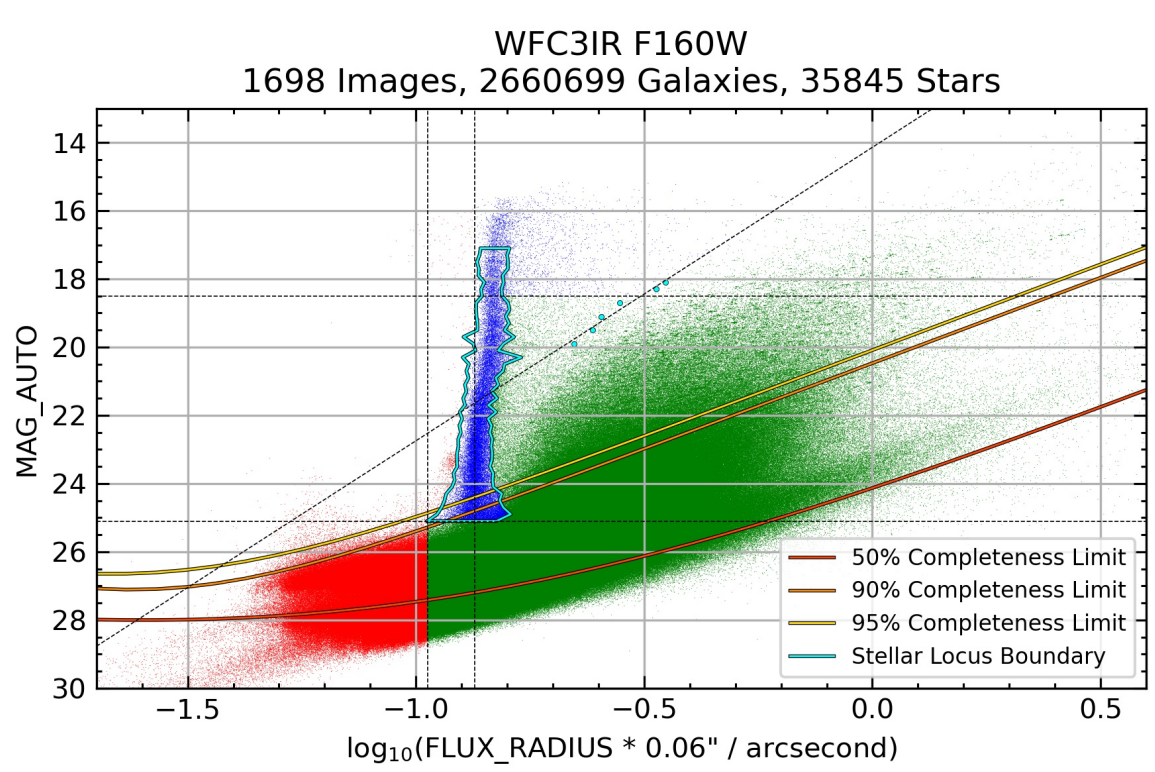

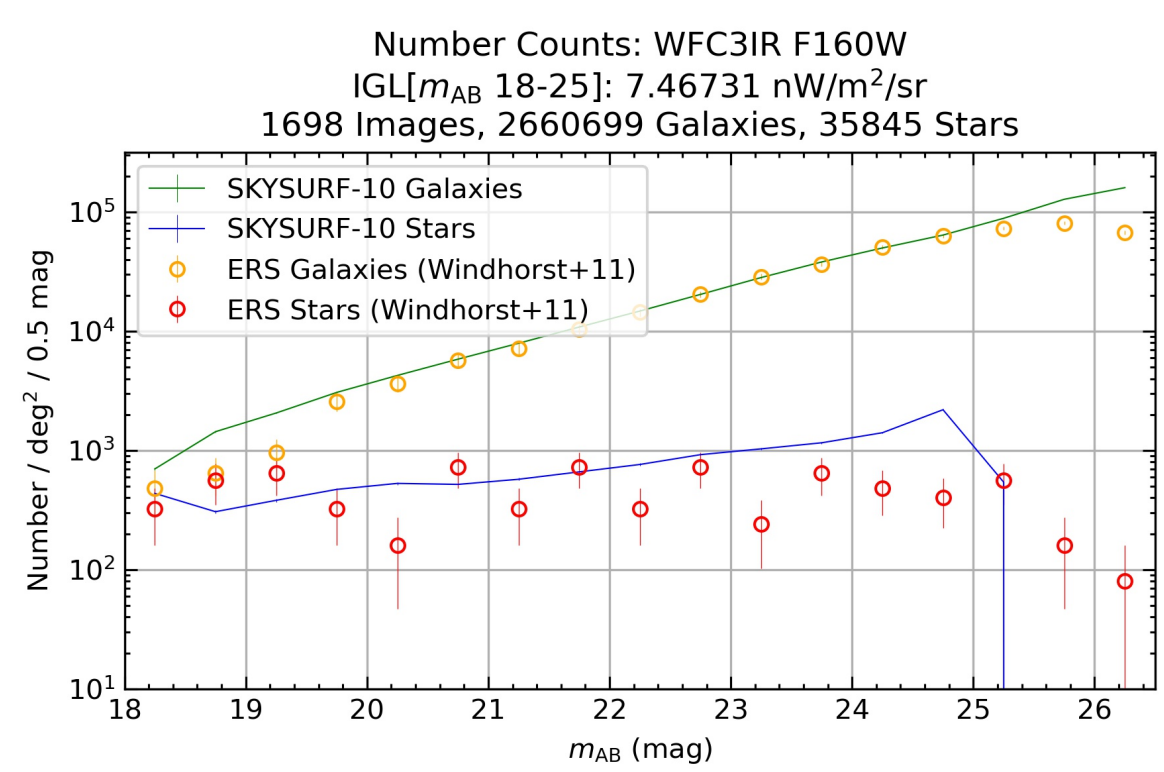

## C. MAGNITUDE VERSUS SIZE PLOTS AND NUMBER COUNTS (MULTI-VISIT)

Here we present the multi-visit ACSWFC F435W, ACSWFC F606W, WFC3UVIS F606W, and WFC3IR F160W magnitude vs size plots and number counts. Plots for other filters are available as an electronic supplement.

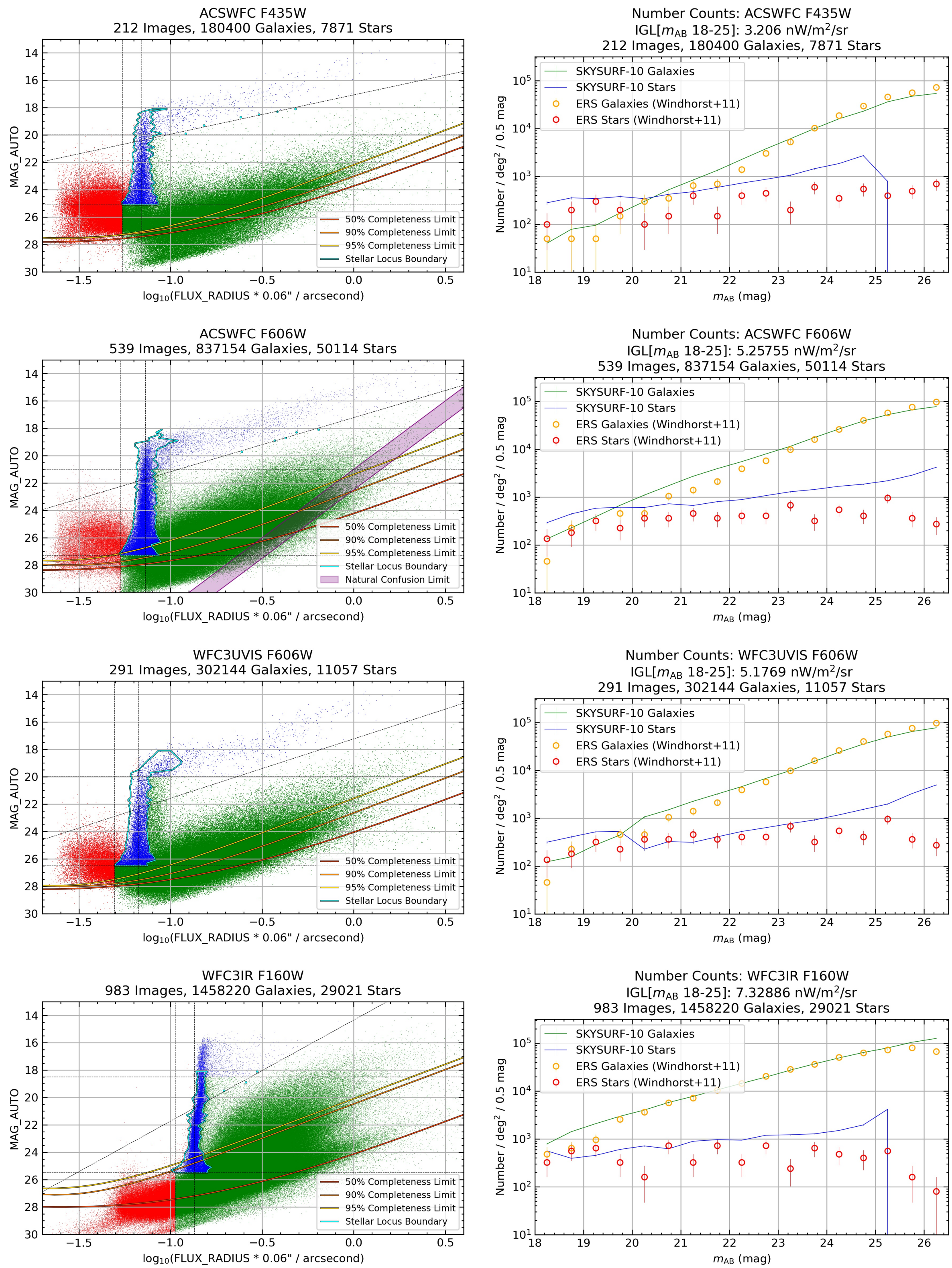

## D. PARTIAL IGL VS BRIGHT GALAXIES PLOTS AND RESIDUALS HISTOGRAMS (SINGLE-VISIT)

Here we present the single-visit ACSWFC F435W, ACSWFC F606W, WFC3UVIS F606W, and WFC3IR F160W partial IGL vs $N_{m_{\rm AB}<20}$ plots and residual histograms. Plots for other filters are available as an electronic supplement.

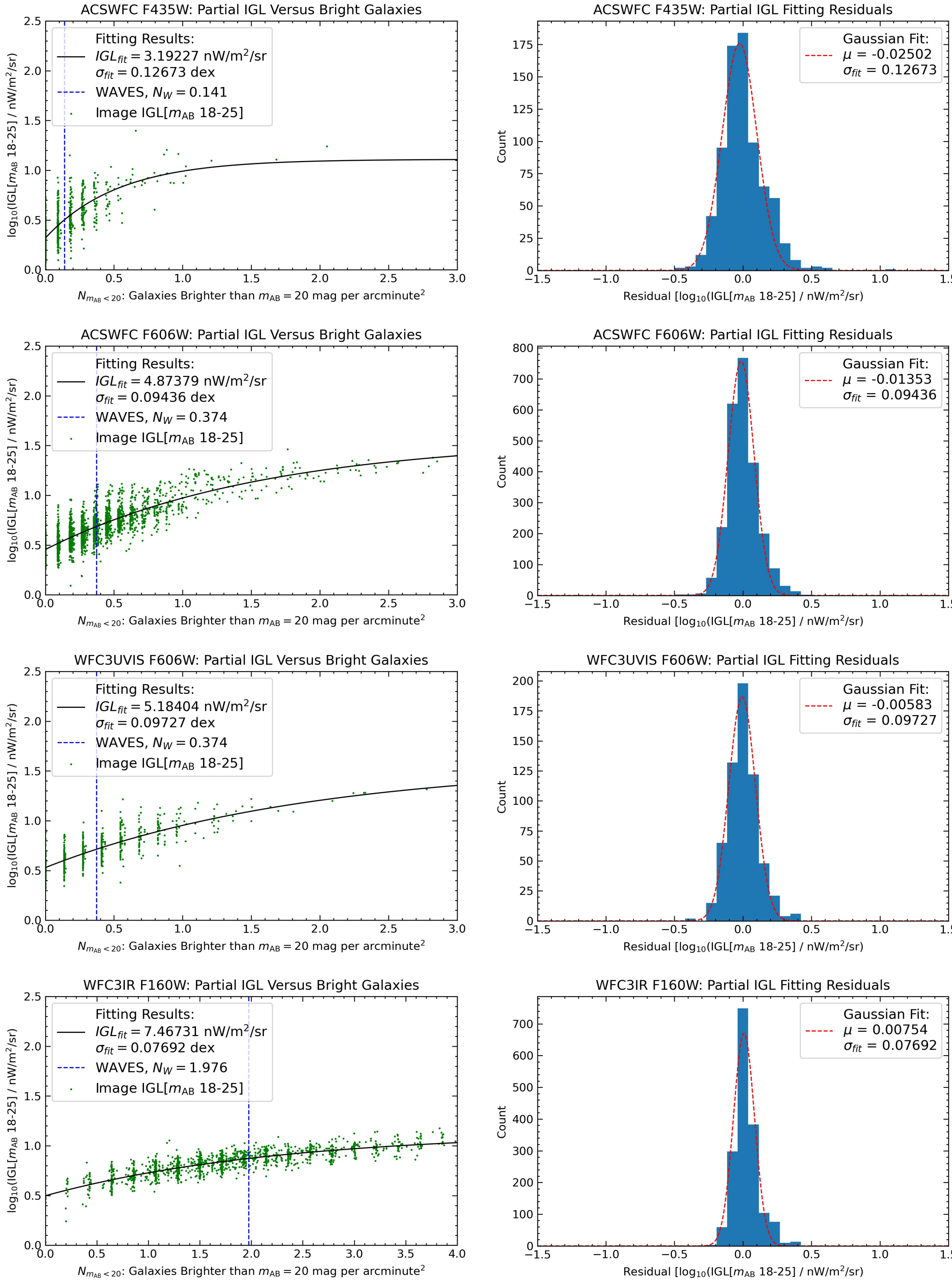

## E. PARTIAL IGL VS BRIGHT GALAXIES PLOTS AND RESIDUALS HISTOGRAMS (MULTI-VISIT)

Here we present the multi-visit ACSWFC F435W, ACSWFC F606W, WFC3UVIS F606W, and WFC3IR F160W partial IGL vs $N_{m_{\rm AB}<20}$ plots and residual histograms. Plots for other filters are available as an electronic supplement.

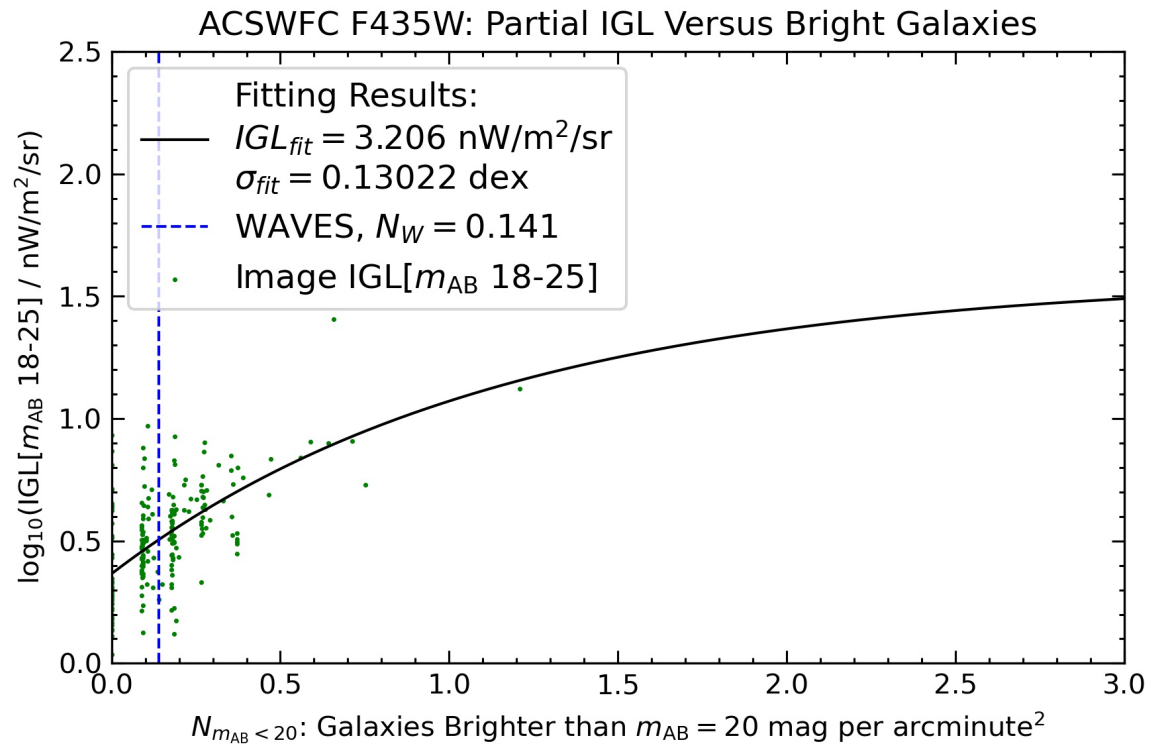

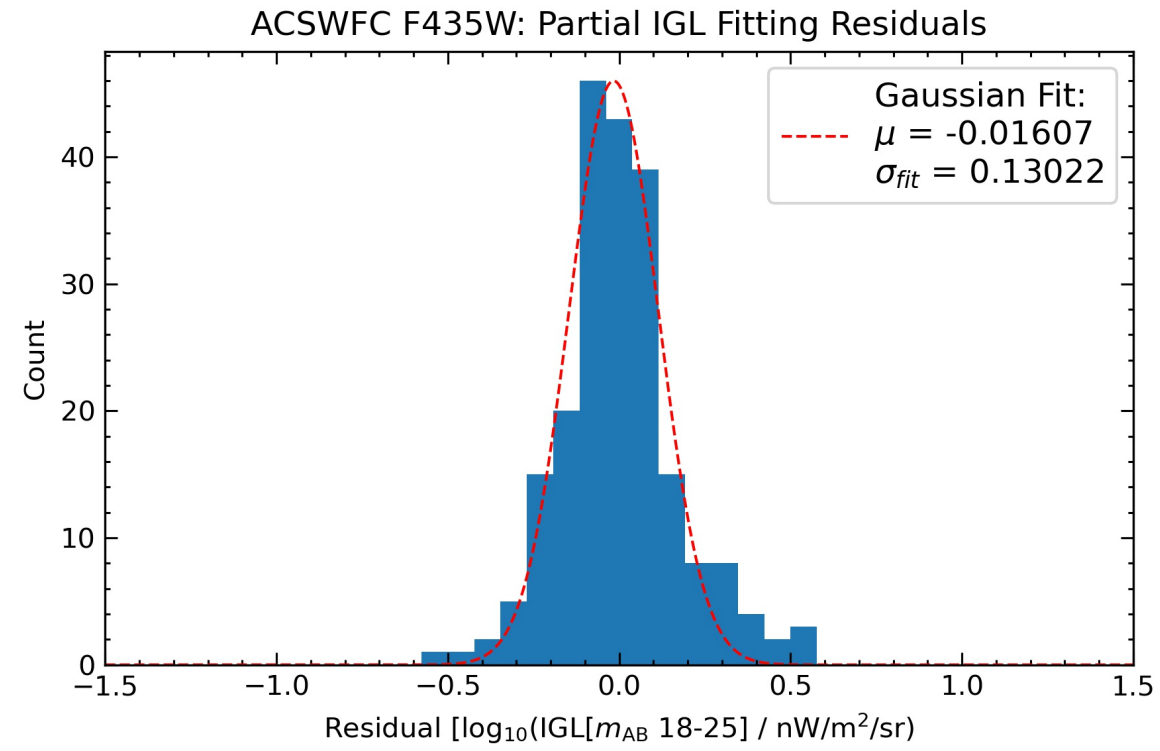

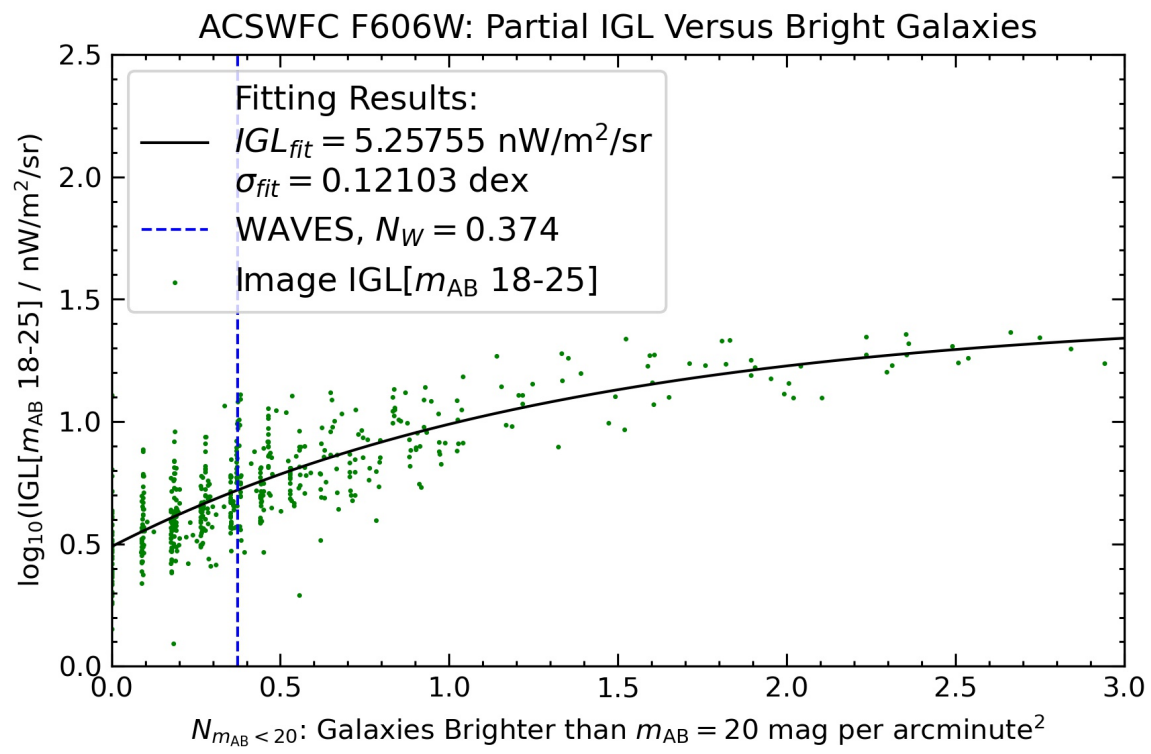

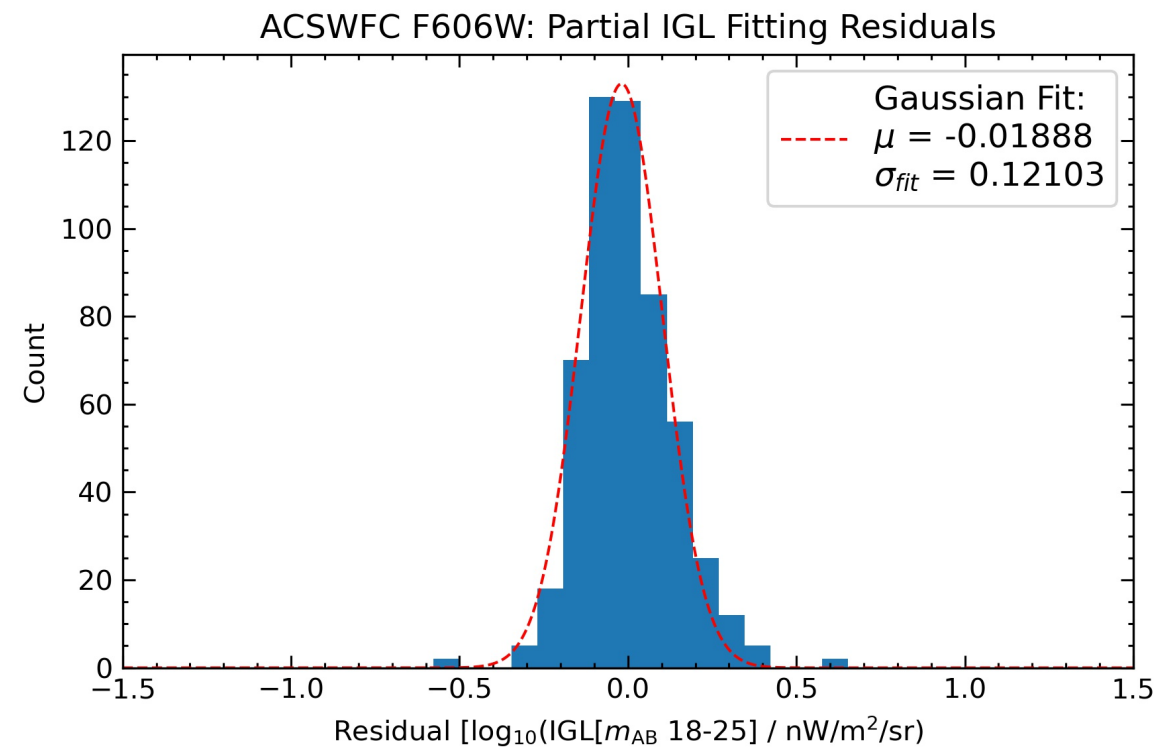

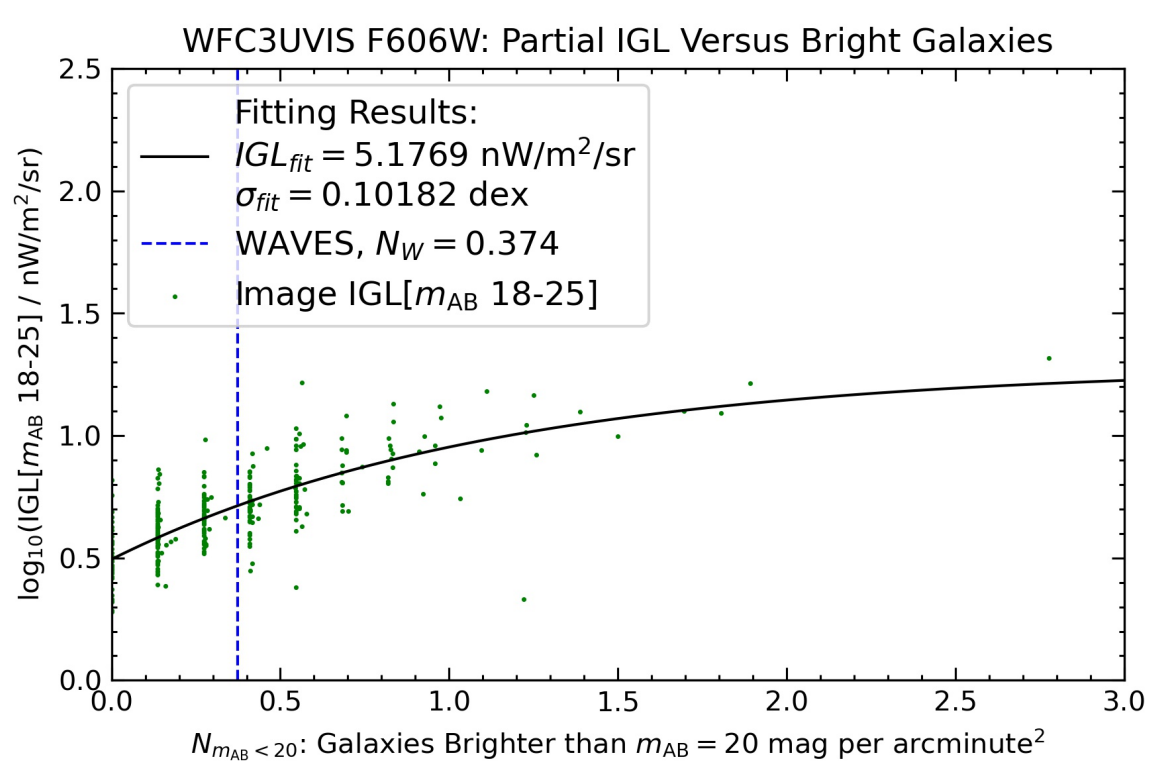

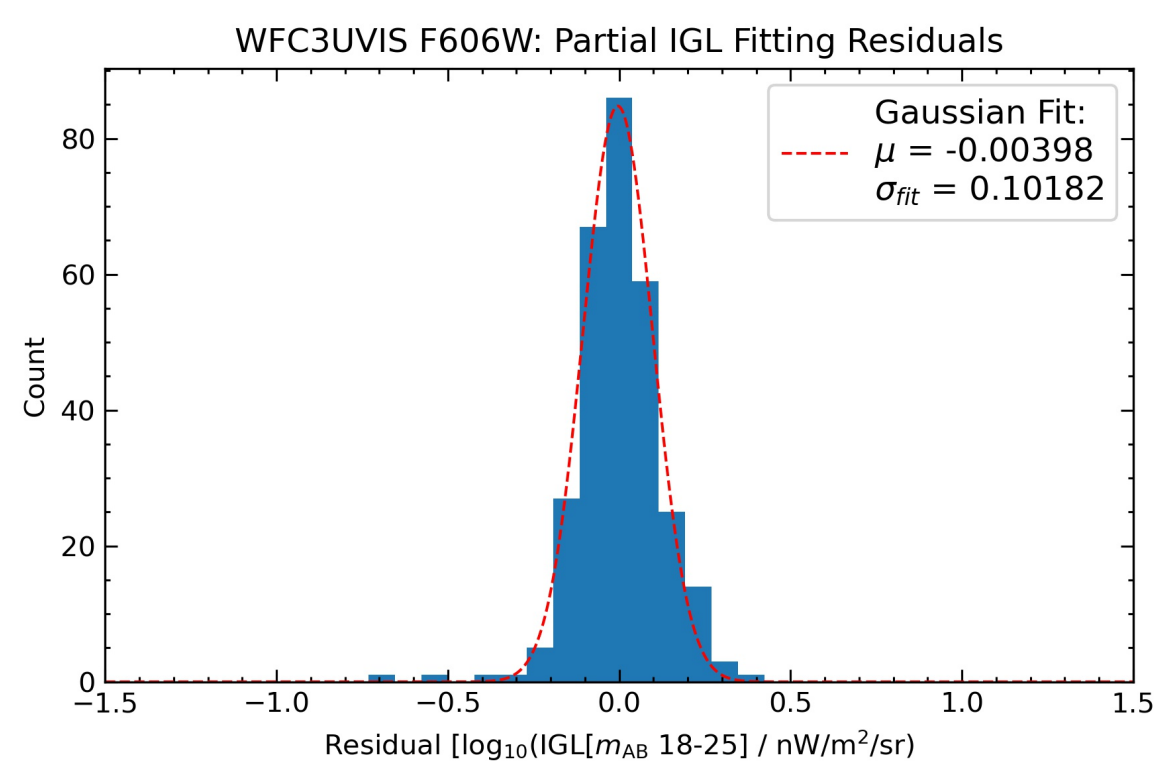

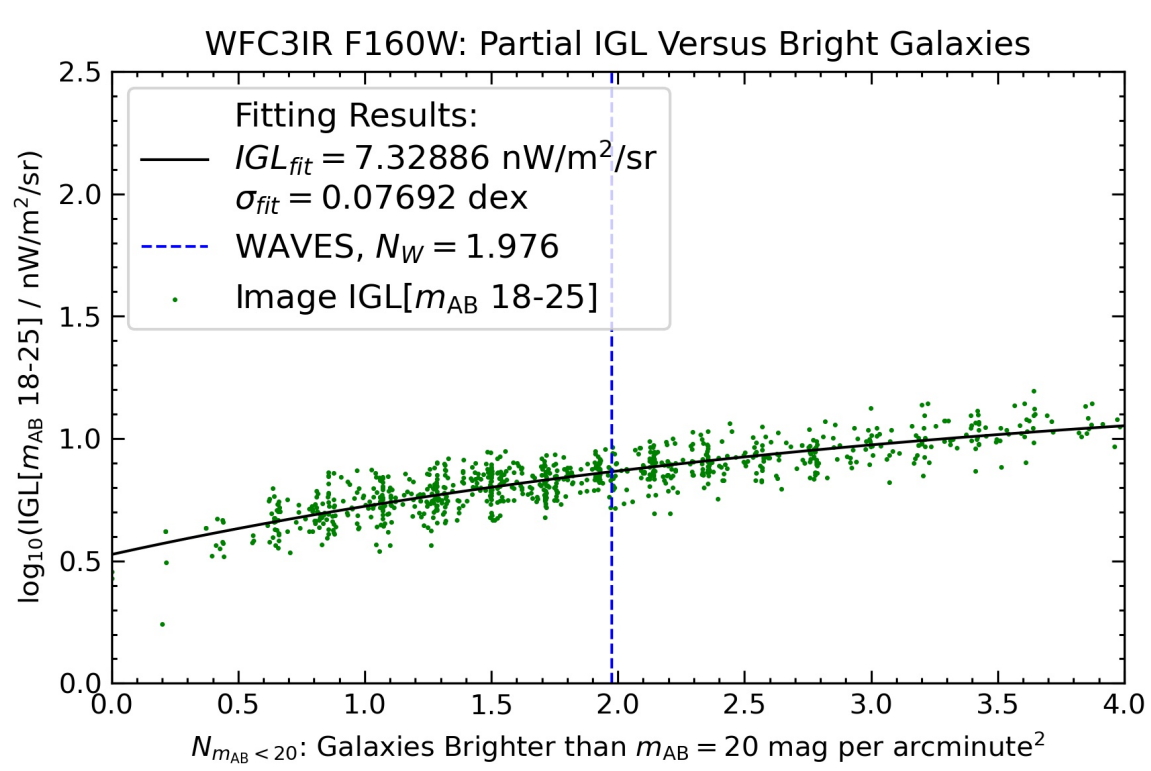

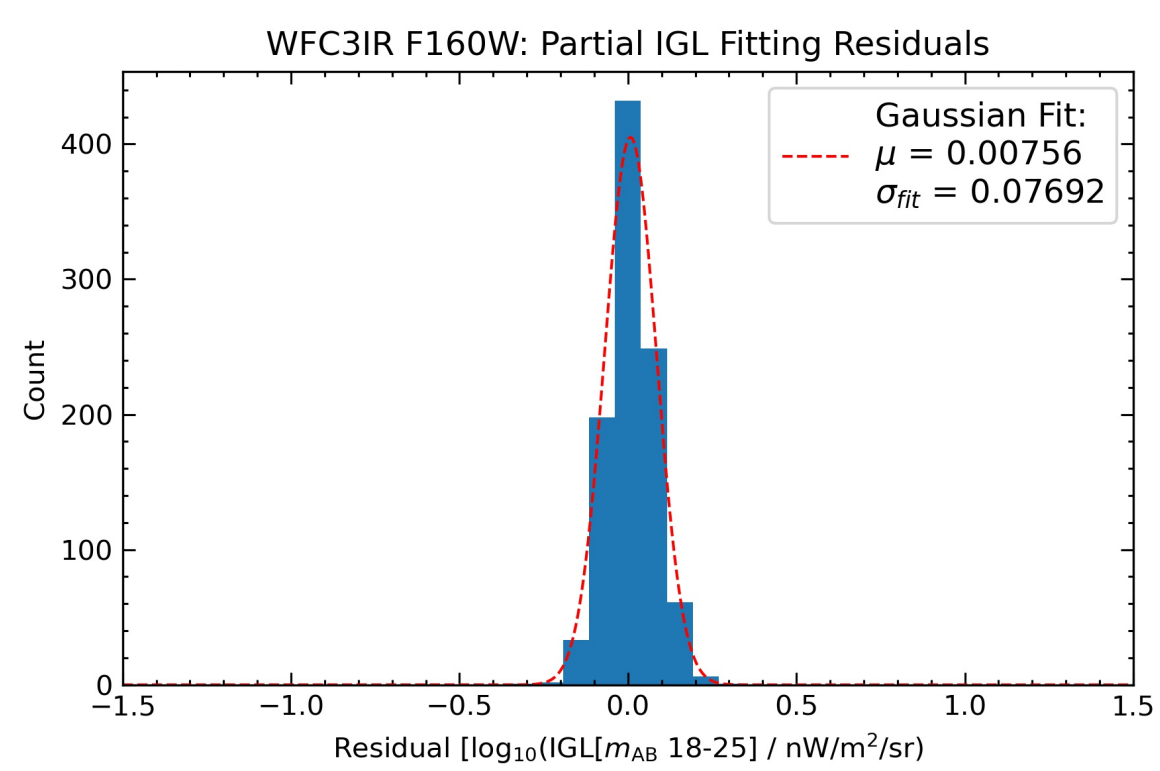

## F. IGL FITTING ANALYSIS

Here we analyze the impact of target selection and IGL fitting bound selection on our partial IGL results. We present plots of partial IGL as a function of randomly selected mosaics, as a function of fitting lower bound, and as a function of fitting upper bound for ACSWFC F606W, ACSWFC F814W, and WFC3IR F160W.

Figure 19 shows that, for a given collection of mosaics, the impact of target selection on the results quickly decreases as the number of available targets increases, varying by a standard deviation ($1\sigma$) as small as $\approx 0.125$ nW/m$^2$/sr for 100 randomly selected targets. This suggests, at minimum, that the majority of our single-visit and multi-visit measurements are internally precise. However, our multi-visit partial IGL measurement for ACSWFC F606W is 0.39 nW/m$^2$/sr greater than the single-visit measurement, which is beyond the level of variation we expect given the results shown in Figure 19. This suggests an additional source of uncertainty rooted in the differences between the single-visit and multi-visit datasets.

The differences that could potentially explain this uncertainty include the specific star-galaxy separation parameters determined, differences in how our weight map modification process impacts the two datasets, and the more stringent filtering criteria applied to the multi-visit dataset. For ACSWFC F606W specifically, applying the star-galaxy-separation parameters of the single-visit dataset to the multi-visit dataset decreased the multi-visit IGL measurement by only 0.05 nW/m$^2$/sr. Furthermore, removing weight map modification from the source detection process actually slightly increases the difference in measurements between the datasets from 0.39 nW/m$^2$/sr to 0.41 nW/m$^2$/sr, suggesting that this is not the main source of the single-visit versus multi-visit measurement difference either.

The remaining plausible explanation is that some fields in the single-visit dataset are simply not contained in the multi-visit dataset, resulting in a different distribution of mosaic IGL measurements between the two datasets. To test this hypothesis, we matched our single-visit ACSWFC F606W mosaics to our multi-visit ACSWFC F606W mosaics using a linking distance ($D_{link}$) equal to the diagonal of one ACSWFC image ($D_{link} = \sqrt{202^2 + 202^2}$ arcseconds) and obtained partial IGL measurements for both matched datasets. This resulted in a partial IGL measurement of 5.37 nW/m$^2$/sr for the single-visit matched data and 5.30 nW/m$^2$/sr for the multi-visit matched data–a measurement difference less than 0.1 nW/m$^2$/sr. This implies that our various methods are being applied in a consistent manner to both datasets and that the main source of uncertainty between the two datasets resides in the non-matched single-visit data. To account for this systematic uncertainty, we added an error quantity, $\sigma_{SVMV}$ (the log-scaled standard deviation of the partial IGL differences between the single-visit and multi-visit datasets), to our uncertainty analysis (see Section 4.4).

With regard to the specific fitting limits of our IGL analysis, Figures 20 and 21 demonstrate that there is a good amount of leeway available to us. Generally speaking, as long as the lower fitting limit is below $N_W$ and the upper fitting limit is above $N_W$, the specific choice of where to draw the boundary lines is inconsequential to the resulting partial IGL measurement. The specific upper bound matters slightly more for WFC3IR, as including very dense fields beyond $N_{m_{\rm AB}<20} = 5$ lowers the resulting partial IGL measurements by $\approx 0.1$ nW/m$^2$/sr, but even this is a relatively minor effect.

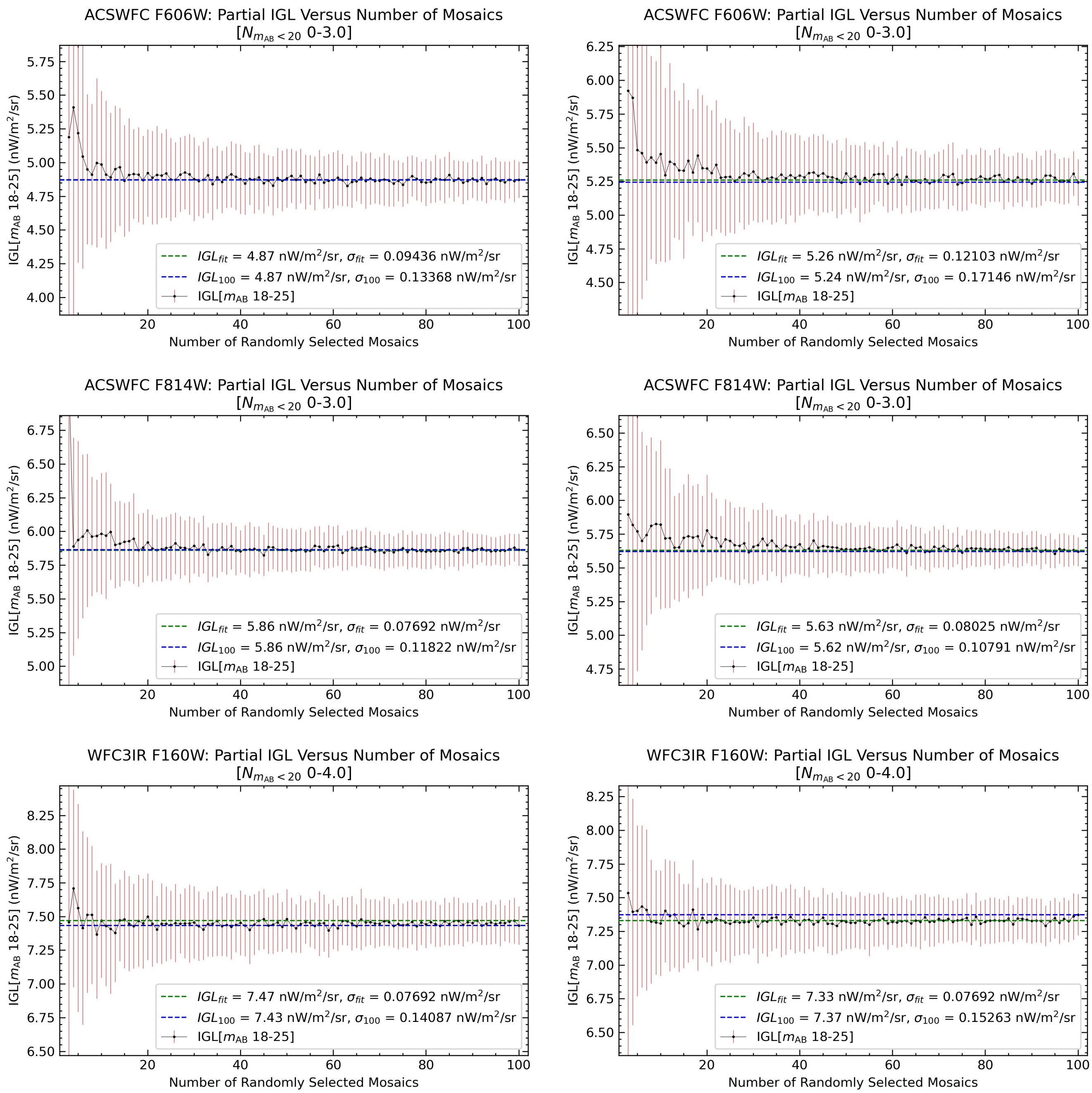

**Figure 19.** Partial IGL measurements versus the number of randomly used mosaics for the single-visit (**Left**) and multi-visit (**Right**) drizzled products of ACSWFC F606W, ACSWFC F814W, and WFC3IR F160W. Each point represents the mean of 100 different partial IGL measurements that resulted from the corresponding number of randomly selected mosaics, and the error bars illustrate the standard deviations ($1\sigma$) about those means.

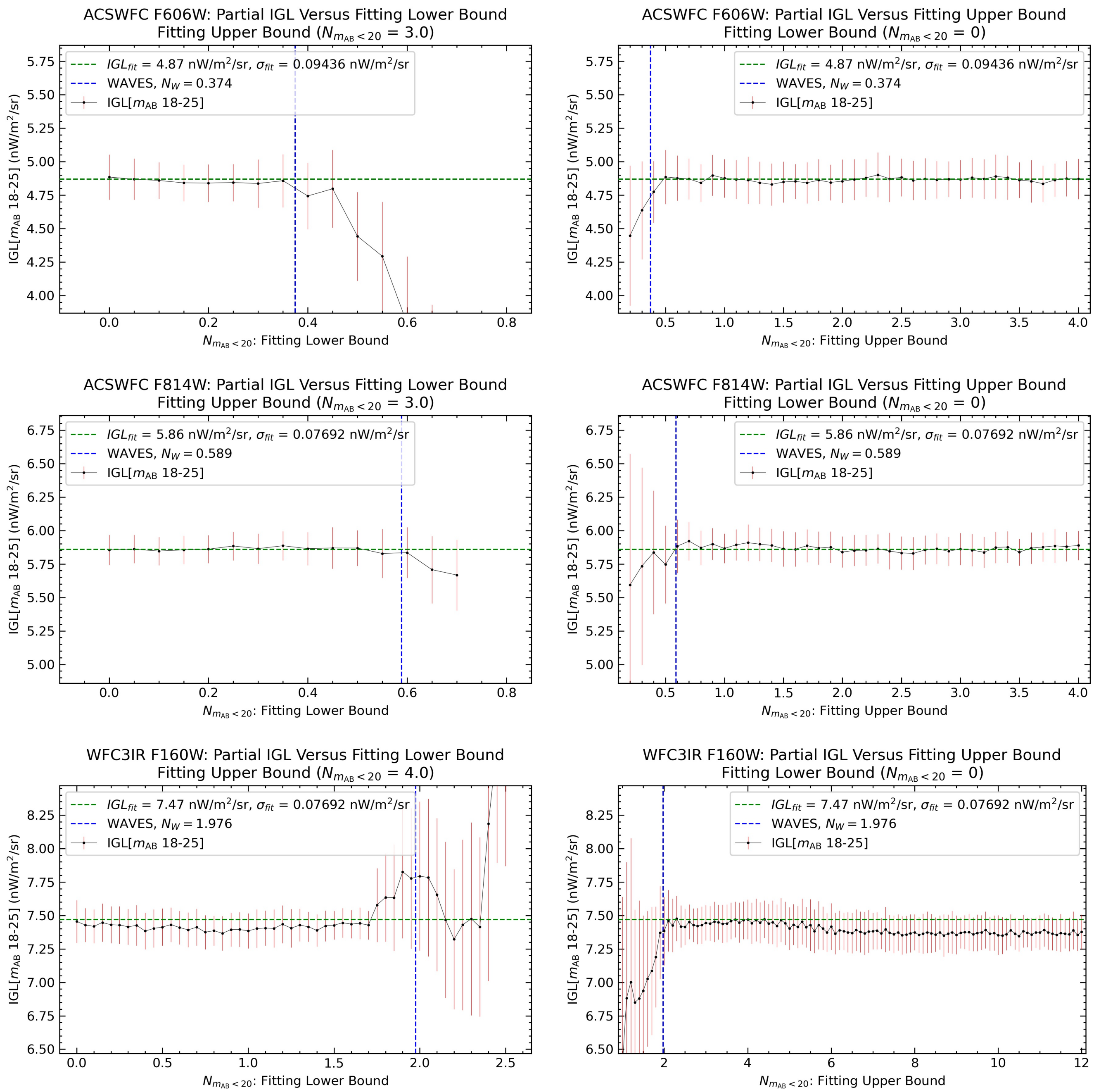

**Figure 20.** Partial IGL measurements versus different $N_{m_{\rm AB}<20}$ fitting lower bounds (**Left**) and fitting upper bounds (**Right**) for the single-visit datasets of ACSWFC F606W, ACSWFC F814W, and WFC3IR F160W. Each point represents the mean of 100 different partial IGL measurements that resulted from 100 randomly selected mosaics sampled from the region between the corresponding fitting lower bound and fitting upper bound, and the error bars illustrate the standard deviations ($1\sigma$) about those means. For the partial IGL versus fitting lower bound plots, the fitting upper bound was held constant at $N_{m_{\rm AB}<20} = 2$ for ACSWFC and $N_{m_{\rm AB}<20} = 4.0$ for WFC3IR while the lower bound was shifted incrementally. Similarly, for the partial IGL versus fitting upper bound plots, the fitting lower bound was held constant at $N_{m_{\rm AB}<20} = 0$ while the upper bound was shifted incrementally.

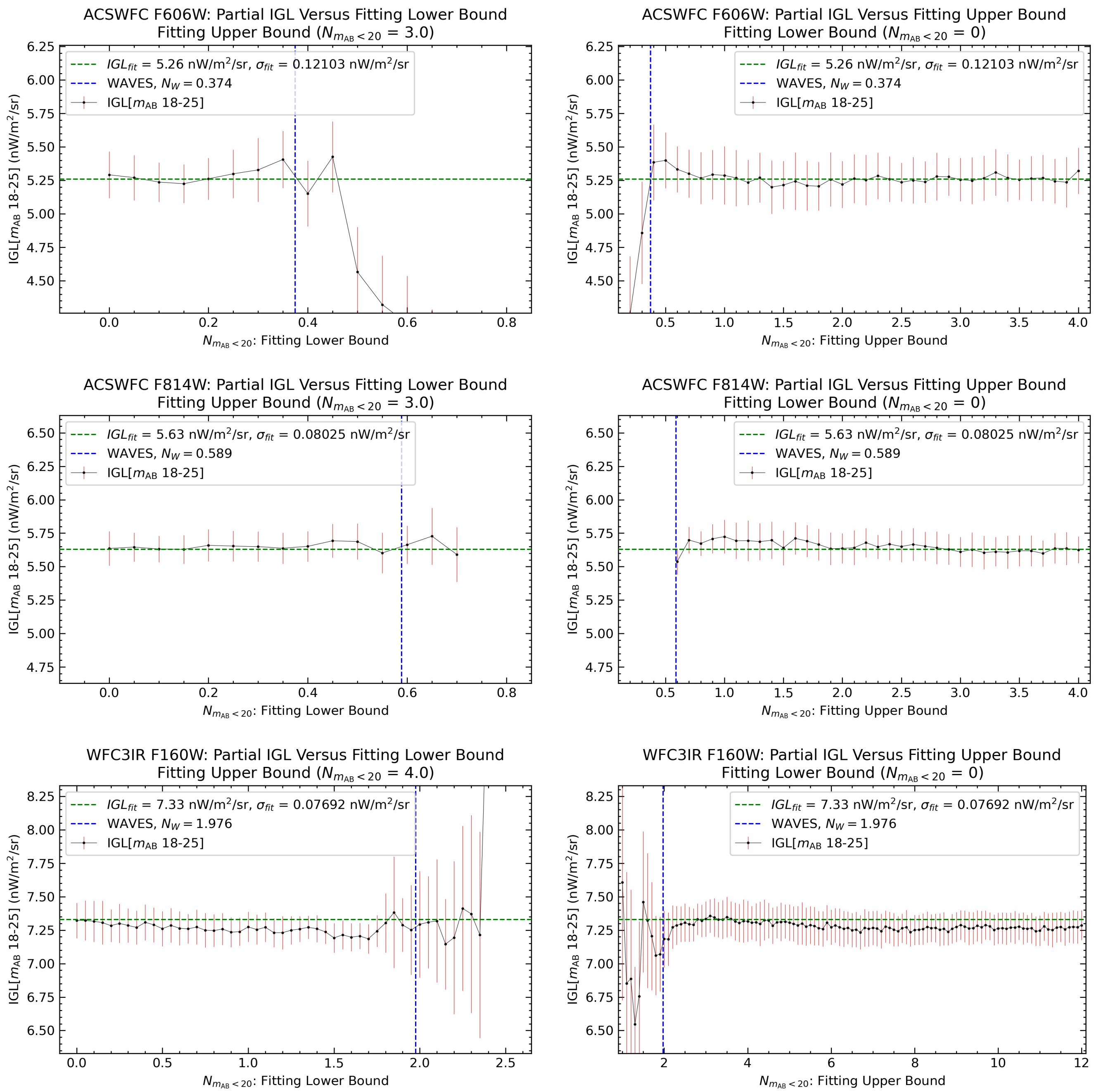

**Figure 21.** Partial IGL measurements versus different $N_{m_{\rm AB}<20}$ fitting lower bounds (**Left**) and fitting upper bounds (**Right**) for the multi-visit datasets of ACSWFC F606W, ACSWFC F814W, and WFC3IR F160W. Each point represents the mean of 100 different partial IGL measurements that resulted from 100 randomly selected mosaics sampled from the region between the corresponding fitting lower bound and fitting upper bound, and the error bars illustrate the standard deviations ($1\sigma$) about those means. For the partial IGL versus fitting lower bound plots, the fitting upper bound was held constant at $N_{m_{\rm AB}<20} = 2$ for ACSWFC and $N_{m_{\rm AB}<20} = 4.0$ for WFC3IR while the lower bound was shifted incrementally. Similarly, for the partial IGL versus fitting upper bound plots, the fitting lower bound was held constant at $N_{m_{\rm AB}<20} = 0$ while the upper bound was shifted incrementally.

## G. PARTIAL IGL FITTING MEASUREMENTS VS IMAGE FILTERING CRITERIA (SINGLE-VISIT)

Here we present the single-visit ACSWFC F435W, ACSWFC F606W, WFC3UVIS F606W, and WFC3IR F160W partial IGL vs sky sub-region cut and partial IGL vs ProFound sky chi-squared cut plots discussed in Section 4.4.

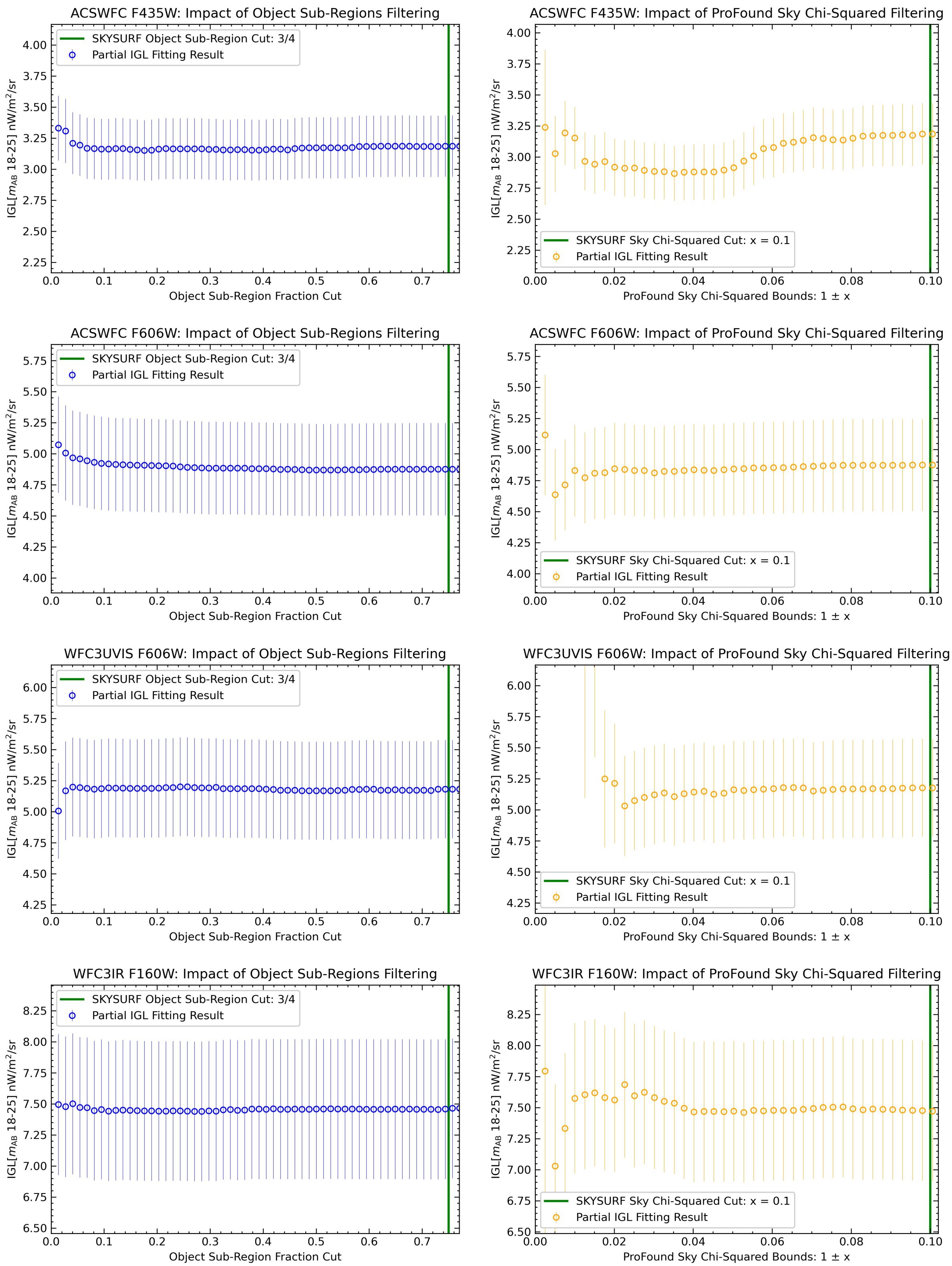

## H. TABLE OF STAR-GALAXY SEPARATION PARAMETERS

Here we present a table of the final star-galaxy separation parameters determined for each filter for the single-visit and multi-visit data, as discussed in Section 3.2.

**Table 4.** Star-galaxy separation parameters used for each instrument+filter (SV = Single-Visit, MV = Multi-Visit).

| Instrument Filter | $h_{sat}$ | $h_{faint}$ | $v_{spurious}$ | $v_{locus}$ | $m_{sat}$ | $b_{sat}$ | Spurious | Stars | Galaxies | Total |
|---|---|---|---|---|---|---|---|---|---|---|
| SV ACSWFC F435W | 20.0 | 26.10 | -1.299 | -1.157 | -3.083 | 17.468 | 134253 | 50606 | 898709 | 1083568 |
| SV ACSWFC F475W | 20.0 | 26.50 | -1.296 | -1.154 | -3.275 | 17.612 | 150410 | 64298 | 742633 | 957341 |
| SV ACSWFC F555W | 20.0 | 26.10 | -1.292 | -1.154 | -2.895 | 17.258 | 47484 | 18822 | 308140 | 374446 |
| SV ACSWFC F606W | 21.0 | 26.70 | -1.298 | -1.146 | -6.133 | 17.200 | 516370 | 243752 | 4502484 | 5262606 |
| SV ACSWFC F625W | 20.0 | 25.90 | -1.292 | -1.148 | -5.876 | 16.386 | 74325 | 21673 | 363779 | 459777 |
| SV ACSWFC F775W | 20.0 | 25.70 | -1.296 | -1.140 | -2.767 | 17.193 | 424814 | 123650 | 2094273 | 2642737 |
| SV ACSWFC F814W | 20.5 | 26.10 | -1.273 | -1.116 | -3.904 | 17.015 | 755252 | 329254 | 7020225 | 8104731 |
| SV ACSWFC F850LP | 18.5 | 24.90 | -1.236 | -1.079 | -5.094 | 15.775 | 504086 | 85012 | 1713303 | 2302401 |
| SV WFC3IR F098M | 18.5 | 25.50 | -1.053 | -0.939 | -1.500 | 17.724 | 30425 | 2116 | 198495 | 231036 |
| SV WFC3IR F105W | 19.0 | 25.10 | -1.044 | -0.923 | -7.619 | 14.071 | 160363 | 15975 | 1491772 | 1668110 |
| SV WFC3IR F110W | 19.0 | 25.90 | -1.025 | -0.912 | -8.333 | 14.315 | 154779 | 29428 | 1309019 | 1493226 |
| SV WFC3IR F125W | 18.5 | 25.30 | -1.019 | -0.902 | -23.333 | 3.382 | 277104 | 18071 | 1592319 | 1887494 |
| SV WFC3IR F140W | 18.5 | 25.30 | -0.979 | -0.889 | -15.000 | 9.198 | 121143 | 12142 | 773219 | 906504 |
| SV WFC3IR F160W | 18.5 | 25.10 | -0.975 | -0.871 | -8.605 | 14.142 | 689153 | 35845 | 2660699 | 3385697 |
| SV WFC3UVIS F336W | 19.0 | 24.90 | -1.318 | -1.190 | -0.800 | 17.940 | 30973 | 26007 | 142778 | 199758 |
| SV WFC3UVIS F390W | 20.0 | 25.75 | -1.359 | -1.205 | -5.000 | 17.139 | 11161 | 5261 | 100124 | 116546 |
| SV WFC3UVIS F438W | 20.0 | 25.75 | -1.329 | -1.201 | -2.000 | 17.877 | 4628 | 5976 | 54194 | 64798 |
| SV WFC3UVIS F475W | 20.0 | 26.30 | -1.313 | -1.180 | -2.621 | 17.463 | 66136 | 16009 | 193785 | 275930 |
| SV WFC3UVIS F475X | 20.0 | 25.75 | -1.264 | -1.182 | -2.500 | 17.581 | 8287 | 3321 | 59105 | 70713 |
| SV WFC3UVIS F555W | 20.0 | 23.10 | -1.232 | -4.285 | -5.600 | 14.788 | 35101 | 2284 | 220673 | 258058 |
| SV WFC3UVIS F606W | 20.0 | 26.50 | -1.305 | -1.176 | -4.339 | 17.232 | 73944 | 34715 | 814302 | 922961 |
| SV WFC3UVIS F625W | 20.0 | 25.75 | -1.291 | -1.180 | -5.000 | 17.292 | 2468 | 2878 | 36492 | 41838 |
| SV WFC3UVIS F775W | 20.0 | 24.75 | -1.319 | -1.172 | -15.000 | 6.831 | 5408 | 1473 | 20468 | 27349 |
| SV WFC3UVIS F814W | 20.0 | 25.30 | -1.285 | -1.155 | -8.000 | 13.917 | 202685 | 50478 | 735740 | 988903 |
| SV WFC3UVIS F850LP | 18.0 | 23.75 | -1.221 | -1.117 | -5.000 | 15.375 | 5524 | 1941 | 8181 | 15646 |
| MV ACSWFC F435W | 20.0 | 25.10 | -1.263 | -1.157 | -2.870 | 17.075 | 43829 | 7871 | 180400 | 232100 |
| MV ACSWFC F475W | 20.0 | 26.90 | -1.298 | -1.151 | -3.094 | 17.318 | 29629 | 16943 | 218428 | 265000 |
| MV ACSWFC F555W | 20.0 | 25.75 | -1.263 | -1.158 | -5.000 | 16.389 | 10920 | 3820 | 90684 | 105424 |
| MV ACSWFC F606W | 21.0 | 27.30 | -1.271 | -1.137 | -3.955 | 17.221 | 29691 | 50114 | 837154 | 916959 |
| MV ACSWFC F625W | 20.0 | 25.90 | -1.295 | -1.148 | -2.000 | 17.568 | 38017 | 13287 | 210582 | 261886 |
| MV ACSWFC F775W | 20.0 | 25.90 | -1.289 | -1.139 | -3.010 | 17.031 | 73594 | 36265 | 673081 | 782940 |
| MV ACSWFC F814W | 20.5 | 26.90 | -1.247 | -1.108 | -6.000 | 14.960 | 5543 | 25150 | 512505 | 543198 |
| MV ACSWFC F850LP | 18.5 | 25.10 | -1.217 | -1.078 | -3.000 | 16.843 | 43346 | 11955 | 253898 | 309199 |
| MV WFC3IR F098M | 18.5 | 25.10 | -1.038 | -0.928 | -8.000 | 13.074 | 33914 | 1957 | 195748 | 231619 |
| MV WFC3IR F105W | 19.0 | 25.30 | -1.042 | -0.920 | -7.000 | 14.727 | 91150 | 9857 | 863415 | 964422 |
| MV WFC3IR F110W | 19.0 | 25.50 | -0.997 | -0.908 | -15.000 | 9.211 | 114649 | 16823 | 944076 | 1075548 |
| MV WFC3IR F125W | 18.5 | 25.50 | -1.020 | -0.899 | -11.667 | 11.725 | 117674 | 13554 | 886582 | 1017810 |
| MV WFC3IR F140W | 18.5 | 25.50 | -1.003 | -0.886 | -3.462 | 16.555 | 113204 | 15642 | 830903 | 959749 |
| MV WFC3IR F160W | 18.5 | 25.50 | -0.974 | -0.870 | -7.381 | 14.346 | 272221 | 29021 | 1458220 | 1759462 |
| MV WFC3UVIS F336W | 19.0 | 24.90 | -1.318 | -1.190 | -0.800 | 17.940 | 1830 | 1649 | 13628 | 17107 |
| MV WFC3UVIS F390W | 20.0 | 25.75 | -1.359 | -1.205 | -5.000 | 17.139 | 1844 | 1726 | 59520 | 63090 |
| MV WFC3UVIS F438W | 20.0 | 25.75 | -1.329 | -1.201 | -2.000 | 17.877 | 1801 | 1407 | 19767 | 22975 |
| MV WFC3UVIS F475W | 20.0 | 26.30 | -1.313 | -1.180 | -2.621 | 17.463 | 9611 | 3385 | 65163 | 78159 |
| MV WFC3UVIS F475X | 20.0 | 25.75 | -1.264 | -1.182 | -2.500 | 17.581 | 7083 | 1550 | 31614 | 40247 |
| MV WFC3UVIS F555W | 20.0 | 23.10 | -1.232 | -4.285 | -5.600 | 14.788 | 6336 | 371 | 17208 | 23915 |
| MV WFC3UVIS F606W | 20.0 | 26.50 | -1.305 | -1.176 | -4.339 | 17.232 | 16609 | 11057 | 302144 | 329810 |
| MV WFC3UVIS F625W | 20.0 | 25.75 | -1.291 | -1.180 | -5.000 | 17.292 | 1527 | 851 | 15105 | 17483 |
| MV WFC3UVIS F775W | 20.0 | 24.75 | -1.319 | -1.172 | -15.000 | 6.831 | 4388 | 1155 | 16805 | 22348 |
| MV WFC3UVIS F814W | 20.0 | 25.30 | -1.285 | -1.155 | -8.000 | 13.917 | 36592 | 14611 | 258886 | 310089 |
| MV WFC3UVIS F850LP | 18.0 | 23.75 | -1.221 | -1.117 | -5.000 | 15.375 | 1338 | 354 | 3038 | 4730 |

## I. TABLE OF PARTIAL IGL MEASUREMENTS AND FITTING PARAMETERS

Here we present a table of our final partial IGL measurements and corresponding uncertainties for each filter for the single-visit and multi-visit data, as discussed in Section 4.

**Table 5.** Partial IGL measurements and fitting parameters for each SKYSURF filter obtained using the methods described in Section 4.1 (SV = Single-Visit, MV = Multi-Visit).

| Instrument Filter | $\lambda_{eff}$ | $IGL_i$ | $IGL_{fit}$ | $a_{fit}$ | $b_{fit}$ | $c_{fit}$ | $\sigma_{fit}$ | $\sigma_{nofit}$ | $E_{fit}$ | $E_{fit\%}$ | $N_W$ | $N_{WE}$ |
|---|---|---|---|---|---|---|---|---|---|---|---|---|
| SV ACSWFC F435W | 0.4342 | 3.78 | 3.19 | 0.791 | 1.893 | 0.319 | 0.127 | 0.220 | 0.2471 | 7.74 | 0.141 | 0.0065 |
| SV ACSWFC F475W | 0.4709 | 5.13 | 3.19 | 1.100 | 0.743 | 0.395 | 0.121 | 0.194 | 0.2438 | 7.64 | 0.141 | 0.0065 |
| SV ACSWFC F555W | 0.5332 | 5.75 | 3.45 | 0.791 | 1.347 | 0.402 | 0.124 | 0.220 | 0.2708 | 7.84 | 0.141 | 0.0065 |
| SV ACSWFC F606W | 0.5809 | 5.51 | 4.87 | 1.112 | 0.626 | 0.456 | 0.094 | 0.198 | 0.3714 | 7.62 | 0.374 | 0.0111 |
| SV ACSWFC F625W | 0.6266 | 6.79 | 4.81 | 1.130 | 0.521 | 0.482 | 0.080 | 0.199 | 0.3686 | 7.66 | 0.374 | 0.0111 |
| SV ACSWFC F775W | 0.7652 | 6.21 | 5.71 | 1.018 | 0.439 | 0.525 | 0.077 | 0.150 | 0.4337 | 7.60 | 0.589 | 0.0159 |
| SV ACSWFC F814W | 0.7973 | 7.50 | 5.86 | 1.135 | 0.367 | 0.547 | 0.077 | 0.159 | 0.4442 | 7.58 | 0.589 | 0.0159 |
| SV ACSWFC F850LP | 0.9005 | 7.13 | 6.28 | 1.049 | 0.288 | 0.584 | 0.077 | 0.123 | 0.4759 | 7.57 | 0.795 | 0.0205 |
| SV WFC3IR F098M | 0.9827 | 7.82 | 6.38 | 1.830 | 0.169 | 0.575 | 0.077 | 0.171 | 0.4968 | 7.78 | 0.795 | 0.0205 |
| SV WFC3IR F105W | 1.0431 | 12.84 | 7.24 | 1.023 | 0.304 | 0.574 | 0.077 | 0.207 | 0.5516 | 7.62 | 1.077 | 0.0257 |
| SV WFC3IR F110W | 1.1201 | 10.82 | 6.83 | 1.003 | 0.280 | 0.573 | 0.077 | 0.179 | 0.5194 | 7.60 | 1.077 | 0.0257 |
| SV WFC3IR F125W | 1.2364 | 10.29 | 6.62 | 0.865 | 0.317 | 0.535 | 0.077 | 0.143 | 0.5023 | 7.59 | 1.268 | 0.0279 |
| SV WFC3IR F140W | 1.3735 | 11.62 | 6.35 | 0.862 | 0.294 | 0.534 | 0.080 | 0.143 | 0.4831 | 7.61 | 1.268 | 0.0279 |
| SV WFC3IR F160W | 1.5278 | 8.63 | 7.47 | 0.642 | 0.444 | 0.498 | 0.077 | 0.122 | 0.5638 | 7.55 | 1.976 | 0.0348 |
| SV WFC3UVIS F336W | 0.3359 | 13.30 | 1.79 | 1.453 | 1.567 | 0.185 | 0.206 | 0.467 | 0.1420 | 7.91 | 0.031 | 0.0030 |
| SV WFC3UVIS F390W | 0.4022 | 3.33 | 2.43 | 3.715 | 0.246 | 0.358 | 0.143 | 0.203 | 0.1918 | 7.89 | 0.031 | 0.0030 |
| SV WFC3UVIS F438W | 0.4323 | 10.56 | 3.29 | 1.292 | 0.988 | 0.350 | 0.169 | 0.390 | 0.2842 | 8.64 | 0.141 | 0.0065 |
| SV WFC3UVIS F475W | 0.4732 | 4.76 | 3.44 | 1.097 | 0.610 | 0.447 | 0.108 | 0.191 | 0.2666 | 7.74 | 0.141 | 0.0065 |
| SV WFC3UVIS F475X | 0.4856 | 5.64 | 3.88 | 61.492 | 0.006 | 0.536 | 0.126 | 0.194 | 0.3235 | 8.33 | 0.141 | 0.0065 |
| SV WFC3UVIS F555W | 0.5236 | 14.21 | 5.39 | 1.229 | 0.559 | 0.499 | 0.148 | 0.293 | 0.4353 | 8.08 | 0.374 | 0.0111 |
| SV WFC3UVIS F606W | 0.5782 | 5.85 | 5.18 | 1.044 | 0.521 | 0.530 | 0.097 | 0.177 | 0.3949 | 7.62 | 0.374 | 0.0111 |
| SV WFC3UVIS F625W | 0.6188 | 36.09 | 4.71 | 1.255 | 0.535 | 0.445 | 0.077 | 0.257 | 0.3806 | 8.08 | 0.374 | 0.0111 |
| SV WFC3UVIS F775W | 0.7613 | 7.92 | 5.51 | 1.469 | 0.231 | 0.554 | 0.119 | 0.181 | 0.4876 | 8.85 | 0.589 | 0.0159 |
| SV WFC3UVIS F814W | 0.7964 | 7.57 | 6.08 | 1.230 | 0.263 | 0.607 | 0.091 | 0.153 | 0.4609 | 7.58 | 0.589 | 0.0159 |
| SV WFC3UVIS F850LP | 0.9154 | 9.98 | 5.49 | 1.453 | 0.203 | 0.523 | 0.077 | 0.155 | 0.4429 | 8.07 | 0.795 | 0.0205 |
| MV ACSWFC F435W | 0.4342 | 3.68 | 3.21 | 1.213 | 0.872 | 0.366 | 0.130 | 0.215 | 0.2526 | 7.88 | 0.141 | 0.0065 |
| MV ACSWFC F475W | 0.4709 | 3.92 | 3.27 | 1.038 | 0.931 | 0.387 | 0.119 | 0.199 | 0.2558 | 7.82 | 0.141 | 0.0065 |
| MV ACSWFC F555W | 0.5332 | 5.55 | 3.72 | 0.850 | 0.979 | 0.461 | 0.116 | 0.201 | 0.3034 | 8.16 | 0.141 | 0.0065 |
| MV ACSWFC F606W | 0.5809 | 6.64 | 5.26 | 0.955 | 0.743 | 0.489 | 0.121 | 0.237 | 0.4047 | 7.70 | 0.374 | 0.0111 |
| MV ACSWFC F625W | 0.6266 | 6.67 | 4.85 | 0.928 | 0.784 | 0.450 | 0.081 | 0.196 | 0.3750 | 7.73 | 0.374 | 0.0111 |
| MV ACSWFC F775W | 0.7652 | 6.27 | 5.68 | 1.309 | 0.307 | 0.538 | 0.077 | 0.161 | 0.4325 | 7.61 | 0.589 | 0.0159 |
| MV ACSWFC F814W | 0.7973 | 8.24 | 5.63 | 0.955 | 0.457 | 0.525 | 0.080 | 0.186 | 0.4306 | 7.65 | 0.589 | 0.0159 |
| MV ACSWFC F850LP | 0.9005 | 8.23 | 6.10 | 1.240 | 0.243 | 0.568 | 0.077 | 0.157 | 0.4670 | 7.65 | 0.795 | 0.0205 |
| MV WFC3IR F098M | 0.9827 | 7.08 | 6.04 | 1.188 | 0.246 | 0.570 | 0.088 | 0.151 | 0.4718 | 7.82 | 0.795 | 0.0205 |
| MV WFC3IR F105W | 1.0431 | 10.12 | 6.94 | 0.975 | 0.354 | 0.532 | 0.090 | 0.194 | 0.5332 | 7.69 | 1.077 | 0.0257 |
| MV WFC3IR F110W | 1.1201 | 11.22 | 6.91 | 0.979 | 0.297 | 0.571 | 0.080 | 0.185 | 0.5266 | 7.62 | 1.077 | 0.0257 |
| MV WFC3IR F125W | 1.2364 | 8.06 | 6.58 | 0.861 | 0.312 | 0.537 | 0.077 | 0.150 | 0.5012 | 7.62 | 1.268 | 0.0279 |
| MV WFC3IR F140W | 1.3735 | 9.56 | 6.28 | 0.962 | 0.247 | 0.539 | 0.077 | 0.149 | 0.4783 | 7.61 | 1.268 | 0.0279 |
| MV WFC3IR F160W | 1.5278 | 8.26 | 7.33 | 0.743 | 0.307 | 0.527 | 0.077 | 0.123 | 0.5545 | 7.57 | 1.976 | 0.0348 |
| MV WFC3UVIS F336W | 0.3359 | 2.97 | 1.41 | 0.964 | 3.142 | 0.060 | 0.161 | 0.352 | 0.1193 | 8.45 | 0.031 | 0.0030 |
| MV WFC3UVIS F390W | 0.4022 | 2.97 | 2.35 | 0.873 | 1.091 | 0.343 | 0.146 | 0.182 | 0.1913 | 8.13 | 0.031 | 0.0030 |
| MV WFC3UVIS F438W | 0.4323 | 2.77 | 2.71 | 0.746 | 1.271 | 0.310 | 0.155 | 0.178 | 0.2522 | 9.32 | 0.141 | 0.0065 |
| MV WFC3UVIS F475W | 0.4732 | 3.96 | 3.42 | 0.440 | 2.419 | 0.407 | 0.124 | 0.168 | 0.2827 | 8.26 | 0.141 | 0.0065 |
| MV WFC3UVIS F475X | 0.4856 | 5.53 | 3.91 | 0.881 | 0.832 | 0.494 | 0.103 | 0.194 | 0.3281 | 8.40 | 0.141 | 0.0065 |
| MV WFC3UVIS F555W | 0.5236 | 6.42 | 4.87 | 0.926 | 0.708 | 0.472 | 0.077 | 0.213 | 0.4079 | 8.37 | 0.374 | 0.0111 |
| MV WFC3UVIS F606W | 0.5782 | 5.49 | 5.18 | 0.789 | 0.871 | 0.495 | 0.102 | 0.189 | 0.3990 | 7.71 | 0.374 | 0.0111 |
| MV WFC3UVIS F625W | 0.6188 | 5.41 | 4.72 | 0.677 | 1.564 | 0.374 | 0.077 | 0.169 | 0.4017 | 8.51 | 0.374 | 0.0111 |
| MV WFC3UVIS F775W | 0.7613 | 8.33 | 5.64 | 0.865 | 0.608 | 0.491 | 0.157 | 0.205 | 0.5756 | 10.21 | 0.589 | 0.0159 |
| MV WFC3UVIS F814W | 0.7964 | 6.98 | 5.80 | 0.823 | 0.554 | 0.534 | 0.087 | 0.164 | 0.4433 | 7.65 | 0.589 | 0.0159 |
| MV WFC3UVIS F850LP | 0.9154 | 8.09 | 6.28 | 0.777 | 1.062 | 0.355 | 0.093 | 0.169 | 0.6095 | 9.70 | 0.795 | 0.0205 |

## J. TABLE OF EXTRAPOLATED IGL MEASUREMENTS

Here we present a table of our final extrapolated IGL estimates and corresponding uncertainties for each filter for the single-visit and multi-visit data, as discussed in Section 4.6.

**Table 6.** Extrapolated (total) IGL Measurements and error measurements for each SKYSURF filter obtained using results from Tompkins et al. (2025).

| Instrument Filter | $\lambda_{eff}$ | $IGL_e$ | $ERR_e$ | $ERR_{e\%}$ | $IGL_{TP}$ | $ERR_{TP}$ | $IGL_T$ | $ERR_T$ | $ERR_{T\%}$ | $IGL_{Ratio}$ |
|---|---|---|---|---|---|---|---|---|---|---|
| SV ACSWFC F435W | 0.4342 | 5.85 | 0.505 | 8.63 | 3.18 | 0.067 | 5.82 | 0.195 | 3.35 | 1.83 |
| SV ACSWFC F475W | 0.4709 | 5.62 | 0.433 | 7.70 | 3.23 | 0.015 | 5.68 | 0.051 | 0.90 | 1.76 |
| SV ACSWFC F555W | 0.5332 | 6.62 | 0.931 | 14.06 | 2.90 | 0.210 | 5.56 | 0.509 | 9.15 | 1.92 |
| SV ACSWFC F606W | 0.5809 | 8.82 | 0.760 | 8.62 | 4.92 | 0.121 | 8.90 | 0.297 | 3.34 | 1.81 |
| SV ACSWFC F625W | 0.6266 | 8.53 | 0.738 | 8.66 | 5.18 | 0.024 | 9.17 | 0.377 | 4.11 | 1.77 |
| SV ACSWFC F775W | 0.7652 | 9.71 | 0.782 | 8.05 | 5.94 | 0.048 | 10.11 | 0.257 | 2.55 | 1.70 |
| SV ACSWFC F814W | 0.7973 | 9.43 | 0.729 | 7.74 | 6.23 | 0.030 | 10.02 | 0.148 | 1.48 | 1.61 |
| SV ACSWFC F850LP | 0.9005 | 10.00 | 0.808 | 8.08 | 7.17 | 0.044 | 11.40 | 0.315 | 2.76 | 1.59 |
| SV WFC3IR F105W | 1.0431 | 11.43 | 0.956 | 8.36 | 7.45 | 0.152 | 11.77 | 0.335 | 2.85 | 1.58 |
| SV WFC3IR F110W | 1.1201 | 10.51 | 1.001 | 9.52 | 7.50 | 0.276 | 11.55 | 0.525 | 4.55 | 1.54 |
| SV WFC3IR F125W | 1.2364 | 10.23 | 0.806 | 7.88 | 7.33 | 0.091 | 11.33 | 0.195 | 1.72 | 1.55 |
| SV WFC3IR F140W | 1.3735 | 9.20 | 0.874 | 9.49 | 7.03 | 0.066 | 10.19 | 0.571 | 5.60 | 1.45 |
| SV WFC3IR F160W | 1.5278 | 11.54 | 0.886 | 7.68 | 7.43 | 0.062 | 11.48 | 0.130 | 1.13 | 1.55 |
| SV WFC3UVIS F336W | 0.3359 | 3.24 | 0.318 | 9.82 | 2.15 | 0.087 | 3.89 | 0.170 | 4.36 | 1.81 |
| SV WFC3UVIS F390W | 0.4022 | 4.97 | 0.622 | 12.52 | 2.16 | 0.044 | 4.41 | 0.419 | 9.51 | 2.04 |
| SV WFC3UVIS F438W | 0.4323 | 6.03 | 0.570 | 9.46 | 3.18 | 0.067 | 5.82 | 0.195 | 3.35 | 1.83 |
| SV WFC3UVIS F475W | 0.4732 | 6.06 | 0.473 | 7.81 | 3.23 | 0.015 | 5.68 | 0.051 | 0.90 | 1.76 |
| SV WFC3UVIS F555W | 0.5236 | 10.33 | 1.449 | 14.03 | 2.90 | 0.210 | 5.56 | 0.509 | 9.15 | 1.92 |
| SV WFC3UVIS F606W | 0.5782 | 9.38 | 0.811 | 8.64 | 4.92 | 0.121 | 8.90 | 0.297 | 3.34 | 1.81 |
| SV WFC3UVIS F625W | 0.6188 | 8.34 | 0.755 | 9.05 | 5.18 | 0.024 | 9.17 | 0.377 | 4.11 | 1.77 |
| SV WFC3UVIS F775W | 0.7613 | 9.37 | 0.866 | 9.25 | 5.94 | 0.048 | 10.11 | 0.257 | 2.55 | 1.70 |
| SV WFC3UVIS F814W | 0.7964 | 9.79 | 0.757 | 7.74 | 6.23 | 0.030 | 10.02 | 0.148 | 1.48 | 1.61 |
| SV WFC3UVIS F850LP | 0.9154 | 8.73 | 0.745 | 8.53 | 7.17 | 0.044 | 11.40 | 0.315 | 2.76 | 1.59 |
| MV ACSWFC F435W | 0.4342 | 5.88 | 0.514 | 8.75 | 3.18 | 0.067 | 5.82 | 0.195 | 3.35 | 1.83 |
| MV ACSWFC F475W | 0.4709 | 5.76 | 0.454 | 7.88 | 3.23 | 0.015 | 5.68 | 0.051 | 0.90 | 1.76 |
| MV ACSWFC F555W | 0.5332 | 7.12 | 1.015 | 14.25 | 2.90 | 0.210 | 5.56 | 0.509 | 9.15 | 1.92 |
| MV ACSWFC F606W | 0.5809 | 9.51 | 0.827 | 8.70 | 4.92 | 0.121 | 8.90 | 0.297 | 3.34 | 1.81 |
| MV ACSWFC F625W | 0.6266 | 8.60 | 0.752 | 8.74 | 5.18 | 0.024 | 9.17 | 0.377 | 4.11 | 1.77 |
| MV ACSWFC F775W | 0.7652 | 9.66 | 0.779 | 8.06 | 5.94 | 0.048 | 10.11 | 0.257 | 2.55 | 1.70 |
| MV ACSWFC F814W | 0.7973 | 9.05 | 0.706 | 7.80 | 6.23 | 0.030 | 10.02 | 0.148 | 1.48 | 1.61 |
| MV ACSWFC F850LP | 0.9005 | 9.71 | 0.792 | 8.15 | 7.17 | 0.044 | 11.40 | 0.315 | 2.76 | 1.59 |
| MV WFC3IR F105W | 1.0431 | 10.96 | 0.924 | 8.43 | 7.45 | 0.152 | 11.77 | 0.335 | 2.85 | 1.58 |
| MV WFC3IR F110W | 1.1201 | 10.63 | 1.018 | 9.57 | 7.50 | 0.276 | 11.55 | 0.525 | 4.55 | 1.54 |
| MV WFC3IR F125W | 1.2364 | 10.18 | 0.805 | 7.91 | 7.33 | 0.091 | 11.33 | 0.195 | 1.72 | 1.55 |
| MV WFC3IR F140W | 1.3735 | 9.11 | 0.864 | 9.48 | 7.03 | 0.066 | 10.19 | 0.571 | 5.60 | 1.45 |
| MV WFC3IR F160W | 1.5278 | 11.33 | 0.872 | 7.70 | 7.43 | 0.062 | 11.48 | 0.130 | 1.13 | 1.55 |
| MV WFC3UVIS F336W | 0.3359 | 2.55 | 0.259 | 10.16 | 2.15 | 0.087 | 3.89 | 0.170 | 4.36 | 1.81 |
| MV WFC3UVIS F390W | 0.4022 | 4.80 | 0.609 | 12.67 | 2.16 | 0.044 | 4.41 | 0.419 | 9.51 | 2.04 |
| MV WFC3UVIS F438W | 0.4323 | 4.96 | 0.502 | 10.13 | 3.18 | 0.067 | 5.82 | 0.195 | 3.35 | 1.83 |
| MV WFC3UVIS F475W | 0.4732 | 6.02 | 0.501 | 8.32 | 3.23 | 0.015 | 5.68 | 0.051 | 0.90 | 1.76 |
| MV WFC3UVIS F555W | 0.5236 | 9.34 | 1.342 | 14.37 | 2.90 | 0.210 | 5.56 | 0.509 | 9.15 | 1.92 |
| MV WFC3UVIS F606W | 0.5782 | 9.37 | 0.817 | 8.73 | 4.92 | 0.121 | 8.90 | 0.297 | 3.34 | 1.81 |
| MV WFC3UVIS F625W | 0.6188 | 8.36 | 0.791 | 9.46 | 5.18 | 0.024 | 9.17 | 0.377 | 4.11 | 1.77 |
| MV WFC3UVIS F775W | 0.7613 | 9.59 | 1.012 | 10.55 | 5.94 | 0.048 | 10.11 | 0.257 | 2.55 | 1.70 |
| MV WFC3UVIS F814W | 0.7964 | 9.33 | 0.728 | 7.80 | 6.23 | 0.030 | 10.02 | 0.148 | 1.48 | 1.61 |
| MV WFC3UVIS F850LP | 0.9154 | 10.00 | 1.010 | 10.10 | 7.17 | 0.044 | 11.40 | 0.315 | 2.76 | 1.59 |

REFERENCES

Astropy Collaboration, Robitaille, T. P., Tollerud, E. J., et al. 2013, A&A, 558, A33, doi: 10.1051/0004-6361/201322068

Astropy Collaboration, Price-Whelan, A. M., Sipőcz, B. M., et al. 2018, AJ, 156, 123, doi: 10.3847/1538-3881/aabc4f

Astropy Collaboration, Price-Whelan, A. M., Lim, P. L., et al. 2022, ApJ, 935, 167, doi: 10.3847/1538-4357/ac7c74

Baldry, I. K., Sullivan, T., Rani, R., & Turner, S. 2021, MNRAS, 500, 1557, doi: 10.1093/mnras/staa3327

Bellstedt, S., Robotham, A. S. G., Driver, S. P., et al. 2020, MNRAS, 498, 5581, doi: 10.1093/mnras/staa2620

Bernstein, R. A. 2007, ApJ, 666, 663, doi: 10.1086/519824

Bertin, E., & Arnouts, S. 1996, Astron. Astrophys. Suppl. Ser., 117, 393, doi: 10.1051/aas:1996164

Bradley, L., Sipőcz, B., Robitaille, T., et al. 2020, astropy/photutils: 1.0.1, 1.0.1, Zenodo, doi: 10.5281/zenodo.4049061

Brammer, G. B., van Dokkum, P. G., Franx, M., et al. 2012, ApJS, 200, 13, doi: 10.1088/0067-0049/200/2/13

Carleton, T., Windhorst, R. A., O'Brien, R., et al. 2022, AJ, 164, 170, doi: 10.3847/1538-3881/ac8d02

Cheng, Y.-T., & Chang, T.-C. 2022, ApJ, 925, 136, doi: 10.3847/1538-4357/ac3aee

Cheng, Y.-T., Arai, T., Bangale, P., et al. 2021, ApJ, 919, 69, doi: 10.3847/1538-4357/ac0f5b

Cooray, A. 2016, Royal Society Open Science, 3, 150555, doi: 10.1098/rsos.150555

Cooray, A., Gong, Y., Smidt, J., & Santos, M. G. 2012, ApJ, 756, 92, doi: 10.1088/0004-637X/756/1/92

Corwin, Harold G., J., Buta, R. J., & de Vaucouleurs, G. 1994, AJ, 108, 2128, doi: 10.1086/117225

Crill, B. P., Werner, M., Akeson, R., et al. 2020, in Society of Photo-Optical Instrumentation Engineers (SPIE) Conference Series, Vol. 11443, Society of Photo-Optical Instrumentation Engineers (SPIE) Conference Series, 114430I, doi: 10.1117/12.2567224

Domínguez, A., Primack, J. R., Rosario, D. J., et al. 2011, MNRAS, 410, 2556, doi: 10.1111/j.1365-2966.2010.17631.x

Driver, S. P., Andrews, S. K., Davies, L. J., et al. 2016, ApJ, 827, 108, doi: 10.3847/0004-637X/827/2/108

Driver, S. P., Andrews, S. K., da Cunha, E., et al. 2018, MNRAS, 475, 2891, doi: 10.1093/mnras/stx2728

Driver, S. P., Liske, J., Davies, L. J. M., et al. 2019, The Messenger, 175, 46, doi: 10.18727/0722-6691/5126

D'Silva, J. C. J., Driver, S. P., Lagos, C. D. P., et al. 2023, MNRAS, 524, 1448, doi: 10.1093/mnras/stad1974

Dwek, E., & Krennrich, F. 2013, Astroparticle Physics, 43, 112, doi: 10.1016/j.astropartphys.2012.09.003

Finke, J. D., Razzaque, S., & Dermer, C. D. 2010, ApJ, 712, 238, doi: 10.1088/0004-637X/712/1/238

Fruchter, A. S., & Hook, R. N. 2002, The Publications of the Astronomical Society of the Pacific, 114, 144, doi: 10.1086/338393

Giavalisco, M., Ferguson, H. C., Koekemoer, A. M., et al. 2004, ApJL, 600, L93, doi: 10.1086/379232

Goisman, Z., Carleton, T., Cohen, S. H., et al. 2025, PASP, 137, 094501, doi: 10.1088/1538-3873/adfab0

Gong, Y., Cooray, A., Mitchell-Wynne, K., et al. 2016, ApJ, 825, 104, doi: 10.3847/0004-637X/825/2/104

Gonzaga, S., Hack, W., Fruchter, A., & Mack, J. 2012, The DrizzlePac Handbook

Graham, A. W., & Driver, S. P. 2005, PASA, 22, 118, doi: 10.1071/AS05001

Gréaux, L., Biteau, J., & Nievas Rosillo, M. 2024, ApJL, 975, L18, doi: 10.3847/2041-8213/ad85c9

Hauser, M. G., & Dwek, E. 2001, ARA&A, 39, 249, doi: 10.1146/annurev.astro.39.1.249

Hill, R., Masui, K. W., & Scott, D. 2018, Applied Spectroscopy, 72, 663, doi: 10.1177/0003702818767133

Kashlinsky, A. 2005, PhR, 409, 361, doi: 10.1016/j.physrep.2004.12.005

Kelsall, T., Weiland, J. L., Franz, B. A., et al. 1998, The Astrophysical Journal, 508, 44, doi: 10.1086/306380

Khaire, V., & Srianand, R. 2015, ApJ, 805, 33, doi: 10.1088/0004-637X/805/1/33

Koekemoer, A. M., Fruchter, A. S., Hook, R. N., & Hack, W. 2003, in HST Calibration Workshop : Hubble after the Installation of the ACS and the NICMOS Cooling System, ed. S. Arribas, A. Koekemoer, & B. Whitmore, 337

Koekemoer, A. M., Faber, S. M., Ferguson, H. C., et al. 2011, ApJS, 197, 36, doi: 10.1088/0067-0049/197/2/36

Korngut, P. M., Kim, M. G., Arai, T., et al. 2022, The Astrophysical Journal, 926, 133, doi: 10.3847/1538-4357/ac44ff

Koushan, S., Driver, S. P., Bellstedt, S., et al. 2021, MNRAS, 503, 2033, doi: 10.1093/mnras/stab540

Kramer, D. M., Carleton, T., Cohen, S. H., et al. 2022, ApJL, 940, L15, doi: 10.3847/2041-8213/ac9cca

Kron, R. G. 1980, ApJS, 43, 305, doi: 10.1086/190669

Lagache, G., Puget, J.-L., & Dole, H. 2005, ARA&A, 43, 727, doi: 10.1146/annurev.astro.43.072103.150606

Landsman, W. B. 1993, in Astronomical Society of the Pacific Conference Series, Vol. 52, Astronomical Data Analysis Software and Systems II, ed. R. J. Hanisch, R. J. V. Brissenden, & J. Barnes, 246

Lauer, T. R., Postman, M., Weaver, H. A., et al. 2021, The Astrophysical Journal, 906, 77, doi: 10.3847/1538-4357/abc881

Lauer, T. R., Postman, M., Spencer, J. R., et al. 2022, ApJL, 927, L8, doi: 10.3847/2041-8213/ac573d

Laur, J., Tempel, E., Tamm, A., et al. 2022, A&A, 668, A8, doi: 10.1051/0004-6361/202243881

Madau, P., & Dickinson, M. 2014, ARA&A, 52, 415, doi: 10.1146/annurev-astro-081811-125615

Matsumoto, T., Tsumura, K., Matsuoka, Y., & Pyo, J. 2018, AJ, 156, 86, doi: 10.3847/1538-3881/aad0f0

Matsuura, S., Arai, T., Bock, J. J., et al. 2017, The Astrophysical Journal, 839, 7, doi: 10.3847/1538-4357/aa6843

McIntyre, I. A., Carleton, T., O'Brien, R., et al. 2025, AJ, 169, 136, doi: 10.3847/1538-3881/ad946b

McVittie, G. C., & Wyatt, S. P. 1959, ApJ, 130, 1, doi: 10.1086/146688

Momcheva, I. G., Brammer, G. B., van Dokkum, P. G., et al. 2016, ApJS, 225, 27, doi: 10.3847/0067-0049/225/2/27

Oke, J. B., & Gunn, J. E. 1983, ApJ, 266, 713, doi: 10.1086/160817

O'Brien, R., Carleton, T., Windhorst, R. A., et al. 2023, The Astronomical Journal, 165, 237, doi: 10.3847/1538-3881/acccee

Partridge, R. B., & Peebles, P. J. E. 1967a, ApJ, 147, 868, doi: 10.1086/149079

—. 1967b, ApJ, 148, 377, doi: 10.1086/149161

Postman, M., Lauer, T. R., Parker, J. W., et al. 2024, ApJ, 972, 95, doi: 10.3847/1538-4357/ad5ffc

Robotham, A. S. G., Bellstedt, S., Lagos, C. d. P., et al. 2020, MNRAS, 495, 905, doi: 10.1093/mnras/staa1116

Robotham, A. S. G., Davies, L. J. M., Driver, S. P., et al. 2018a, MNRAS, 476, 3137, doi: 10.1093/mnras/sty440

—. 2018b, MNRAS, 476, 3137, doi: 10.1093/mnras/sty440

Rodrigo, C., & Solano, E. 2020, in XIV.0 Scientific Meeting (virtual) of the Spanish Astronomical Society, 182

Rodrigo, C., Solano, E., & Bayo, A. 2012, SVO Filter Profile Service Version 1.0, IVOA Working Draft 15 October 2012, doi: 10.5479/ADS/bib/2012ivoa.rept.1015R

Saldana-Lopez, A., Domínguez, A., Pérez-González, P. G., et al. 2021, MNRAS, 507, 5144, doi: 10.1093/mnras/stab2393

Schlafly, E. F., & Finkbeiner, D. P. 2011, ApJ, 737, 103, doi: 10.1088/0004-637X/737/2/103

Schlegel, D. J., Finkbeiner, D. P., & Davis, M. 1998a, The Astrophysical Journal, 500, 525, doi: 10.1086/305772

—. 1998b, ApJ, 500, 525, doi: 10.1086/305772

Spergel, D., Gehrels, N., Breckinridge, J., et al. 2013, arXiv e-prints, arXiv:1305.5422, doi: 10.48550/arXiv.1305.5422

Symons, T., Zemcov, M., Cooray, A., Lisse, C., & Poppe, A. R. 2023, ApJ, 945, 45, doi: 10.3847/1538-4357/acaa37

Tempel, E., Laur, J., Jones, Z. R., et al. 2025, arXiv e-prints, arXiv:2503.24039, doi: 10.48550/arXiv.2503.24039

Tompkins, S. A., Driver, S. P., Robotham, A. S. G., et al. 2025, arXiv e-prints, arXiv:2507.03412, doi: 10.48550/arXiv.2507.03412

Virtanen, P., Gommers, R., Oliphant, T. E., et al. 2020, Nature Methods, 17, 261, doi: 10.1038/s41592-019-0686-2

Windhorst, R. A., Cohen, S. H., Hathi, N. P., et al. 2011, ApJS, 193, 27, doi: 10.1088/0067-0049/193/2/27

Windhorst, R. A., Carleton, T., O'Brien, R., et al. 2022, The Astronomical Journal, 164, 141, doi: 10.3847/1538-3881/ac82af

Windhorst, R. A., Adams, N. J., Arendt, R. G., et al. 2024, DARK-SKY: Constrain Zodiacal Light & diffuse Extragalactic Background Light from Archival JWST images, JWST Proposal. Cycle 3, ID. #4695

Wright, E. L. 1998, ApJ, 496, 1, doi: 10.1086/305345

—. 2001, ApJ, 553, 538, doi: 10.1086/320942